\documentclass[twocolumn,showpacs,preprintnumbers,amsmath,amssymb]{revtex4}
\usepackage{epsfig}
\usepackage{dcolumn}
\usepackage{bm}
\usepackage{amsthm}

\usepackage{amsmath}
\usepackage{slashed}

\newcommand{\remove}[1]{}

\newcommand{\cM}{{\cal M}}

\def\be{\begin{equation}}
\def\ee{\end{equation}}

\newcommand{\beq}{\begin{equation}}
\newcommand{\eeq}{\end{equation}}
\newcommand{\beqa}{\begin{eqnarray}}
\newcommand{\eeqa}{\end{eqnarray}}

\renewcommand{\pl}{\partial}

\newcommand{\ii}{{\rm i}}

\renewcommand{\vr}{{\bf r}}

\newcommand{\bea}{\begin{array}}
\newcommand{\ea}{\end{array}}

\begin{document}

\title{Quantum Field Theory of K-mouflage}

\author{Philippe Brax}
\affiliation{Institut de Physique Th\'eorique,\\
Universit\'e Paris-Saclay CEA, CNRS, F-91191 Gif-sur-Yvette, C\'edex, France\\}

\author{Patrick Valageas}
\affiliation{Institut de Physique Th\'eorique,\\
Universit\'e Paris-Saclay CEA, CNRS, F-91191 Gif-sur-Yvette, C\'edex, France\\}
\vspace{.2 cm}

\date{\today}
\vspace{.2 cm}

\begin{abstract}

We consider K-mouflage models, which are K-essence  theories coupled to matter. 
We analyze their quantum properties and in particular the quantum corrections 
to the classical Lagrangian.
We setup the renormalization program for these models and show that, contrary 
to renormalizable field theories where renormalization by infinite counterterms can be 
performed in one step, K-mouflage theories involve a recursive construction whereby 
each set of counterterms introduces new divergent quantum contributions which in turn 
must be subtracted by new counterterms. This tower of counterterms can be in principle 
constructed step by step by recursion and allows one to calculate the finite renormalized 
action of the model. In particular, it can be checked that the classical action is not 
renormalized and that the finite corrections to the renormalized action contain only 
higher-derivative operators. 
We concentrate then on the regime where calculability is ensured, i.e., when the corrections 
to the classical action are negligible. We establish an operational criterion for classicality 
and show that this is satisfied in cosmological and astrophysical situations for (healthy) 
K-mouflage models which pass the Solar System tests. These results rely on perturbation 
theory around a background and are  only valid when the background configuration is 
quantum stable. We analyze the quantum stability of astrophysical and cosmological 
backgrounds and find that models that pass the Solar System tests are quantum stable.
We then consider the possible embedding of the K-mouflage models in an Ultra-Violet 
completion. We find that the healthy models which pass the Solar System tests all violate 
the positivity constraint which would follow from the unitarity of the putative UV completion, 
implying that these healthy K-mouflage theories have no UV completion. 
We then analyze their behavior at high energy, and we find that the classicality criterion is 
satisfied in the vicinity of a high-energy collision, implying that the classical K-mouflage theory 
can be applied in this context. Moreover, the classical description becomes more accurate as 
the energy increases, in a way compatible with the classicalization concept.

\end{abstract}

 \vskip2pc

\maketitle

\section{Introduction}
\label{sec:Introduction}

Scalar-tensor theories motivated by the discovery of the acceleration of the expansion 
\cite{Riess:1998cb,Perlmutter:1998np} of the Universe suffer from severe gravitational 
problems in the Solar System \cite{Will:2001mx} unless the scalar
interaction is screened \cite{Khoury:2010xi}. 
Four mechanisms are now known, and the list seems complete for conformally coupled 
scalar fields: the chameleon \cite{Khoury:2003aq,Khoury:2003rn,Brax:2004qh}
\footnote{On quasi-linear scales ultra-local models behave as chameleon models
with large mass and coupling, although they have a different behavior on subgalactic
scales \cite{Brax:2016}.}, 
Damour-Polyakov \cite{Damour:1994zq}, Vainshtein \cite{Vainshtein:1972sx}, 
and K-mouflage mechanisms \cite{Babichev:2009ee}. 
Their classification follows from the requirement of preserving second-order equations 
of motion for the scalar field, and  they can be seen as restrictions on the second derivatives, 
first derivatives, and the value itself of the Newton potential in the presence of matter 
\cite{Khoury:2013tda,Brax:2014a}. 
All these theories involve nonlinearities in either their scalar potential or the (generalized) 
kinetic terms. Locally, Newtonian gravity is retrieved thanks to the relevant role played by 
the nonlinearities. This property is also a drawback of these models as the quantum 
corrections are not guaranteed to preserve the form of the nonlinearities required to screen 
the scalar field locally. For instance, in chameleon models, the scalar potential can be 
largely modified by quantum corrections in dense environments \cite{Upadhye:2012vh}. 
For K-mouflage and theories with the Vainshtein property like Galileons,  
the nonlinear kinetic terms become dominant when the scalar field is screened. 
This may cast a doubt on the validity of these models as nonrenormalizable operators 
play a fundamental role there. Fortunately, in the K-mouflage and Galileon cases 
nonrenormalization theorems \cite{Nicolis:2008in, deRham2014} have been obtained 
whereby the quantum corrections in a background where the scalar field is screened are 
under control. In fact, they do not affect the classical Lagrangian and only add finite 
corrections to the classical action that involve higher-order derivatives.
In this paper, we reconsider this issue and set up the renormalization program for 
K-mouflage theories. Starting from the classical Lagrangian, we construct
a first set of infinite counterterms which cancel the quantum divergences obtained in 
perturbation theory around a classical background. In renormalizable theories, this is all there 
is to do, and this yields the renormalized action. For K-mouflage, the counterterms introduce 
new vertices in the perturbative series which in turn lead to new divergences which require 
the introduction of other counterterms. The whole procedure carries on recursively. 
Using this approach, we find that indeed the classical Lagrangian is not renormalized 
because the quantum corrections depend on higher-order derivatives than the original 
Lagrangian. This whole process is unwieldy and is not guaranteed to converge. 
What we find is that there is a classicality criterion which ensures that the quantum 
corrections are negligible. We then focus on ``healthy'' K-mouflage theories  which pass the 
Solar System tests \cite{Barreira2015}, i.e., models with no ghosts and no gradient instability 
\cite{Brax:2014a} that also satisfy the stringent quantitative constraints of the
Solar System (this implies that when the argument of the kinetic function goes to
$-\infty$ the kinetic function becomes very close to linear). 
We find that the quantum corrections calculated in the cosmological or astrophysical 
backgrounds receive negligible quantum corrections. This whole approach relies on 
perturbation theory and is only valid when the background is quantum stable. 
The healthy K-mouflage theories which pass the Solar System tests show no such 
quantum instabilities.
Our approach should be compared to Ref.~\cite{deRham2014} where the nonrenormalization 
theorem was first stated. The classicality criterion that we obtain is similar to the conditions 
obtained there, which were then applied to the Dirac-Born-Infeld (DBI) models, 
although our approach is different. 
Our explicit construction of the renormalization procedure allows us to highlight how the 
renormalized action can be constructed in a step by step way.  Here we do not insist on the 
DBI models as they are either ruled out by local tests or fail to screen the effects of the 
scalar field in the early Universe. On the contrary, we insist on considering ``realistic'' 
models, which pass the Solar System tests while giving a realistic cosmology.

Even though the K-mouflage theories are not renormalizable quantum field theories, 
in the classical regime they are calculable theories with negligible quantum corrections. 
One may wonder what happens when one pushes the K-mouflage theories outside their 
classical regime at sufficiently high energy. The natural expectation would be that some kind 
of UV completion ought to be existing. It turns out that for healthy K-mouflage theories which 
pass the Solar System tests this is not the case, as they violate the positivity criterion of 
scattering amplitudes which follows from unitarity 
\cite{Adams:2006aa,Joyce:2014kja,Goon:2016ihr}. 
This implies that these K-mouflage models, contrary to the example given 
in \cite{Kaloper:2014vqa}, cannot be UV completed in a traditional way. 
This justifies our approach to treat the classical K-mouflage Lagrangian as our fundamental 
field theory that we renormalize step by step. Indeed another procedure used 
in \cite{deRham2014} would be to start from a UV completed theory at high energy 
and study the naturality of the K-mouflage action at low energy using the functional 
renormalization group \cite{Wetterich1993}. In the cases that we consider, this top-down 
approach is inoperative as the starting point of the renormalization group at high energy,
i.e., the UV completion, does not exist. All that is left is the bottom-up approach 
that we present here, where the classical Lagrangian is renormalized at low energy. 
Moreover, in the classical regime, the quantum corrections are negligible, and one can 
deduce reliable low-energy predictions from such models.

The absence of UV completion raises the obvious question of what happens when the 
theory is probed at higher and higher energies. We then consider the high-energy regime 
of the theory, and we find that at high energy in scalar collisions there is a large region 
of space around the interaction point where the classicality criterion applies and the 
K-mouflage theory behaves classically. Moreover, the classicality region grows with 
the energy in agreement with the classicalization picture \cite{Dvali:2010jz}. 
We also consider the fermion - antifermion annihilation processes via a scalar 
intermediate state, and we find that the cross section can always be calculated in the 
classical regime up to very high energy.

In section II, we recall facts about K-mouflage theories, and then in section III 
we discuss the quantum corrections of such models. 
In section IV, we consider the quantum stability of the background configurations. 
In section V, we analyze the UV completion of these theories and their behavior 
at high energy for scalar scattering.
In section VI, we study the interactions between fermions and the scalar.
We then conclude. 
We have added two Appendices on the one-loop quantum corrections and on matter loops.

\section{K-mouflage}
\label{sec:K-mouflage}

\subsection{K-mouflage action}
\label{sec:action}

We consider the K-mouflage models defined by the Einstein-frame action
\be
S= \int d^4 x \; \sqrt{-g} \left[ \frac{M_{\rm Pl}^2}{2} R + {\cal M}^4 K \right]
+ S_{\rm m}(\psi_i, A^2(\varphi) g_{\mu\nu}) .
\label{S-def}
\ee
The fields $\psi_i$ are matter fields governed by the matter action $S_{\rm m}$,
which involves the Jordan-frame metric $A^2(\varphi) g_{\mu\nu}$ that
explicitly depends on a scalar field $\varphi$ through the coupling function
$A(\varphi)$.
The scalar-field Lagrangian only depends on its rescaled kinetic term,
\be
\chi = - \frac{g^{\mu\nu}\partial_\mu \varphi \partial_\nu\varphi}{2{\cal M}^4} ,
\label{chi-def}
\ee
through the kinetic function $K(\chi)$, and ${\cal M}$ is a scale related to the dark-energy
scale.

The coupling function is chosen, for instance, as
\be
\beta > 0 : \;\;\;  A(\varphi) = e^{\beta\varphi/M_{\rm Pl}}
\simeq 1 + \frac{\beta\varphi}{M_{\rm Pl}} .
\label{A-def}
\ee
We can take $\beta>0$ without loss of generality (its sign is absorbed by the change
of sign of the scalar field). In practice, $\beta \lesssim 0.1$ and
$|\beta\varphi/M_{\rm Pl}| \lesssim 1$, so that only the first-order term of the expansion
of $A(\varphi)$ matters and the exact nonlinear form of $A(\varphi)$ does not significantly
modify the results.

Canonically normalized models are defined by
\be
K_{\rm canonical}(\chi) = \chi
\ee
and K-mouflage models involve nonlinear functions of $\chi$.
However, we consider models with the low-$\chi$ expansion
\be
\chi \rightarrow 0 : \;\;\; K(\chi) = -1 + \chi + ...
\label{low-chi}
\ee
where the dots stand for higher-order terms.
The constant term, $-1$, corresponds to the cosmological constant, giving a constant
dark-energy density at late times equal to ${\cal M}^4$. The linear term, $\chi$,
gives a canonically-normalized field in the weak-field limit, with a multiplicative factor of
$1$ that sets the normalization of the field $\varphi$.
In addition, well-behaved K-mouflage models must obey the constraints
\be
K' > 0 , \;\;\; K'+2\chi K'' > 0 ,
\label{ghosts}
\ee
which ensure that there are no ghosts nor small-scale instabilities.
Here and in the following, we note $K' = dK/d\chi$ and $K'' = d^2K/d\chi^2$.

From the action (\ref{S-def}), the scalar field obeys the nonlinear Klein-Gordon
equation
\beq
\frac{1}{\sqrt{-g}} \pl_{\mu} \left[ \sqrt{-g} K' \pl^{\mu} \varphi \right] =
- \frac{\beta}{M_{\rm Pl}} T ,
\label{KG-matter}
\eeq
where $T = T^{\mu}_{\mu}$ is the trace of the matter energy-momentum tensor.

The nonlinear kinetic terms give rise to a screening mechanism that provides
a convergence to the standard cosmology at high redshift, when $\chi \rightarrow +\infty$,
and to General Relativity on small scales, such as in the Solar System,
where $\chi \rightarrow -\infty$. This screening requires that $K' \gg 1$ in the nonlinear
regime, which suppresses the scalar-field fluctuations and the fifth force, as can be seen
from the nonlinear Klein-Gordon equation (\ref{KG-matter}).
Therefore, as field theories, these models involve an arbitrary number of
higher-derivative terms.
This may be problematic and lead to inconsistencies at the quantum level.
In the following, we study how the K-mouflage models are affected by quantum corrections
around a background value $\bar \chi$ and how the K-mouflage models respect
the unitarity of the S-matrix. These models share very similar properties with Galileon
theories for  which the arguments developed here have already been mostly applied.
Here, we consider K-mouflage models where the field-theoretic arguments are simpler
due to the form of the Lagrangian which involves only powers of $\partial_\mu \varphi$.

\subsection{Examples of kinetic functions $K(\chi)$}
\label{sec:K-chi}

\begin{figure}
\begin{center}
\epsfxsize=8.5 cm \epsfysize=5.8 cm {\epsfbox{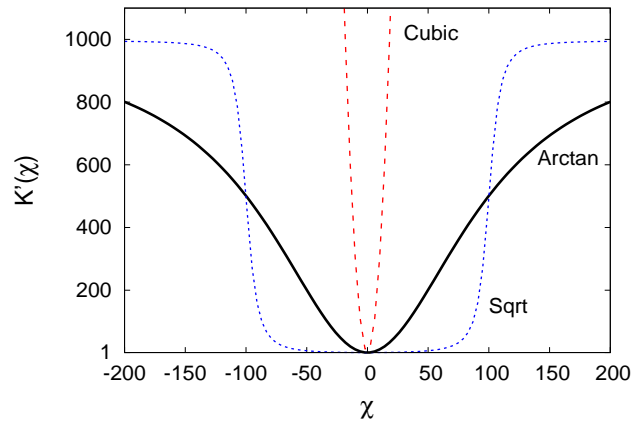}}
\end{center}
\caption{Derivative of the kinetic function, $K'(\chi)$, for the Cubic, Arctan and
Sqrt (square-root) models.}
\label{fig_Kp_chi}
\end{figure}

In the numerical computations, we consider for illustration the ``Cubic'', ``Arctan''
and ``Sqrt'' (square-root) models defined by
\beqa
\mbox{``Cubic'':} &&  K = -1 + \chi + K_0 \chi^3 , \nonumber \\
&& K'= 1+3 K_0 \chi^2 , \;\;\; \mbox{with} \;\;\; K_0 = 1 ,
\label{Cubic-def}
\eeqa
\beqa
\lefteqn{\mbox{``Arctan'':} \;\;\; K = -1 + \chi + K_* [ \chi - \chi_* \arctan(\chi/\chi_*) ] , }
\nonumber \\
&& K' = 1 + \frac{K_* \chi^2}{\chi_*^2+\chi^2} ,
\;\;\; \mbox{with} \;\;\; K_* = 10^3, \;\; \chi_*=10^2 , \;\;\;\;
\label{Arctan-def}
\eeqa
\beqa
\lefteqn{\mbox{``Sqrt'':} \;\;\; K = -1 + \left( 1 +
\frac{K_* \chi_*}{\sqrt{\sigma_*^2+\chi_*^2}} \right) \chi } \nonumber \\
&& + \frac{K_*}{2} \left( \sqrt{\sigma_*^2+(\chi-\chi_*)^2}
- \sqrt{\sigma_*^2+(\chi+\chi_*)^2} \right) ,
\nonumber \\
&& \;\;\; \mbox{with} \;\;\; K_* = 10^3, \;\; \chi_*=10^2 , \sigma_*=10 ,
\;\;\;\; \nonumber \\
&& K' = 1 + \frac{K_* \chi_*}{\sqrt{\sigma_*^2+\chi_*^2}}
+ \frac{K_* (\chi-\chi_*)}{2\sqrt{\sigma_*^2+(\chi-\chi_*)^2}}  \nonumber \\
&& - \frac{K_* (\chi+\chi_*)}{2\sqrt{\sigma_*^2+(\chi+\chi_*)^2}} .
\label{Sqrt-def}
\eeqa

They all satisfy the general constraints (\ref{ghosts}) and provide realistic cosmological
scenarios. However, only the Arctan and Sqrt models obey the Solar System
constraint associated with the measurement of the perihelion of the Moon,
which mainly requires that $\chi K'' \ll K'$ in the small-scale nonlinear limit
$\chi \rightarrow -\infty$.
This means that $K'$ is almost constant for $\chi \lesssim - 10^6$.
Then, the Cubic model should be seen as a simple phenomenological model for the
weak-field ($|\chi| \lesssim 1$) and the cosmological ($\chi > 0$) regimes, with an
unspecified tail for large and negative $\chi$.
Thus, the highly nonlinear large-$\chi$ behavior of these models satisfies:
\be
\mbox{``Cubic'':} \;\; K' \sim K'' \chi \sim K''' \chi^2 , \;\;
K^{(n)}=0 \;\; \mbox{for} \;\; n \geq 4 ,
\label{no-gap-Cubic}
\ee
and
\be
\mbox{``Arctan'' and ``Sqrt'':} \;\; K' \gg K'' \chi \sim K''' \chi^2 \sim K^{(n)} \chi^{n-1} .
\label{gap-Arctan}
\ee

Because in most analyses it is the first derivative $K'(\chi)$ that matters, we show
the behavior of $K'(\chi)$ for these models in Fig.~\ref{fig_Kp_chi}.
For the Cubic model $K'(\chi)$ goes to infinity as $|\chi| \rightarrow +\infty$, whereas
it goes to the constant $K_*=1000$ for the Arctan and Sqrt models,
with the transition from the low-$\chi$ regime (\ref{low-chi}) at $\chi_*=100$.
For the Arctan model this transition is slow, with a width of order $\chi_*$,
while for the Sqrt model it is fast, with a width of order $\sigma_*=10$.
For simplicity, the three examples (\ref{Cubic-def})-(\ref{Sqrt-def}) for $K'(\chi)$
are even functions of $\chi$, but this is not necessary, and we could consider functions
which combine two different behaviors on the semiaxis $\chi>0$ and $\chi<0$.

\section{Quantum Corrections}
\label{sec:Quantum-corrections}

\subsection{Renormalization of K-mouflage}
\label{sec:renormalization-notations}

K-mouflage theories involve nonrenormalizable interactions of higher order and need to be
understood as  effective field theories of some sort. Nonetheless, we will show in the following subsections
that, in a certain sense which will be made clearer step by step, the K-mouflage theories
satisfy some remarkable field-theoretic properties. In particular, we will show that the classical
(tree-level) Lagrangian ${\cal L}_{\rm classical}$ is not renormalized by quantum corrections but
the full renormalized effective action $\Gamma_{\rm renorm}[\varphi]$ acquires
an infinite number of finite corrections which all depend on higher derivatives of $\chi$.

More precisely, to introduce our notations, let us write the bare action
$S_{\rm bare}[\varphi]$ as
\beqa
S_{\rm bare}[\varphi] & = & \int d^4 x \, {\cal L}_{\rm bare}(\varphi) \\
& = & \int d^4 x \, \left[ {\cal L}_{\rm classical}(\varphi) + \Delta {\cal L}(\varphi) \right] ,
\label{S-renorm-def}
\eeqa
where ${\cal L}_{\rm classical}$ is the scalar-field Lagrangian given by Eq.(\ref{S-def}),
\be
{\cal L}_{\rm classical}(\varphi) = {\cal M}^4 K(\chi) ,
\label{L0-def}
\ee
and $\Delta {\cal L}(\varphi)$ is the counterterm Lagrangian that is needed to make the
renormalized effective action $\Gamma_{\rm renorm}[\varphi]$ finite, which we write as
\beqa
\Gamma_{\rm renorm}[\varphi] & = & \int d^4 x \, \pounds_{\rm renorm}(\varphi) \\
& = & \int d^4 x \, \left[ \pounds_{\rm classical}(\varphi) + \Delta\pounds(\varphi) \right] .
\label{Gamma-renorm-def}
\eeqa
Hereafter, we use the symbol $\pounds$ to distinguish the integrand of the effective
action $\Gamma$ from the integrand ${\cal L}$ of the action $S$.
In this section, we consider the Minkowski metric, with $\sqrt{-g}=1$, and we focus
on the scalar-field Lagrangian, as we study the quantum corrections associated with
the K-mouflage scalar field and we investigate the renormalization of its ultra-violet
divergences (see Appendix B for more details about matter loops).
At tree order, we simply have
\beqa
\mbox{tree order:} && \;\; \Gamma_{\rm renorm}^{(0)} = S_{\rm bare}^{(0)} 
= S_{\rm classical} , \\
&& \hspace{-1cm} \pounds_{\rm renorm}^{(0)} = {\cal L}_{\rm bare}^{(0)} 
= \pounds_{\rm classical} = {\cal L}_{\rm classical} .
\label{renorm-level0}
\eeqa
In particular, we defined $\pounds_{\rm classical} (\varphi)$ in 
Eq.(\ref{Gamma-renorm-def}) by
$\pounds_{\rm classical} = {\cal L}_{\rm classical}$, so that $\Delta\pounds$ corresponds 
both to the radiative corrections due to the quantum fluctuations of the scalar field and 
to the counterterm Lagrangian $\Delta{\cal L}$ in Eq.(\ref{S-renorm-def}).

In renormalizable theories, the counterterm Lagrangian $\Delta{\cal L}$ contains
a finite number of operators. Usually, one actually defines the initial Lagrangian of the
theory ${\cal L}_{\rm classical}$ by including all renormalizable operators, which are 
generated by radiative corrections, using the symmetries of the theory to simplify 
the analysis, so that there is no need for 
$\Delta {\cal L}$ and ${\cal L}_{\rm bare} = {\cal L}_{\rm classical}$.
More precisely,
in ``bare perturbation theory'', one starts from such a complete Lagrangian
${\cal L}_{\rm bare}$, expressed in terms of the bare parameters (such as the mass
$m_0$ and coupling $\lambda_0$)
that are actually infinite if we do not use a regularization method (such as a hard cutoff
$\Lambda$ or better dimensional regularization). Then, one computes amplitudes, correlation functions, or $\Gamma$,
from the Feynman diagrams associated with ${\cal L}_{\rm bare}$, up to some perturbative order
(e.g., in powers of some coupling constant); derives the physical mass $m$
and coupling constant $\lambda$ (up to that order); and finally eliminates
$m_0$ and $\lambda_0$ in favor of $m$ and $\lambda$ in the expressions of the amplitudes.
The resulting expressions must then be finite when the regularization is removed.
In the alternative approach known as ``renormalized perturbation theory'',
one splits the bare Lagrangian as
${\cal L}_{\rm bare} = {\cal L}^{\rm renorm} + \Delta{\cal L}$, where ${\cal L}^{\rm renorm}$
is written in terms of the renormalized parameters $m$ and $\lambda$.
Then, one computes the Feynman diagrams associated with the renormalized
Lagrangian ${\cal L}^{\rm renorm}$, to which is added the piece $\Delta{\cal L}$
that appears as additional vertices.
After the counterterm $\Delta{\cal L}$ is adjusted according to the chosen renormalization conditions,
the amplitudes must remain finite when the regularization is removed.
These two approaches are equivalent and only differ by their bookkeeping rules.

In our case, matters are somewhat more complex because the theory is not renormalizable.
Thus, the counterterm $\Delta{\cal L}$ contains an infinite number of operators, and
it is not possible to write a simple explicit expression for the complete bare Lagrangian
${\cal L}_{\rm bare}$, or the renormalized Lagrangian ${\cal L}^{\rm renorm}$. This is
why we keep the distinction between ${\cal L}_{\rm classical}$, defined in
Eq.(\ref{L0-def}), which provides a simple explicit expression that defines the theory at the
classical level (when we do not worry about quantum corrections), and the bare Lagrangian
${\cal L}_{\rm bare}$.
Fortunately, we can still obtain some predictions for the K-mouflage theory in a
``classical regime'' where  the radiative contributions
$\Delta\pounds$ are negligible and the effective action is  dominated by the operators included
in the initial classical Lagrangian ${\cal L}_{\rm classical}$ of Eq.(\ref{L0-def}).
Moreover, it happens that these operators are not renormalized, so that in the classical
regime we obtain
\be
\mbox{classical regime:} \;\;\;\;
\Gamma_{\rm renorm}[\varphi] \simeq \int d^4 x \, {\cal L}_{\rm classical}(\varphi) .
\label{Gamma-classical}
\eeq

In the following sections, we present a recursive algorithm which allows us to construct
both the bare action (\ref{S-renorm-def}) and the renormalized effective action
(\ref{Gamma-renorm-def}) and to estimate the range of the classical regime
(\ref{Gamma-classical}).
We recall here that, while the renormalized effective action $\Gamma_{\rm renorm}$ is finite,
the bare action $S_{\rm bare}$ is infinite (if we remove the regularization).
Starting from the tree-level Lagrangian ${\cal L}^{(0)}={\cal L}_{\rm classical}$,
we obtain in a first step ``(1)'' the radiative corrections to the effective action $\Gamma$ as
\be
{\cal L}^{(0)} \rightarrow \S\Gamma^{(1)} = \int d^4 x \; \S\pounds^{(1)} =
\int d^4 x \, \sum_{L=1}^{\infty} \S\pounds_L^{(1)} ,
\label{delta-Gamma-1-def}
\ee
where we sum over all loop orders $L$. Here, the superscript ``(1)'' denotes that we are
at the first step of the recursive algorithm, and the symbol $\S$ denotes the sum of
Feynman diagrams that give the radiative contributions in a loop expansion.
These contributions can be generated from a perturbative expansion around a background
configuration $\bar\varphi$ that allows us to define Feynman rules for the
K-mouflage theories.
Using these Feynman rules, the radiative contribution $\S\Gamma^{(1)}$
to the effective action is obtained by summing the infinite tower of one-particle-irreducible
vacuum diagrams of the model defined from ${\cal L}^{(0)}={\cal L}_{\rm classical}$.
These diagrams involve slowly varying functions of time and space as coupling constants
(through $\bar\varphi$) and show  ultra-violet divergences corresponding
to short-distance singularities on scales much smaller than the scale of variation
of the background $\bar\varphi$.
We can regularize these integrals using dimensional regularization for instance.
Then, we build up order by order in the loop expansion a series of counterterms
$\Delta{\cal L}_L^{(1)}$ that cancel the divergent parts of the loop diagrams
(\ref{delta-Gamma-1-def}), leaving only finite parts involving powers of $\ln(m_\phi/\mu)$,
where $\mu$ is the renormalization scale and $m_\phi$ is the mass of the normalized
scalar field. So we have, at the end of the first step, the bare action
\be
S^{(1)} =  \int d^4 x \, \left[ {\cal L}_{\rm classical} + \Delta{\cal L}^{(1)} \right] ,
\label{S(1)-def}
\ee
and the renormalized effective action
\beqa
\Gamma^{(1)} & = & \int d^4 x \, \left[ {\cal L}_{\rm classical} + \Delta\pounds^{(1)} \right] \\
& = & \int d^4 x \, \left[ {\cal L}_{\rm classical} + \S\pounds^{(1)} + \Delta{\cal L}^{(1)} \right] ,
\eeqa
where $\S\pounds_L^{(1)}$ are the regularized and otherwise infinite $L$-loop
contributions to the effective action, which can be explicitly calculated using the Feynman
rules defined from the previous-level Lagrangian ${\cal L}^{(0)}={\cal L}_{\rm classical}$, and
$\Delta{\cal L}^{(1)}_L$ are the counterterms that cancel all the divergent poles in powers
of $1/(d-4)$ when calculating in dimensional regularization in dimension $d$.
The result $\Delta\pounds^{(1)}_L = \S\pounds^{(1)}_L + \Delta{\cal L}^{(1)}_L$
is the $L$-loop finite contribution to the renormalized effective action $\Gamma^{(1)}$.

However, this renormalization procedure does not stop after this first step. Indeed, we have
only used the Feynman rules generated from ${\cal L}^{(0)}={\cal L}_{\rm classical}$ to obtain
the renormalized effective action $\Gamma^{(1)}$.
We should also take into account the radiative contributions introduced by the new
counterterm Lagrangian $\Delta{\cal L}^{(1)}$ of the action $S^{(1)}$.
Therefore, in the second step ``(2)'' of the algorithm, starting from the previous-level action
$S^{(1)}$ of Eq.(\ref{S(1)-def}), we derive as in Eq.(\ref{delta-Gamma-1-def}) the
radiative corrections $\S\pounds^{(2)}$ to the effective action $\Gamma^{(2)}$
that arise from the new terms $\Delta{\cal L}^{(1)}$ of the Lagrangian.
Again, these diagrams show ultra-violet divergences that are canceled
by introducing a new counterterm $\Delta{\cal L}^{(2)}$ to the action $S^{(2)}$.
In principle, we have an infinite sequence of the form
\beqa
&& ... \rightarrow \left\{ \begin{array}{l} \Delta{\cal L}^{(n)} \\ \Delta\pounds^{(n)}
\end{array} \right\} \rightarrow \S\pounds^{(n+1)} \rightarrow \Delta{\cal L}^{(n+1)}
\rightarrow \nonumber \\
&& \left\{ \begin{array}{l} \Delta{\cal L}^{(n+1)} \\
\Delta\pounds^{(n+1)}= \S\pounds^{(n+1)} + \Delta{\cal L}^{(n+1)} \end{array} \right\}
\rightarrow .... \;\;\;
\label{sequence-renormalization}
\eeqa
which provides a series of approximations for the  actions
(\ref{S-renorm-def}) and (\ref{Gamma-renorm-def}), which are given by
\be
\Delta{\cal L} = \Delta{\cal L}^{(1)} + \Delta{\cal L}^{(2)} + ... , \;\;
\Delta\pounds = \Delta\pounds^{(1)} + \Delta\pounds^{(2)} + ...
\label{series-DeltaL-Deltapounds}
\ee

In the case of standard renormalizable theories, this procedure can actually be made to stop
after one step. Indeed, starting with a complete bare Lagrangian ${\cal L}_{\rm classical}$,
the counterterm $\Delta{\cal L}^{(1)}$ needed in Eq.(\ref{S(1)-def}) to make the effective
action $\Gamma^{(1)}$ finite has the same form as ${\cal L}_{\rm classical}$.
Then, the algorithm (\ref{series-DeltaL-Deltapounds}) simply provides a series of approximations
to the bare parameters $\{m_0,\lambda_0,..\}$ of the bare Lagrangian (in terms of the
renormalized ones) that can be resummed in one step by solving an implicit equation
(i.e., finding the fixed point of the algorithm).
In our case, we need to go up to two steps at least. Indeed, we shall find that the
counterterm $\Delta{\cal L}^{(1)}$ of Eq.(\ref{S(1)-def}) does not have the same form as
${\cal L}_{\rm classical}$.
Whereas ${\cal L}_{\rm classical}$ is only a function of $\chi$, see Eq.(\ref{L0-def}),
$\Delta{\cal L}^{(1)}$ and $\Delta\pounds^{(1)}$ have the form
\beqa
&& \Delta{\cal L}^{(1)} = m_{\phi}^4 \, F(m_{\phi}/\mu,\chi)  ,
\;\; \mbox{with} \;\; m_{\phi}^2 \supset \partial\chi ,
\label{Delta-L(1)-F}
\\
&& \Delta\pounds^{(1)} = m_{\phi}^4 \, {\cal F}(m_{\phi}/\mu,\chi)  ,
\label{Delta-pounds(1)-F}
\eeqa
where the functions $F$ and ${\cal F}$ can be constructed order by order in perturbation
theory.
By equation (\ref{Delta-L(1)-F}), we mean that $\Delta{\cal L}^{(1)}$ contains derivatives
of $\chi$ and vanishes if these derivatives are set to zero. This implies that
$\Delta{\cal L}^{(1)}$ contains no term of the same form as ${\cal L}_{\rm classical}$, so that at this
level the initial bare Lagrangian ${\cal L}_{\rm classical}$ is not renormalized and $\Delta{\cal L}^{(1)}$
brings in an infinite series of new operators.
Therefore, this step cannot be absorbed by a redefinition of the parameters of ${\cal L}_{\rm classical}$.
At the second step, we shall find that the counterterm $\Delta{\cal L}^{(2)}$
has the same form as $\Delta{\cal L}^{(1)}$: it also involves
derivatives of $\chi$, with either one, zero, or two derivatives for each factor $\chi$
(but subleading corrections can introduce a higher number of derivatives).
Therefore, we find that the initial bare Lagrangian ${\cal L}_{\rm classical}$ is not
renormalized by this procedure and that the counterterms are ``stable'' after the first step.
We shall not prove that the operators generated at step ``(n+1)'' are exactly the same as
those already obtained at the previous step ``(n)'', but they show the same form
at leading order, and their amplitude decreases as we move to later steps of the
algorithm. Moreover, if we make an approximation such as truncating the list of operators
to a finite number (e.g., we disregard small-scale details), we are back to the case of
renormalizable theories, and we can stop the procedure at the second step by looking for
the fixed point $\Delta{\cal L}$ of the procedure, such that the associated effective action
is finite.

In practice, we do not compute the explicit form of the counterterms
$\Delta{\cal L}^{(n)}$ and the effective-action integrands
$\Delta\pounds^{(n)}$. We simply study their functional form as in
Eqs.(\ref{Delta-L(1)-F})-(\ref{Delta-pounds(1)-F})
and their scalings, to ensure that they do not renormalize ${\cal L}_{\rm classical}$
and to estimate the range of the classical regime (\ref{Gamma-classical}) where they can
be neglected.
In this regime, the Klein-Gordon equation associated with the tree-level Lagrangian
${\cal L}_{\rm classical}$ is not modified, and the classical configuration $\bar \chi$ is not altered by
the quantum corrections.
Therefore, we can compute accurate and reliable predictions.
Outside this classical regime, the theory is not predictive in the sense that we would need
to take into account the infinite number of operators brought by the radiative contributions
and the renormalization procedure, which is not practical.
Fortunately, we shall find that the classical regime (\ref{Gamma-classical}) applies to all
configurations of interest, both in the cosmological large-scale context and in the small-scale
astrophysical context.

In the following, we give a detailed account of this procedure.
First, in section~\ref{sec:Perturbations} we present the basis of perturbation theory
within background quantization, at the first step of the procedure on the initial bare
Lagrangian ${\cal L}^{(0)}= {\cal L}_{\rm classical}$.
Next, in section~\ref{sec:Quadratic}, we obtain the one-loop contribution
$\Delta\pounds^{(1)}_1$ to the first-level effective action $\Gamma^{(1)}$.
This allows us to check that this contribution is of the form (\ref{Delta-pounds(1)-F}),
i.e., it does not renormalize ${\cal L}_{\rm classical}$ but brings new operators, and we derive
a one-loop first-level condition for the classical regime (\ref{Gamma-classical}).
We also check that this classicality condition is met in relevant cosmological and
astrophysical backgrounds, so that the theory is not already ruled out at this level.
Then, in section~\ref{sec:Higher-orders}, we obtain the higher-loop contributions
$\Delta\pounds_L^{(1)}$ to $\Gamma^{(1)}$ and we check that they do not spoil the
one-loop classicality criterion.
We push the analysis to the second step $\Gamma^{(2)}$ in section~\ref{sec:second-step},
and we extend to all steps of the renormalization algorithm in
section~\ref{sec:beyond-second-step}.
We compare our procedure with previous works in section~\ref{sec:de-Rham}.
Finally, in section~\ref{sec:quantum-nonlinear}, we estimate the location of the
quantum-classical transition and discuss its existence in cosmological and
astrophysical backgrounds.

\subsection{Perturbation theory}
\label{sec:Perturbations}

In the spirit of background quantization, we consider K-mouflage models in a background
$\bar\varphi(t,\vr)$ in Minkowski spacetime and expand
\be
\varphi= \bar{\varphi} + \tilde{\phi} ,
\label{background-split}
\ee
where $\tilde \phi$ has quantum fluctuations around a zero mean.
From Eq.(\ref{chi-def}), the rescaled kinetic term reads as
\be
\chi = \bar{\chi} + \delta\chi \;\;\; \mbox{with} \;\;\;
\delta\chi = - \frac{1}{{\cal M}^4} \partial^{\mu}\bar\varphi \partial_\mu\tilde\phi
- \frac{1}{2{\cal M}^4} \partial^{\mu}\tilde\phi \partial_\mu\tilde\phi ,
\ee
and the scalar-field Lagrangian reads as
\be
{\cal L}^{(0)} = \bar{\cal L}^{(0)} + \delta{\cal L}^{(0)}
\ee
with
\be
\delta{\cal L}^{(0)} = {\cal M}^4 \left[ \bar{K}' \delta\chi + \frac{\bar{K}''}{2} (\delta\chi)^2
+ .. \right] .
\label{delta-L-def-1}
\ee
Here, the superscript ``(0)'' recalls that we are at the zeroth level (\ref{renorm-level0})
of the renormalization procedure, and the $\delta$ in $\delta{\cal L}$ denotes the
background-fluctuation split (\ref{background-split}).

Since we expand around the background $\bar\varphi$, which is the solution of the equations
of motion at the classical level, the linear term in $\tilde\phi$ vanishes, and the quadratic
Lagrangian reads as
\be
\delta {\cal L}^{(0)}_2 = - \frac{\bar{K}'}{2} \partial^{\mu}\tilde\phi \partial_\mu\tilde\phi
+ \frac{\bar{K}''}{2{\cal M}^4} \left( \partial^{\mu}\bar\varphi \partial_\mu\tilde\phi \right)^2 .
\label{L2-def-1}
\ee
It turns out that in many cases $\bar\chi \bar{K}'' \ll \bar{K}'$ (this holds in the
weak-field regime, where $\bar\chi \ll 1$, and in the highly nonlinear regime for
the Arctan and Sqrt models).
Then, to recover a canonical kinetic term, we define the rescaled scalar field $\phi$ by
\be
\tilde\phi =  \frac{\phi}{\sqrt{\bar{K}'}} ,
\label{phi-def}
\ee
where we used that $\bar{K}'>0$ for well-behaved K-mouflage theories (i.e., without ghosts
and instabilities), and the quadratic Lagrangian becomes, after one integration by parts,
\be
\delta {\cal L}^{(0)}_2 \simeq - \frac{1}{2} \partial^{\mu}\phi \partial_{\mu}\phi
- \frac{1}{2} m_{\phi}^2 \phi^2 ,
\label{L2-def-2}
\ee
where we used the approximation $\bar{\chi} \bar{K}'' \ll \bar{K}'$, and we defined
\be
m_{\phi}^2 = \partial_{\mu} \left( \frac{\partial^{\mu} \bar{K}'}{2\bar{K}'} \right)
+ \frac{\partial^\mu {\bar K}' \partial_{\mu} {\bar K}'}{4\bar{K}'^2} .
\label{m-def}
\ee
In terms of the rescaled field $\phi$, the kinetic term $\chi$ reads as
\be
\chi = \bar{\chi} + \delta\chi \;\;\; \mbox{with} \;\;\;
\delta\chi = - \frac{X^{\mu} D_{\mu}\phi}{{\cal M}_K^2}
- \frac{D^{\mu}\phi D_{\mu}\phi}{2{\cal M}_K^4} ,
\label{dchi-phi}
\ee
where we have defined
\be
X_{\mu} = \frac{\partial_{\mu}\bar\varphi}{{\cal M}^2} , \;\;\;
D_{\mu} \phi = \left( \partial_{\mu} - \frac{\partial_{\mu} \bar{K}'}{2\bar{K}'} \right) \phi , \;\;\;
{\cal M}_K = {\cal M} \, \bar{K}'^{1/4}  .
\label{Xmu-Dmu-Mk-def}
\ee
With Eq.(\ref{delta-L-def-1}), this provides an expansion of the K-mouflage action in
powers of $D_{\mu}\phi$.
The effective cutoff scale ${\cal M}_K$ can be significantly higher than ${\cal M}$ in the
nonlinear regime, $\bar{K}' \gg 1$, associated with small-scale systems, such as the
Solar System or astrophysical objects, or the cosmological background in the early Universe.

From Eq.(\ref{m-def}), we obtain the scaling
\beq
m_{\phi}^2 \sim  \frac{\bar{K}''\bar{\chi}}{\bar{K}'} \, \bar{p}^2 ,
\label{mphi-pbar}
\eeq
where $\bar{p}$ is the 4-momentum of the typical scale of variation for the background and we assumed
$K''' \chi^2 \lesssim K'' \chi$ and $K'' \chi \lesssim K'$, which holds for the models
introduced in section~\ref{sec:K-chi} and more generally for models where
$K'$ behaves as a growing power law or a constant with inverse power-law corrections
in the nonlinear regime $| \chi | \rightarrow \infty$.
In the Cubic model $K''' \chi^2 \sim K''\chi \sim K'$ in the nonlinear regime and
$m_{\phi} \sim \bar p$, but for the Arctan and Sqrt models $K''\chi \ll K'$ and
$m_{\phi} \ll \bar p$.

\subsection{Quadratic action and one-loop quantum contributions}
\label{sec:Quadratic}

We now study the quantum corrections to the effective action
$\Gamma[\bar\varphi]$ of the background field generated by the fluctuations
$\phi$, using the background-field approach.
At tree level, the effective action is simply
$\Gamma^{(0)}[\bar\varphi] = S[\bar\varphi] = \int d^4 x \, {\cal L}_{\rm classical}(\bar\varphi)$,
where ${\cal L}_{\rm classical}(\bar\varphi) = {\cal M}^4 K(\bar\chi)$ is the K-mouflage Lagrangian
introduced in Eq.(\ref{S-def}); see also Eq.(\ref{renorm-level0}).
$L$-loop contributions, $\S\Gamma^{(1)}_{L} = \int d^4x \, \S\pounds^{(1)}_L$
as in Eq.(\ref{delta-Gamma-1-def}), can be obtained by summing the vacuum one-particle-irreducible 
Feynman diagrams defined by the Lagrangian $\delta {\cal L}^{(0)}$ of
Eq.(\ref{delta-L-def-1}) over the fluctuation field $\phi$, with the propagator and vertices
parametrized by the background field $\bar\varphi$.
In this section, we focus on the one-loop contribution, $\S\pounds^{(1)}_{\rm 1-loop}$,
 associated with the determinant of the quadratic Lagrangian $\delta {\cal L}^{(0)}_2$ in $\phi$.
This corresponds to a truncation of the Lagrangian $\delta {\cal L}^{(0)}$ at quadratic order.
We consider higher-loop corrections in section~\ref{sec:Higher-orders} below.

\subsubsection{One-loop contribution $\Delta\pounds^{(1)}_{\rm 1-loop}$
to the effective action $\Gamma^{(1)}$}
\label{sec:one-loop}

We first consider the one-loop quantum corrections generated by the quadratic action
(\ref{L2-def-1}).
From Eqs.(\ref{delta-L-def-1}) and (\ref{dchi-phi}), at second order the Lagrangian reads as
\be
\delta {\cal L}^{(0)}_2 = - \frac{1}{2} D^{\mu}\phi D_{\mu}\phi + \frac{\bar{K}''}{2\bar{K}'}
( X^{\mu} D_{\mu}\phi )^2 .
\label{L2-1}
\ee
Since $\bar{\chi} = - X^{\mu} X_{\mu}/2$,
the second term is smaller than the first term by a factor of order
$\bar{\chi}\bar{K}''/\bar{K}'$,
which is typically much smaller than unity. Indeed, this factor vanishes in the low-redshift
cosmological background, where $\bar{\chi} \ll 1$ and $\bar{K} \simeq -1$ plays the role
of the cosmological constant. It is also very small (below $10^{-4}$) in the small-scale limit,
$\bar{\chi} \rightarrow -\infty$, because of the constraint associated with the measure of the
perihelion of the Moon.
This also means that the speed $c_s$ of scalar waves is close to the speed of light,
as we have $c_s^2=(\bar{K}'+2\bar{\chi}\bar{K}'')/\bar{K}'$ around small-scale
astrophysical backgrounds and $c_s^2=\bar{K}'/(\bar{K}'+2\bar{\chi}\bar{K}'')$
around the cosmological background.
Then, if we neglect the second term we recover the approximate quadratic Lagrangian
(\ref{L2-def-2}).
On the other hand, if we keep the second term in Eq.(\ref{L2-1}) the full quadratic Lagrangian
reads as
\be
\delta {\cal L}^{(0)}_2 = - \left[ \eta^{\mu\nu} - \frac{\bar{K}''}{\bar{K}'}
X^{\mu} X^{\nu} \right] \frac{\partial_{\mu}\phi \partial_{\nu}\phi}{2}
- \bar{m}_{\phi}^2  \frac{\phi^2}{2}
\label{L2-exact}
\ee
with
\be
\bar{m}_{\phi}^2 = m_{\phi}^2 - \partial_{\mu} \left( \frac{\bar{K}''}{\bar{K}'}
X^{\mu} \frac{X^{\nu} \partial_{\nu} \bar{K}'}{2\bar{K}'} \right)
- \frac{\bar{K}''}{\bar{K}'} \left( \frac{X^{\mu} \partial_{\mu}\bar{K}'}{2\bar{K}'} \right)^2 .
\label{m2phi-bar}
\ee
The second and third terms in the total squared-mass $\bar{m}_{\phi}^2$ are of order
$(\bar{\chi}\bar{K}''/\bar{K}')^2 \bar{p}^2$ and $(\bar{\chi}\bar{K}''/\bar{K}')^3 \bar{p}^2$,
whereas the first term in Eq.(\ref{m-def}) is of order $(\bar{\chi}\bar{K}''/\bar{K}') \bar{p}^2$,
as seen in Eq.(\ref{mphi-pbar}).

As recalled above, $\bar{\chi}\bar{K}''/\bar{K}'$ is small in both the linear and highly nonlinear
regimes for the Arctan and Sqrt models, but it is of order unity in the transition regime for
$\bar\chi \sim 1$.
For polynomial or power-law kinetic functions $K$, which, however, do not satisfy
Solar System constraints, it is of order unity in the nonlinear regime.
Then, for  dimensional analysis and orders of magnitude estimates, we can focus on the
simpler quadratic Lagrangian (\ref{L2-def-2}).
This allows us to calculate the one-loop corrections in dimensional regularization;
see Appendix~\ref{app:one-loop}. Thus, we obtain \cite{Schwartz2013}
\be
\S\pounds_{\rm 1-loop}^{(1)}( \bar\varphi) = m_{\phi}^4 \; I_1 ,
\label{deltaL-one-loop}
\ee
for the sum of the one-loop Feynman diagrams, where $I_1$ is a dimensionless factor that
diverges when the spacetime dimension $d \rightarrow 4$. This leads to the introduction
of a counterterm in the Lagrangian,
\be
\Delta{\cal L}_{\rm 1-loop}^{(1)}( \bar\varphi) = m_{\phi}^4 \; J_1 ,
\label{DeltaL-one-loop}
\ee
which cancels the ultra-violet divergences (\ref{deltaL-one-loop}) in the one-loop effective
action, and we obtain for the finite part \cite{Schwartz2013}
\be
\Gamma^{(1)}_{\rm 1-loop} = \int d^4x \, \Delta\pounds^{(1)}_{\rm 1-loop}
= \int d^4x \, \frac{m_{\phi}^4}{64 \pi^2}
\ln \left( \frac{\mu^2}{m_{\phi}^2} \right) ,
\label{Gamma-1-1loop}
\ee
where $\mu$ is the renormalization scale and
$\Delta\pounds^{(1)}_{\rm 1-loop} = \S\pounds_{\rm 1-loop}^{(1)}
+ \Delta{\cal L}_{\rm 1-loop}^{(1)}$, see also Appendix~\ref{app:one-loop}
and Eq.(\ref{rhov-renorm}) for an alternative explicit computation.
The one-loop Lagrangian counterterm $\Delta{\cal L}_{\rm 1-loop}^{(1)}$ and the
one-loop effective-action integrand $\Delta\pounds^{(1)}_{\rm 1-loop}$ involve derivatives of
$\chi$ through $m_{\phi}$, as seen from the expression (\ref{m-def}), whereas the tree-level
Lagrangian ${\cal L}_{\rm classical}={\cal M}^4 K(\chi)$ is a simple function of $\chi$.
This implies that the K-mouflage Lagrangian ${\cal L}_{\rm classical}$ is not renormalized
at one-loop order and remains unchanged in both the renormalized action
$S^{(1)}$ and the effective action $\Gamma^{(1)}$, up to this one-loop order.
Moreover, the amplitude of the one-loop correction is small provided
\be
\Delta\pounds_{\rm 1-loop}^{(1)} \ll {\cal L}_{\rm classical} \;\; \mbox{if:} \;\;\;\;
m_{\phi}^4 \ll {\cal M}^4 \bar{K}' \bar{\chi} .
\label{one-loop-small-1}
\ee
Here and in the following, we compare the one-loop contribution with
${\cal M}^4 \bar{K}'\bar{\chi}$ instead of ${\cal M}^4 \bar{K}$ to set apart the
cosmological constant contribution $\bar{K}(0)=-1$, as $\bar{K}'\bar{\chi} \sim \bar{K}$
in the nonlinear regime but $\bar{K}'\bar{\chi} \ll \bar{K}$ in the linear regime,
and we write
\be
{\cal L}^{(0)} = {\cal L}_{\rm classical} \sim {\cal M}^4 \bar{K}'\bar{\chi} .
\label{L0-approx-noLambda}
\ee
Using Eq.(\ref{mphi-pbar}), the condition (\ref{one-loop-small-1}) reads as
\be
\Delta\pounds_{\rm 1-loop}^{(1)} \ll {\cal L}_{\rm classical} \;\; \mbox{if:} \;\;\;\;
 \bar{p} \ll {\cal M} \, (\bar{K}' \bar{\chi})^{1/4} \left( \frac{\bar{K}'}{\bar{K}''\bar{\chi}} \right)^{1/2} .
\label{one-loop-small-2}
\ee

These results remain valid for the exact quadratic Lagrangian (\ref{L2-exact}).
Indeed, the additional contributions to the mass $m_{\phi}^2$ in Eq.(\ref{L2-exact})
also involve derivatives of $\bar\chi$, so that the operators of the bare Lagrangian $K$
remain unrenormalized. In terms of the scaling analysis, the condition (\ref{one-loop-small-2})
remains valid as the additional terms in Eq.(\ref{L2-exact}) are either subdominant or of the
same order as those in Eq.(\ref{L2-def-2}).

\subsubsection{Validity of the classical regime according to the one-loop criterion}
\label{sec:classical-one-loop}

A necessary condition for the validity of the classical regime, where the dynamics of the
K-mouflage system are described by the classical equations of motion defined by the
action (\ref{S-def}), is that the one-loop corrections to the effective action,
$\Gamma = S + \Gamma_{\rm 1-loop} + ....$, are negligible.
This condition is given by the equations (\ref{one-loop-small-1}) and
(\ref{one-loop-small-2}) above. In this section, we check that this constraint is
satisfied for cosmological and small-scale backgrounds.
\newline

\paragraph{Cosmological background}
\label{sec:one-loop-cosmos}
\mbox{}\newline

In the cosmological background, we have $\bar{p} \sim H$, where $H$ is the Hubble expansion
rate. The scale ${\cal M}$ is set by the dark-energy density today,
\be
{\cal M}^4 \simeq \bar{\rho}_{\rm de 0} \;\;\; \mbox{whence} \;\;\;
{\cal M} \sim 2.3 \times 10^{-12} \; \rm{GeV} ,
\label{M-scale-de0}
\ee
and the condition (\ref{one-loop-small-2}) reads as
\be
H \ll 2.3 \times 10^{-12} \; (\bar{K}' \bar{\chi})^{1/4}
\left( \frac{\bar{K}'}{\bar{K}''\bar{\chi}} \right)^{1/2} \; \rm{GeV} .
\label{one-loop-H}
\ee
The Hubble rate grows with the redshift, and at Big Bang Nucleosynthesis (BBN) we have
$H_{\rm BBN} \lesssim 10^{-23} \, \rm{GeV}$, which is much smaller than ${\cal M}$.
The factor $(\bar{K}'\bar{\chi})^{1/4} ( \bar{K}'/\bar{K}''\bar{\chi})^{1/2}$ is greater than unity
in both the linear and nonlinear regimes, and it is of order unity at the transition
$\bar\chi \sim 1$.
Therefore, the condition (\ref{one-loop-H}) is satisfied from $z=0$ up to redshifts much
higher than $z_{\rm BBN}$.
\newline

\paragraph{Astrophysical background}
\label{sec:one-loop-static}
\mbox{}\newline

Around a compact object of mass $M$, such as a star, the background scalar field
$\bar\varphi$ enters the nonlinear regime at the K-mouflage radius $R_K$
\cite{Brax:2014c},
\be
R_K = \left( \frac{\beta M}{4\pi M_{\rm Pl} {\cal M}^2} \right)^{1/2} .
\label{RK-def}
\ee
Writing $\bar{p} \sim 1/r$ and substituting the expression (\ref{RK-def})  for  ${\cal M}$,
the condition (\ref{one-loop-small-2}) becomes
\be
\frac{r}{R_K} \gg 4.7\times 10^{-20} \left( \frac{M_{\odot}}{\beta M} \right)^{1/2}
(\bar{K}'\bar{\chi})^{-1/4} \left( \frac{\bar{K}''\bar{\chi}}{\bar{K}'}\right)^{1/2} .
\label{one-loop-static-1}
\ee
The factor $(\bar{K}'\bar{\chi})^{-1/4} (\bar{K}''\bar{\chi}/\bar{K}')^{1/2}$ is smaller than
unity in both the linear and nonlinear regimes, and it is of order unity at the transition
$\bar{\chi} \sim 1$.
We have $R_K \sim 0.04 h^{-1}{\rm Mpc}$ for clusters of galaxies,
$R_K \sim 4 h^{-1}{\rm kpc}$ for galaxies, and $R_K \sim 1000 \, {\rm A.U.}$ for the Sun.
Therefore, the condition (\ref{one-loop-static-1}) is satisfied by a large margin
well below the K-mouflage radius of screened systems and on all
astrophysical and cosmological scales.

\subsection{Higher-order contributions $\Delta\pounds^{(1)}_L$ to the effective action
$\Gamma^{(1)}$}
\label{sec:Higher-orders}

\subsubsection{Radiative contribution $\S\pounds^{(1)}_L$ associated with
$L$-loop Feynman diagrams}
\label{sec:radiative-L-loop}

We can generalize the quadratic results by including the higher-order corrections.
From Eqs.(\ref{delta-L-def-1}) and (\ref{dchi-phi}), we can write the Lagrangian in perturbation
theory as
\be
\delta {\cal L}^{(0)} = {\cal M}^4 \sum_{n=1}^{\infty} \frac{\bar{K}^{(n)}}{n!}
\left( - \frac{X^\mu D_\mu \phi}{{\cal M}_K^2}
- \frac{D^{\mu}\phi D_{\mu}\phi}{2 {\cal M}_K^4} \right)^n ,
\label{deltaL-phi-n}
\ee
where $\bar{K}^{(n)}=d^n K/d\chi^n(\bar\chi)$.
This series can be rearranged order by order in $D\phi$.
As an order of magnitude estimate, the series behaves like
\be
\delta{\cal L}^{(0)} \sim {\cal M}^4 \sum_{m= 2}^{\infty} c_m \frac{(D\phi)^m}{{\cal M}^{2m}_K}
\label{deltaL-cm-def}
\ee
with
\be
c_m = \sum_{n=[m/2]_+}^{m} \frac{\bar{K}^{(n)} \bar{\chi}^{n-m/2}}{(2n-m)! (m-n)!} ,
\label{cm-def}
\ee
where $[m/2]_+$ is the closest integer greater than or equal to $m/2$.
In the linear regime, where $\bar{\chi} \rightarrow 0$, the series is dominated by the
even orders, with $n=m/2$ in the sum over $n$ for $c_m$, and we have
\be
\mbox{linear regime,} \;\; \bar\chi\rightarrow 0 : \;\;\;
c_m \sim \frac{\bar{K}^{(m/2)}(0)}{(m/2)!} \;\;\;
\mbox{for} \;\;\; m \;\; \rm{even} .
\label{cm-linear}
\ee
In the nonlinear regime, $| \bar\chi | \rightarrow \infty$, we assume the scalings
\beqa
| \bar{\chi} | \rightarrow \infty, \;\;\; n \geq 2 : && \;\;\;
\bar{K}^{(n)} \sim \bar{K}'' \bar{\chi}^{2-n} , \nonumber \\
\;\;\; \mbox{whence} && \;\;\; c_m \sim \bar{K}'' \bar{\chi}^{2-m/2} .
\label{Kn-K2}
\eeqa
This corresponds to models where $K''$ is a power law at large $\chi$,
with either a positive or negative exponent, and this includes both cases
such as the Cubic model with a positive exponent (although in the case of the Cubic
model with an integer exponent, the derivatives vanish for $n \geq 4$) and
cases such as the Arctan and Sqrt models where $K'$ goes to a constant $K_*$
at large $| \chi |$, with $(K'-K_*)$ decreasing as an inverse power law of $\chi$.
Such models are required to satisfy Solar System constraints, more precisely the
measurement of the perihelion of the Moon (which applies in the range $\chi \sim - 10^{6}$).

The effective action $\Gamma^{(1)}$ is obtained order by order in a loop expansion by
summing all the vacuum one-particle-irreducible Feynman diagrams built using
the previous perturbative expansion. The quadratic part of the action gives the propagators,
which depend on the mass $m_\phi^2$, and the higher-order terms give the vertices.
In short, each line of the Feynman diagrams carries a propagator $(p^2 +m_\phi^2)^{-1}$,
vertices bring factors of the type
\be
\cM^4 c_m \prod_{s=1}^{m} \frac{p_s}{\cM^2_K} ,
\label{vertex-Fourier}
\ee
and each loop involves an integral over the momenta $\int d^4 p_{\ell}$.
At the $L$-th loop order, we have the estimate
\be
\S\pounds_L^{(1)} \sim \int \prod_{\ell=1}^L d^4 p_{\ell} \prod_{n=1}^N \frac{1}{p^2_n +m_{\phi}^2}
\prod_{v=1}^V \cM^4 c_{m_v} \left( \prod_{s=1}^{m_v} \frac{p_s}{\cM_K^2} \right) ,
\label{Gamma-def}
\ee
for a diagram with $L$ loops (i.e., independent momenta), $N$ propagators,
and $V$ vertices (each vertex $v$ having $m_v$ legs).
Using the Euler relation $L=N-V+1$ and the sum $\sum_v m_v = 2N$,
this can be rewritten as
\be
\S\pounds_L^{(1)} \sim \frac{{\cal M}^4}{\bar{K}'^N}
\left( \frac{m_{\phi}^4}{{\cal M}^4} \right)^{L} \left( \prod_{v=1}^V c_{m_v} \right) \, I_L ,
\label{deltaL-nloop}
\ee
where $I_L$ is a dimensionless integral that needs to be regularized as it is formally infinite, 
for instance using dimensional regularization,
\be
I_L = \int \prod_{\ell=1}^L d^4 x_{\ell} \prod_{n=1}^N \frac{1}{x^2_n +1} \prod_{v=1}^V
\left( \prod_{s=1}^{m_v} x_s \right) .
\ee
The scaling (\ref{deltaL-nloop}) corresponds to setting all wave numbers $p$ in
Eq.(\ref{Gamma-def}) to $p \sim m_{\phi}$.
This is because the diagrams diverge with ultraviolet divergences at large $p$,
corresponding to coincident points in real space [typically associated with singular factors
of the form $\delta_D(0)^n$]. Then, the infrared dependence of the propagator and vertices
(e.g., their slow evolution with cosmic time) is irrelevant for the renormalization of these
integrals.

\subsubsection{Renormalization by the counterterms $\Delta{\cal L}^{(1)}_L$}
\label{sec:renormalization-L-loop}

From Eq.(\ref{deltaL-nloop}), we can see that at each loop order, to renormalize the
effective action $\Gamma^{(1)}$, that is, to remove the ultraviolet divergences,
we must introduce counterterms in the action $S^{(1)}$ of the form
\be
\Delta {\cal L}_L^{(1)} \sim {\cal M}^4 \left( \frac{m_{\phi}^4}{{\cal M}^4} \right)^{\!L}
\;\; \sum_{V=0}^{2(L-1)} \bar{K}'^{1-L-V} \left( \prod_{v=1}^V c_{m_v} \right) J_L ,
\label{Delta-L-counterterm}
\ee
where we used the property $2 N= \sum_{v} m_v \geq 3 V$, which implies $V \leq 2 (L-1)$.
The coefficients $J_L$ are divergent [e.g., they contain poles over $1/(d-4)$
in dimensional regularization] and cancel the divergent integrals $I_L$,
so that $(I_L+J_L)$ are finite, of order unity, and contain powers of
$\ln(m_{\phi}^2/\mu^2)$, where $\mu$ is the renormalization scale.
Equations (\ref{deltaL-nloop}) and (\ref{Delta-L-counterterm}) generalize to $L$-loop order
the one-loop results (\ref{deltaL-one-loop}) and (\ref{DeltaL-one-loop}), while
Eq.(\ref{Gamma-1-1loop}) generalizes as
\beqa
\Delta\pounds^{(1)}_L  \! & = & \! \S\pounds_L^{(1)} + \Delta {\cal L}_L^{(1)} \nonumber \\
& \sim & \! {\cal M}^4 \left( \frac{m_{\phi}^4}{{\cal M}^4} \right)^{\! L} \;\;
\sum_{V=0}^{2(L-1)} \bar{K}'^{1-L-V} \left( \prod_{v=1}^V c_{m_v} \right) . \;\;\;\;\;\;
\label{pounds-L-loop-Gamma1}
\eeqa
We can now estimate the magnitude of these higher-order corrections to the
effective action $\Gamma^{(1)}$.

Using Eq.(\ref{cm-linear}), in the linear regime we have
\be
\Delta\pounds_L^{(1)} \sim \cM^4 \left( \frac{m_\phi^4}{\cM^4} \right)^{\! L}  \;\;\; \mbox{and}
\;\;\; \frac{\Delta\pounds_L^{(1)}}{\Delta\pounds_1^{(1)}} \sim
\left( \frac{m_\phi^4}{\cM^4} \right)^{\! L-1} ,
\ee
where we used $\Delta\pounds_1^{(1)} \sim m_\phi^4$ for the one-loop contribution
(\ref{Gamma-1-1loop}).
Then, in the regime (\ref{one-loop-small-1}) where the one-loop contribution is small
as compared with the tree-level action, which implies $m_{\phi} \ll {\cal M}$
(because ${\bar K}' \simeq 1$ and $\bar\chi \ll 1$ in the linear regime),
we find that the higher-order corrections are even more negligible.

Using Eq.(\ref{Kn-K2}) and $N=V+L-1$, in the nonlinear regime we have
\be
\Delta\pounds_L^{(1)} \sim \cM^4 \bar{K}' \bar\chi
\left( \frac{m_\phi^4}{\cM^4 \bar{K}' \bar\chi } \right)^{\! L}
\left( \frac{\bar{K}'' \bar\chi}{{\bar K}'} \right)^V  .
\label{L-L-loop-nonlinear}
\ee
Then, in the regime (\ref{one-loop-small-1}) where the one-loop contribution
(\ref{Gamma-1-1loop}) is small, we obtain
\be
L \geq 2 : \;\;\; \frac{\Delta\pounds_L^{(1)}}{\Delta\pounds_1^{(1)}} \ll
\left( \frac{\bar{K}'' \bar\chi}{{\bar K}'} \right)^V \lesssim 1.
\label{L-L-loop-L-one-loop-nonlinear}
\ee
Since $\bar K'' \bar\chi / \bar K' \lesssim 1$, we again find that the one-loop criterion
implies that the $L$-loop corrections are negligible.

Therefore, in both the linear and nonlinear regimes, the higher-order quantum
contributions (\ref{pounds-L-loop-Gamma1}) to the effective action $\Gamma^{(1)}$
are negligible as soon as the one-loop contribution (\ref{Gamma-1-1loop})
is small as compared with the tree-level Lagrangian, which gives
\be
\Delta\pounds_L^{(1)} \ll {\cal L}_{\rm classical} \;\; \mbox{for all} \;\; L \geq 1 \;\; \mbox{if:} \;\;\;\;
m_{\phi}^4 \ll {\cal M}^4 \bar{K}' \bar{\chi} .
\label{Lagrangian-small-all-L}
\ee

As for the one-loop result (\ref{Gamma-1-1loop}),
these higher-order contributions also show a remarkable property: all
orders involve derivatives of $\chi$, through the prefactors $m_{\phi}$, whereas the
tree-level Lagrangian ${\cal L}_{\rm classical} = {\cal M}^4 K(\chi)$ does not involve derivatives
of $\chi$.
As already announced in section~\ref{sec:renormalization-notations} and in
Eq.(\ref{Delta-pounds(1)-F}), this implies that the K-mouflage bare Lagrangian is not
renormalized in the sense that higher-order corrections to the effective action only bring
new terms that involve derivatives of $\chi$.
We present in section~\ref{sec:Non-renormalization} below another derivation of this result.

\subsection{Second-step effective action $\Gamma^{(2)}$}
\label{sec:second-step}

As explained in section~\ref{sec:renormalization-notations},
the counterterm $\Delta{\cal L}^{(1)}$ introduced after the first step ``(1)''
in the K-mouflage renormalized action $S^{(1)}$ cancels the diverging quantum correction
$\S\pounds^{(1)}$ to the effective action $\Gamma^{(1)}$, but it also brings new
vertices, which in turn generate new quantum corrections to the effective action $\Gamma$,
as the latter is now defined from the renormalized Lagrangian
${\cal L}_{\rm classical}+ \Delta{\cal L} ^{(1)}$.
This leads to the second step ``(2)'' of the sequence described in
section~\ref{sec:renormalization-notations} and in (\ref{sequence-renormalization}).
We now study the new contributions that appear in this second step of the renormalization
procedure.

\subsubsection{New vertices brought by the counterterms $\Delta {\cal L}^{(1)}_L$}
\label{sec:quantum-from-counterterms}

It is convenient to compare the vertices generated by the counterterms
$\Delta{\cal L}^{(1)}_L$ to those associated with the bare Lagrangian
${\cal L}^{(0)} = {\cal L}_{\rm classical}$ through the expansions of these Lagrangians in powers
of $\delta\chi$, as in Eq.(\ref{deltaL-phi-n}).
Thus, we write for both the bare and the counterterm Lagrangians
\be
\delta {\cal L}^{(0)} \sim \sum_n {\cal L}^{(0)}_{(n)} (\delta\chi)^n , \;\;\;
\delta (\Delta {\cal L}^{(1)}_L) \sim \sum_n (\Delta {\cal L}_L^{(1)})_{(n)} (\delta\chi)^n .
\label{deltaLn-deltaDeltaLn}
\ee
The symbol ``$\delta$'' again refers to the background-fluctuation split
(\ref{background-split}), as in Eq.(\ref{delta-L-def-1}), and the subscript ``(n)'' denotes
the coefficient of order $n$ in the expansion in powers of $\delta\chi$, as in
Eq.(\ref{deltaL-phi-n}).
We now wish to check whether $(\Delta {\cal L}_L^{(1)})_{(n)} \ll {\cal L}^{(0)}_{(n)}$.
If this is the case, then the new vertices are negligible with respect to those associated
with the initial bare Lagrangian, which were given in Eq.(\ref{vertex-Fourier}).

From Eq.(\ref{deltaL-phi-n}), we have
\be
n = 1 : \;\; {\cal L}^{(0)}_{(1)} = {\cal M}^4 \bar{K}' \gtrsim  {\cal M}^4 \bar{K}'' \bar{\chi} ,
\label{Ln-n=1-bare}
\ee
and
\be
n \geq 2: \;\; {\cal L}^{(0)}_{(n)} \sim {\cal M}^4 \bar{K}^{(n)} \sim
{\cal M}^4 \bar{K}'' \bar{\chi}^{2-n} .
\label{Ln-bare}
\ee
Here, we separated the first- and higher-order derivative cases because, while for models
such as the Cubic model, where $K'$ is a power law at large $\chi$, the scaling
(\ref{Ln-bare}) also applies to $n=1$ and the inequality (\ref{Ln-n=1-bare}) is saturated,
for models such as the Arctan and Sqrt models, where there is a gap between $K'$ and
higher derivatives, see Eq.(\ref{gap-Arctan}), $K' \gg K''\chi$ and the scaling
(\ref{Ln-bare}) only applies to $n \geq 2$ while the inequality (\ref{Ln-n=1-bare}) is
very far from being saturated.
From Eq.(\ref{Delta-L-counterterm}), we obtain in the nonlinear regime,
as in Eq.(\ref{L-L-loop-nonlinear}),
\be
(\Delta {\cal L}^{(1)}_L)_{(n)} \sim  \frac{\cM^4\bar{K}' \bar\chi}{\bar\chi^n}
\left( \frac{m_\phi^4}{\cM^4 \bar{K}' \bar\chi} \right)^{\! L}
\left( \frac{\bar{K}'' \bar\chi}{{\bar K}'} \right)^V
\left( \frac{\partial}{\bar{p}} \right)^{\ell} ,
\label{Ln-counterterm}
\ee
with $0 \leq \ell \leq 2 n$. The first factor $\bar\chi^{-n}$ comes from
$\partial/\partial\chi \sim \bar\chi^{-1}$, as in Eq.(\ref{Ln-bare}), whereas the new factor
$(\partial/\bar{p})^{\ell}$ comes from the terms $m_{\phi}$ that also involve derivatives of
$\chi$. From Eq.(\ref{m-def}), the squared mass $m_{\phi}^2$ contains first- and 
second-derivative terms $\partial\chi$ and $\partial^2\chi$. When we consider the fluctuations
$\chi = \bar\chi+\delta\chi$, these terms can generate factors $\partial\delta\chi$ and
$\partial^2\delta\chi$.
In the Feynman diagrams where $p \sim m_{\phi}$ as in Eq.(\ref{deltaL-nloop}), these
derivative terms yield powers of
$m_{\phi}/\bar{p} \sim \sqrt{\bar{K''} \bar\chi/\bar{K}'} \lesssim 1$,
from Eq.(\ref{mphi-pbar}).
Therefore, the dominant behavior is set by the terms $\ell =0$.
Then, we obtain
\be
n=1 : \;\;\; \frac{(\Delta {\cal L}^{(1)}_L)_{(1)}}{{\cal L}^{(0)}_{(1)} } \sim
\left( \frac{m_\phi^4}{\cM^4 \bar{K}' \bar\chi} \right)^{\! L}
\left( \frac{\bar{K}'' \bar\chi}{{\bar K}'} \right)^V ,
\ee
and
\be
n \geq 2 : \;\;\; \frac{(\Delta {\cal L}^{(1)}_L)_{(n)}}{{\cal L}^{(0)}_{(n)} } \sim
\left( \frac{m_{\phi}^4}{{\cal M}^4 \bar{K}' \bar\chi} \right)^{\! L}
\left( \frac{\bar{K}''\bar\chi}{\bar{K}'}\right)^{V-1}  ,
\ee
where we used the first equality (\ref{Ln-n=1-bare}) and the scaling (\ref{Ln-bare}).
In the regime (\ref{Lagrangian-small-all-L}), where the loop contributions to the
effective action $\Gamma^{(1)}$ at the first step ``(1)'' were small, we have for $V \geq 1$
(i.e., for the counterterms associated with two-loop diagrams and beyond)
\be
n \geq 1, \;\;\; V \geq 1: \;\;\; \frac{(\Delta {\cal L}^{(1)}_L)_{(n)}}{{\cal L}^{(0)}_{(n)} } \ll
\left( \frac{\bar{K}''\bar\chi}{\bar{K}'}\right)^{V-1} \lesssim 1 ,
\label{Ln-vertices-V}
\ee
where we used that $\bar{K}''\bar\chi/\bar{K}' \lesssim 1$ for all models considered in this
paper.

Therefore, the counterterms $\Delta{\cal L}^{(1)}_L$ generated beyond one-loop order,
$V \geq 1$, yield vertices that are negligible as compared with the bare
ones (\ref{vertex-Fourier}). Then, the quantum corrections to the effective action
$\Gamma^{(2)}$ due to these counterterms are negligible, as compared with
those derived in section~\ref{sec:Higher-orders} at the previous step for $\Gamma^{(1)}$. Thus, they do not invalidate the estimates of the quantum effects that we have obtained
in section~\ref{sec:Higher-orders} and the classicality condition (\ref{Lagrangian-small-all-L}).

The one-loop term gives
\be
L=1, \;\; V=0 : \;\;\; (\Delta {\cal L}^{(1)}_1)_{(n)} \sim \frac{m_{\phi}^4}{\bar\chi^n} ,
\label{Ln-1loop-ampli}
\ee
and
\be
n = 1 : \;\; \frac{(\Delta {\cal L}^{(1)}_1)_{(1)}}{{\cal L}^{(0)}_{(1)} }
\sim \frac{m_{\phi}^4}{{\cal M}^4 \bar{K}' \bar\chi} \ll 1 ,
\label{L1-1loop-ampli}
\ee
\be
n \ge 2 : \;\; \frac{(\Delta {\cal L}^{(1)}_1)_{(n)}}{{\cal L}^{(0)}_{(n)} } \sim
\frac{m_{\phi}^4}{{\cal M}^4 \bar{K}' \bar\chi} \;
\frac{\bar{K}'}{\bar{K}''\bar\chi}  .
\label{Ln2-1loop-ampli}
\ee
We can see that the new vertices generated by the one-loop counterterm are negligible
over the full regime (\ref{Lagrangian-small-all-L}) provided
$\bar{K}' \lesssim \bar{K}'' \bar{\chi}$.
This holds for the Cubic model and power-law models, where $\bar{K}''\bar\chi \sim \bar{K}'$,
but this is not valid for the Arctan and Sqrt models where
$\bar{K}''\bar\chi \ll \bar{K}'$. Therefore, in such cases we need a more careful analysis.

\subsubsection{Further analysis for the case ${\bar K}'' \bar\chi \ll {\bar K}'$}
\label{sec:Arctan-Sqrt-quantum-from-counterterms}

In fact, even in the case ${\bar K}'' \bar\chi \ll {\bar K}'$, the amplification
(\ref{Ln2-1loop-ampli}) of the vertices (for $n \geq 2$) is not sufficient to spoil the result
(\ref{Lagrangian-small-all-L}). This is because the initial quantum corrections at two-loop
order and beyond were negligible to start with, see Eq.(\ref{L-L-loop-L-one-loop-nonlinear}),
so that their amplification by the factor (\ref{Ln2-1loop-ampli}) is not sufficient to make them
harmful.

Because the expansions (\ref{deltaLn-deltaDeltaLn}) in $\delta\chi$ mix different
powers in the field $\phi$, as $\delta\chi$ contains linear and quadratic powers of
$\phi$ in Eq.(\ref{dchi-phi}), we first consider with care the linear and quadratic terms
in $\phi$.
At order $n=1$, the result (\ref{L1-1loop-ampli}) implies that in the regime
(\ref{Lagrangian-small-all-L}) the factors associated with $(\Delta {\cal L}_1)_{(1)}$
are negligible.
This means that the term in $\Delta{\cal L}^{(1)}$ that is linear
in $\phi$ is negligible, as compared with the tree-level  one, and that the contribution
of $(\Delta {\cal L}^{(1)})_{(1)} \delta\chi$ to the quadratic term $\phi^2$ is also negligible.
The remaining quadratic term in $\phi$ that could be harmful comes from the contribution
$(\Delta{\cal L}^{(1)}_1)_{(2)} (\delta\chi)^2$, which yields
\be
(\Delta{\cal L}^{(1)}_1)_{(2)} (\delta\chi)^2 \supset \frac{m_{\phi}^4}{\bar\chi^2}
\left( \frac{X^{\mu}D_{\mu}\phi}{{\cal M}_K^2} \right)^2
\sim \frac{m_{\phi}^4}{{\cal M}^4 \bar{K}' \bar\chi} ( D \phi )^2 ,
\ee
and the comparison with Eq.(\ref{L2-1}) shows that this quadratic term is also negligible
with the tree-level one in the regime (\ref{Lagrangian-small-all-L}).
This improvement over what the result (\ref{Ln2-1loop-ampli}) would suggest comes from
the fact that the quadratic term in $\phi$ in the tree-level Lagrangian
includes a contribution from the $n=1$ term in the expansion (\ref{deltaL-phi-n}),
which is much greater than the contribution from the $n=2$ term when
${\bar K}'' \bar\chi \ll {\bar K}'$.
Therefore, both the linear and quadratic terms in $\phi$ are only modified by negligible
corrections from the one-loop (and beyond) counterterm Lagrangian $\Delta {\cal L}^{(1)}$,
\be
\delta ( \Delta{\cal L}^{(1)})_2 \ll \delta{\cal L}^{(0)}_2 ,
\label{Delta-L2-phi}
\ee
where the quadratic part of the bare Lagrangian was given in Eq.(\ref{L2-exact}).
Thus, the background field $\bar\varphi$ and the mass $m_{\phi}^2$ are not modified.

\subsubsection{Quantum corrections to the effective action $\Gamma^{(2)}$}
\label{sec:quantum-Gamma2}

From Eq.(\ref{Ln2-1loop-ampli}), the cubic and higher-order vertices
in $\phi$ are multiplied at most by a factor $m_{\phi}^4/{\cal M}^4\bar{K}''\bar{\chi}^2$.
Because the linear and quadratic terms in $\phi$ are not modified,
the contributions to the effective action $\Gamma^{(2)}$ generated at $L$-loop order
by the previous-step counterterm $\Delta{\cal L}^{(1)}$ can be obtained from
Eq.(\ref{deltaL-nloop}), which we multiply by a factor
$m_{\phi}^4/{\cal M}^4\bar{K}''\bar{\chi}^2$  for each new vertex appearing in the Feynman
diagrams.
Moreover, as we only consider leading orders and dimensional estimates
in this analysis, the new Feynman diagrams associated with these
new vertices only start at two-loop level, because the one-loop diagrams, which only involve
the quadratic part $\delta{\cal L}_2$ of the Lagrangian,  have not changed and were
already taken into account at the first step ``(1)'' of the algorithm.
This gives
\beqa
L \geq 2 : \;\; \S\pounds^{(2)}_L & \sim & \cM^4 \bar{K}' \bar\chi
\left( \frac{m_\phi^4}{\cM^4 \bar{K}' \bar\chi } \right)^{L}
\left( \frac{\bar{K}'' \bar\chi}{{\bar K}'} \right)^V \nonumber \\
&& \times \left( \frac{m_\phi^4}{\cM^4 \bar{K}'' \bar\chi^2} \right)^{\! V_1} \; I_L ,
\label{new-Feynman-Arctan}
\eeqa
where $1\leq V_1 \leq V$ is the number of new vertices, associated with the
Lagrangian counterterm $\Delta{\cal L}^{(1)}$. Indeed, we have to take into account
the fact that new contributions also arise from combinations of the old vertices,
that were included in the bare Lagrangian ${\cal L}_{\rm classical}$, with the new vertices
associated with $\Delta{\cal L}^{(1)}$.
Here, $I_L$ are again divergent dimensionless integrals, which we regularize by
dimensional regularization.
This leads to new counterterms $\Delta{\cal L}^{(2)}_L$ to the Lagrangian, which cancel
these divergences, so that the contribution
$\Delta\pounds^{(2)}_L=\S\pounds^{(2)}_L+\Delta{\cal L}^{(2)}_L$ to the second-step
effective action $\Gamma^{(2)}$ is finite.
Then, we obtain from Eq.(\ref{new-Feynman-Arctan})
\beqa
L \geq 2 : \;\; \Delta\pounds^{(2)}_L & \sim & \cM^4 \bar{K}' \bar\chi
\left( \frac{m_\phi^4}{\cM^4 \bar{K}' \bar\chi } \right)^{L+V_1}  \nonumber \\
&& \times \left( \frac{\bar{K}'' \bar\chi}{{\bar K}'} \right)^{V-V_1} .
\label{pounds-2}
\eeqa
Using Eq.(\ref{L0-approx-noLambda}) and $K''\chi/K' \lesssim 1$,
we obtain in the regime (\ref{Lagrangian-small-all-L})
\be
m_{\phi}^4 \ll {\cal M}^4 \bar{K}' \bar{\chi} : \;\;\;
\Delta\pounds^{(2)}_L \ll {\cal L}_{\rm classical} \;\; \mbox{for all} \;\; L .
\label{L+DL-L0-Arctan}
\ee
Thus, the comparison of Eq.(\ref{pounds-2}) with Eq.(\ref{L-L-loop-nonlinear}) shows that
the two-loop and higher-order contributions generated at the second step ``(2)'' of the
algorithm are greater than those generated at the first step ``(1)'', when we approach
the classicality
boundary (\ref{Lagrangian-small-all-L}), in the case of models where $K''\chi/K' \ll 1$.
[For models where $K''\chi/K' \sim 1$ in the nonlinear regime, they have the same order of
magnitude as we approach the classicality boundary (\ref{Lagrangian-small-all-L}).]
This is due to the last amplification factor in Eq.(\ref{Ln2-1loop-ampli}).
However, even in that case, the contributions $\Delta\pounds^{(2)}_L$ to the effective action
$\Gamma^{(2)}$ remain negligible with respect to the initial bare contribution ${\cal L}_{\rm classical}$
over the full regime (\ref{Lagrangian-small-all-L}).
On the other hand, if $K''\chi/K' \sim 1$, the contributions generated at the
second step ``(2)'' are suppressed with respect to those that were generated at the first step
``(1)'' by the additional factor $(m_{\phi}^4/{\cal M}^4 \bar{K}' \bar\chi)^{V_1}$.

\subsection{Beyond the second step of the renormalization algorithm}
\label{sec:beyond-second-step}

\subsubsection{Series of quantum corrections and classical regime}
\label{sec:series-of-counterterms}

The result (\ref{L+DL-L0-Arctan}) is not
sufficient to ensure that the classical equations of motion derived from the tree-level
Lagrangian ${\cal L}_{\rm classical}$ are valid in the regime (\ref{Lagrangian-small-all-L}).
Indeed, the counterterm $\Delta{\cal L}^{(2)}$ introduced to renormalize
$\S\pounds^{(2)}$ will give rise to new Feynman diagrams, which will also require new
counterterms $\Delta {\cal L}^{(3)}$, and so on, so that we follow the infinite sequence
(\ref{sequence-renormalization}) described in section~\ref{sec:renormalization-notations}.
We now need to check that this series does not blow up at later steps.
In the analysis presented in the previous sections, we estimated the contributions associated
with different loop orders in the perturbative Feynman diagrams, but for our current purpose,
we only need the estimate of the global contributions $\Delta{\cal L}^{(n)}$ and
$\Delta\pounds^{(n)}$ at step ``(n)'' of the algorithm, in the regime
$m_{\phi}^4 \ll {\cal M}^4 \bar{K}'\bar\chi$.

At step ``(0)'', i.e., at the tree-level, we start with the bare Lagrangian, and we have as
in Eq.(\ref{renorm-level0}),
\be
{\cal L}^{(0)} = \pounds^{(0)} = {\cal L}_{\rm classical} \sim {\cM^4} \bar{K}' \bar\chi .
\label{step-0}
\ee
At the end of the first step, ``(1)'', we have taken into account the radiative contributions associated
with the Lagrangian ${\cal L}^{(0)}$. As we have seen in Eq.(\ref{L-L-loop-nonlinear}),
they are dominated by the one-loop diagrams and we obtain
\be
\Delta{\cal L}^{(1)} \sim \Delta\pounds^{(1)} \sim  \cM^4 \bar{K}' \bar\chi
\left( \frac{m_\phi^4}{\cM^4 \bar{K}' \bar\chi } \right) .
\ee
At the end of the second step, ``(2)'', we have taken into account the new radiative contributions
brought by the new counterterm $\Delta{\cal L}^{(1)}$ to the Lagrangian.
From Eq.(\ref{pounds-2}), they are now set by the two-loop diagrams
(because there are no new one-loop diagrams at leading order) and we obtain
\be
\Delta{\cal L}^{(2)} \sim \Delta\pounds^{(2)} \lesssim  \cM^4 \bar{K}' \bar\chi
\left( \frac{m_\phi^4}{\cM^4 \bar{K}' \bar\chi } \right)^3 ,
\ee
where we used $L \geq 2$, $V_1 \geq 1$, $V-V_1 \geq 0$ and
$\bar{K}''\bar\chi/\bar{K}' \lesssim 1$.
Thus, at the end of the second step, we obtain a new counterterm $\Delta{\cal L}^{(2)}$
that scales at most like the one obtained at the end of the first step, $\Delta{\cal L}^{(1)}$,
multiplied by a small factor $(m_\phi^4/\cM^4 \bar{K}' \bar\chi)^2 \ll 1$.
Then, performing the same analysis for the third step as we presented for the second step,
we shall find again that the quadratic part of the Lagrangian is not modified,
$\delta(\Delta{\cal L}^{(2)})_2 \ll \delta{\cal L}^{(0)}_2$ as in Eq.(\ref{Delta-L2-phi}),
while the new vertices are decreased by a factor $(m_\phi^4/\cM^4 \bar{K}' \bar\chi)^2$
with respect to the new vertices that had appeared at the second step.
This means that Eq.(\ref{pounds-2}) becomes at the third step
\beqa
L \geq 2 : \;\; \Delta\pounds^{(3)}_L & \lesssim & \cM^4 \bar{K}' \bar\chi
\left( \frac{m_\phi^4}{\cM^4 \bar{K}' \bar\chi } \right)^{L+V_1+3V_2}  \nonumber \\
&& \times \left( \frac{\bar{K}'' \bar\chi}{{\bar K}'} \right)^{V-V_1-V_2} ,
\label{pounds-3}
\eeqa
where $V_2 \geq 1$ is the number of vertices associated with $\Delta{\cal L}^{(2)}$,
$V_1$ is the number of vertices associated with $\Delta{\cal L}^{(1)}$, and
$V \geq V_1+V_2$ is the total number of vertices.
This yields
\be
\Delta{\cal L}^{(3)} \sim \Delta\pounds^{(3)} \lesssim  \cM^4 \bar{K}' \bar\chi
\left( \frac{m_\phi^4}{\cM^4 \bar{K}' \bar\chi } \right)^5 ,
\ee
which corresponds to $L=2, V_2=1, V_1=0$.
By recursion, we obtain at the step ``(n)'':
\be
\Delta{\cal L}^{(n)} \sim \Delta\pounds^{(n)} \lesssim  \cM^4 \bar{K}' \bar\chi
\left( \frac{m_\phi^4}{\cM^4 \bar{K}' \bar\chi } \right)^{2n-1} ,
\label{DeltaL-pounds-n-recursion}
\ee
which decreases with the order $n$ when $m_\phi^4 \ll \cM^4 \bar{K}' \bar\chi$.

Therefore, we find that at all steps of the algorithm (\ref{sequence-renormalization})
presented in section~\ref{sec:renormalization-notations}, in the regime
(\ref{Lagrangian-small-all-L}) the quantum contributions to the effective action
$\Gamma$ remain negligible with respect to the tree-level result, as in
Eq.(\ref{Gamma-classical}), for all models considered in this paper.
Moreover, we again find that the bare operators associated with ${\cal L}_{\rm classical}$ are not
renormalized, because the additional small quantum corrections involve derivatives
$\partial\chi$ through $m^2_{\phi}$ and vanish if we set $\partial\chi=0$.
In addition, because the terms such as Eq.(\ref{Ln-counterterm}) with $\ell \geq 1$
are subdominant, we find that at leading order the form of the quantum contributions
to the effective action is stable. As in Eq.(\ref{Delta-pounds(1)-F}), they behave as
$m_{\phi}^4 {\cal F}(m_{\phi}/\mu,\chi)$, in the sense that they vanish if we set
$\partial\chi$ to zero and that there are only zero, one or two derivatives for each $\chi$
(i.e., we do not generate increasingly high-order derivatives
$\partial^{\ell}\chi$, if we restrict to the dominant terms).

To conclude this analysis, we find that in all cases, both when ${\bar K}''\bar\chi \ll {\bar K}'$
and when ${\bar K}''\bar\chi \sim {\bar K}'$, the classicality condition remains given by
the lowest-order result (\ref{one-loop-small-1}), even when we take into account higher-order
loop contributions and the counterterms introduced by the renormalization procedure,
\be
\mbox{classical regime:} \;\;\;\;
m_{\phi}^4 \ll {\cal M}^4 \bar{K}' \bar{\chi} ,
\label{classical-regime-1}
\ee
which again also reads as
\be
\mbox{classical regime:} \;\;\;\;
\bar{p} \ll {\cal M} \, (\bar{K}' \bar{\chi})^{1/4} \left( \frac{\bar{K}'}{\bar{K}''\bar{\chi}} \right)^{1/2} .
\label{classical-regime-2}
\ee
When this condition is satisfied the K-mouflage theory is well described by its classical
equations of motion and its tree-level Lagrangian.

\subsubsection{Nonrenormalization of the tree-level Lagrangian}
\label{sec:Non-renormalization}

We have seen above that the bare-Lagrangian operators are
not renormalized at one-loop order, as the perturbative diagrams involve derivatives of
$\bar\chi$ whereas the bare Lagrangian is only a function of $\bar\chi$.
This generalizes to higher orders of perturbation theory as we have already mentioned. This can be confirmed independently.
Indeed, let us consider the case where the background vectors $X_{\mu}$ defined in
Eq.(\ref{Xmu-Dmu-Mk-def}) are constant, which also implies that the background kinetic term
$\bar\chi=-X^{\mu}X_{\mu}/2$ is constant, so that their derivatives vanish.
Then, the quadratic Lagrangian (\ref{L2-exact}) reads as
\beqa
\mbox{if  }  X_{\mu} \mbox{ is constant:} \;\; \delta{\cal L}^{(0)}_2 \! & \!\! = &
- \left( \! \eta^{\mu\nu} - \frac{\bar{K}''}{\bar{K}'} X^{\mu} X^{\nu} \! \right)
\frac{\partial_{\mu}\phi\partial_{\nu}\phi}{2} \nonumber \\
& = & - \frac{1}{2} \tilde\partial^{\mu} \phi \tilde\partial_{\mu}\phi ,
\eeqa
where in the last expression we made the global change of coordinates
$x \rightarrow \tilde{x}$ that diagonalizes the constant symmetric matrix
$(\eta^{\mu\nu}-\bar{K}''X^{\mu}X^{\nu}/\bar{K}')$.
In particular, the quadratic mass term vanishes and the propagator reads as
$G_0(\tilde{x}_1,\tilde{x}_2)= \int d^4 \tilde{p} \, e^{\ii \tilde{p} \cdot \tilde{x}} / \tilde{p}^2$.
From Eq.(\ref{Xmu-Dmu-Mk-def}), we also have $D_{\mu}\phi=\partial_{\mu}\phi$,
and the Lagrangian (\ref{deltaL-cm-def}) takes the form
\be
\delta{\cal L}_{\rm classical} \sim {\cal M}^4 \sum_{m=2}^{\infty} \tilde{c}_m
\frac{(\tilde\partial\phi)^m}{{\cal M}_K^{2m}} ,
\ee
where the $\tilde{c}_m$ and the scale ${\cal M}_K$ are constant factors.
Therefore, the Feynman diagrams (\ref{Gamma-def}) are integrals over integer power laws,
such as $\Gamma \sim \int_0^{\infty} d\tilde{p} \, \tilde{p}^n$,
which correspond in configuration space to integrals over powers of the Dirac distribution
such as $\Gamma \sim \int d^4\tilde{x}_1 d^4\tilde{x}_2 \delta_D(\tilde{x}_1-\tilde{x}_2)^m$.
As in the one-loop computation, we consider such divergent contributions, associated
with massless theories, to vanish in the dimensional regularization.
Then, we find that there are no physical high-order corrections to the effective action
and we have $\Gamma[\bar\varphi] = S[\bar\varphi]$ in the case where the background
vectors $X_{\mu}$ are constant.
[This also means that the algorithm (\ref{series-DeltaL-Deltapounds}) stops at the
first step.]
This implies that the quantum corrections, estimated in Eqs.(\ref{deltaL-one-loop}) and
(\ref{deltaL-nloop}), involve the derivatives $\partial_{\nu}X_{\mu}$.
Therefore, the bare Lagrangian, which only involves the kinetic term
$\bar\chi= - X^{\mu}X_{\mu}/2$ without any dependence on its derivatives, is not
renormalized at any order of perturbation theory.

\subsection{Comparison with previous works}
\label{sec:de-Rham}

We can now compare the results presented in the previous sections with other works.
The impact of quantum corrections on such K-mouflage models was investigated
in \cite{deRham2014}, considering both the generic case and the Dirac-Born-Infeld (DBI)
cases \cite{Tseytlin1999,Alishahiha2004,Martin2008}
\beqa
{\rm DBI}^- : \;\; K(\chi) = -\sqrt{1-\chi} , \label{DBI-def} \\
{\rm DBI}^+ : \;\;  K(\chi) = \sqrt{1+\chi} - 2 ,
\label{DBI+def}
\eeqa
where the choice of sign depends on whether we wish the model to provide nonlinear
screening in the cosmological regime (${\rm DBI}^-$) or in the small-scale astrophysical
regime (${\rm DBI}^-$).
First, considering the one-loop contribution $\Delta\pounds^{(1)}_{\rm 1-loop}$
to the effective action as in section~\ref{sec:one-loop}, they obtain the one-loop criterion
(\ref{one-loop-small-1}), which they write in the form
\be
\left|\frac{\partial^2 \bar{K}'}{\bar{K}'}\right|^2 \ll {\cal M}^4  \bar{K} , \;\;\;
\left|\frac{\partial \bar{K}'}{\bar{K}'}\right|^4 \ll {\cal M}^4  \bar{K} .
\label{deRham-1loop}
\ee
This is equivalent to our result (\ref{one-loop-small-1}). On the right-hand side of
Eq.(\ref{one-loop-small-1}), we replaced $K$ by $\bar{K}'\bar\chi$ to remove the
cosmological constant in the low-$\bar\chi$ regime (i.e. to remove the irrelevant constant term
from the comparison in the expansion $K=-1+\chi+....$). On the left-hand side,
the  first constraint (\ref{deRham-1loop}) arises from the first term in Eq.(\ref{m-def}),
while the second constraint (\ref{deRham-1loop}) arises from both the first and second
terms in Eq.(\ref{m-def}).
In generic cases where $K'' \sim K'/\chi$, such as the Cubic model studied in this paper,
these two terms and the two conditions (\ref{deRham-1loop}) are of the same order
and $m_{\phi} \sim \bar{p}$ in Eq.(\ref{mphi-pbar}).
In the DBI models, in the nonlinear regime where we approach the square-root singularities
(\ref{DBI-def})-(\ref{DBI+def}), we have $K'' \gg K'/\chi$ and $m_{\phi} \gg \bar{p}$,
but the two conditions (\ref{deRham-1loop}) remain of the same order (because of the
$K'''$ term included in the first condition).
In fact, we do not consider such models in this paper, because we have seen in
\cite{Brax:2014c} that they are ruled out by Solar System or cosmological constraints.
Instead, we consider the opposite models where $K'' \ll K'/\chi$ and $m_{\phi} \ll \bar{p}$,
such as the Arctan and Sqrt models defined in Eqs.(\ref{Arctan-def}) and (\ref{Sqrt-def}).
There, the first condition (\ref{deRham-1loop}) dominates over the second one, and
we have in Eq.(\ref{m-def})
$m_{\phi}^2 \simeq \bar{K}''/2\bar{K}' (\partial_{\mu}\partial^{\mu}\bar\chi)
\sim (\bar{K}''\bar\chi/2\bar{K'}) \bar{p}^2$.

As in \cite{deRham2014}, we have checked in section~\ref{sec:classical-one-loop}
that one-loop
quantum corrections are negligible in cosmological and astrophysical backgrounds
(up to very high redshifts and down to very small scales).
In addition, we have checked that this remains true for the new class of models,
such as the Arctan and Sqrt models, that obeys Solar System constraints.
This shows that one can use the tree-level classical equations of motion far in the
nonlinear screening regime, according to this one-loop criterion.

Next, we have developed in sections~\ref{sec:Higher-orders}-\ref{sec:beyond-second-step}
a systematic study of higher-order quantum corrections.
In this approach, we consider the K-mouflage models as quantum field theories for which 
we construct step by step the renormalized effective action. In the process, we introduce
counterterms which cancel the logarithmic divergences and remove the power-law 
divergences using dimensional regularization. We do this as we insist on the independence 
of the renormalized action of any cutoff which could be introduced to regularize the theory. 
The remaining logarithmic corrections in the renormalized effective action depend on the
renormalization scale $\mu$, and we could have written down renormalization group 
equations for the coupling constants of all the operators in the effective action. 
We are not preoccupied with the existence of either an infrared or ultraviolet fixed point 
for this action. Indeed, we are only interested in the ``classical'' regime where the corrections 
to the tree-level Lagrangian are small.

Hence, we are following an approach different
from that  in \cite{deRham2014}.
These authors present an ``exact renormalization group'' (ERG) approach,
using Wetterich's formalism \cite{Wetterich1993} where the cutoff scale is explicit and the 
issue of ``naturalness'' is envisaged, i.e., whether the tree-level K-mouflage action appears 
naturally without fine-tuning when one integrates out fluctuations from high energy 
to low energy.
There, one considers that the classical bare action  $S[\varphi]$ of Eq.(\ref{S-def})
[i.e., the bare Lagrangian ${\cal L}_{\rm classical}$] defines the effective action
$\Gamma_{\Lambda_c}[\varphi]=S[\varphi]$ at a high cutoff scale $\Lambda_c$,
which is much greater than the strong-coupling scale ${\cal M}$.
Thus, the K-mouflage theory is only an effective field theory (EFT), which does not hold up to
arbitrarily large momenta, but only up to a cutoff scale $\Lambda_c$ beyond which one
must take into account new physics (e.g., new degrees of freedom).
Then, if we freeze all momenta below some auxiliary infrared scale $\kappa$
(which is technically performed by adding a Gaussian infrared regulator
$\varphi\cdot R_{\kappa} \cdot \varphi$ to the action that gives a very large mass
to the modes $k \ll \kappa$ while leaving the higher modes unaffected), when
$\kappa=\Lambda_c$ we must have $\Gamma_{\kappa}=S$ (by the definition of $S$
as the effective action at the scale $\Lambda_c$), while when $\kappa$ decreases
we gradually take into account the fluctuations of the modes $\kappa < k < \Lambda_c$,
which make the effective action $\Gamma_{\kappa}$ depart from the initial value
$\Gamma_{\Lambda_c}=S$.
In the limit $\kappa\rightarrow 0$, one recovers the exact effective action of the theory,
where we include the contributions of all modes below $\Lambda_c$.

Of course, the ERG equation is a nonlinear functional equation that usually cannot
be explicitly solved. Thus, \cite{deRham2014} uses a derivative expansion of the effective
action, writing
\be
\Gamma_{\kappa} = \int d^4x \, {\cal M}^4 \left[ K_{\kappa}(\chi)
+ \mbox{higher derivative terms} \right] ,
\label{Gamma-kappa}
\ee
to obtain the evolution of $\Gamma_{\kappa}$ with $\kappa$ at leading order,
and next takes into account higher-derivative terms (in the regime where they are
subdominant and within some approximations).
Requiring that these corrections to the effective action are small, as compared with the
initial value $\Gamma_{\Lambda_c}=S$, gives a condition of the form
\be
\frac{\Lambda_c^4}{\bar{K}'} \left( 1 + \frac{\Lambda_c^4}{\bar\chi \bar{K}'^2{\cal M}^4}
+ ... \right) \ll {\cal M}^4 \bar{K} ,
\label{Wetterich-condition}
\ee
where the left-hand side is the quantum correction in the regime where $\bar{K}' \gg 1$.
Then, far in the screening regime, when $\bar{K}' \rightarrow \infty$, the quantum
contribution is small even though $\Lambda_c \gg {\cal M}$, and the higher-order
terms in Eq.(\ref{Wetterich-condition}) are increasingly suppressed.
For instance, assuming $\Lambda_c \sim 1 \, {\rm eV}$, while
${\cal M} \sim 10^{-3} \, {\rm eV}$ as it corresponds to the dark-energy scale,
see Eq.(\ref{M-scale-de0}), they find that quantum contributions are small down to the
Solar System (but not close to compact objects like the Sun and planets).

The approach presented in this paper follows a different strategy. Instead of computing
the flow of $\Gamma_{\kappa}$ from a high-energy cutoff scale $\Lambda_c$,
we extend the one-loop computation of the radiative contributions to $\Gamma$
through a perturbative approach, where we sum all Feynman diagrams
(at the level of dimensional analysis since we cannot compute explicitly all these
diagrams). A first difference with the approach of \cite{deRham2014}
is that we do not introduce
a cutoff scale $\Lambda_c$. More precisely, such a cutoff scale, which could be introduced
to regularize the ultraviolet divergences of the Feynman diagrams, disappears through
a renormalization procedure (we use dimensional renormalization).
In particular, as in the one-loop computation (\ref{Gamma-1-1loop}), the power-law
divergences are removed and we are left with logarithms $\ln(m_{\phi}^2/\mu^2)$,
whereas the higher-order results of \cite{deRham2014}
are actually dominated by these
power-law divergences; see Eq.(\ref{Wetterich-condition}). This is why our classicality
criterion (\ref{classical-regime-1}), which coincides with the one-loop result
(\ref{one-loop-small-1}) (because in this regime higher-order corrections are actually
increasingly small), is different from their result (\ref{Wetterich-condition}).
Our renormalization procedure also requires the introduction of the sequential
algorithm (\ref{sequence-renormalization}) to take into account the counterterms that
are generated by this renormalization.
A second difference is that, because the terms associated with power-law divergences
are cancelled, the bare Lagrangian ${\cal L}_{\rm classical}$ is not renormalized. Contrary to
the higher-order computation presented in \cite{deRham2014}, associated with
Eqs.(\ref{Gamma-kappa}) and (\ref{Wetterich-condition}),
within our framework the nonrenormalization
of the bare operators obtained at one loop in Eq.(\ref{Gamma-1-1loop})
(where we have also cancelled power-law divergences) extends to higher orders;
see Eq.(\ref{DeltaL-pounds-n-recursion}) and section~\ref{sec:Non-renormalization}.
Nevertheless, we can note that in both frameworks the quantum corrections become
increasingly negligible in the nonlinear screening regime, $\bar{K}' \rightarrow \infty$.
Thus, the quantum corrections do not spoil the K-mouflage screening mechanism,
and one can make accurate predictions on relevant cosmological and astrophysical
scales.

More generally, our approach follows a bottom-up perspective where the initial K-mouflage 
theory is given, and one tries to ascertain the maximal domain of validity of its predictions 
in the quantum domain, i.e. when quantum corrections are taken into account. 
This differs from the approach of \cite{deRham2014} where a top-down approach is 
advocated. The latter assumes that the K-mouflage theory makes sense at high energy and 
requires that its extension to lower energies receives negligible corrections. The former gives 
a criterion of classicality which allows one to guarantee that this is a consistent 
approximation. 
We will make use of this criterion to probe the high-energy behavior of 
K-mouflage, and we will check that the classical Lagrangian allows us to make quantitative
predictions in all cases of practical interest.
In fact, we will find in section~\ref{sec:quantum-collision} that the theory becomes 
increasingly classical at high energy in the scalar sector.
This means that we do not need to interpret the K-mouflage model as an EFT and
that the K-mouflage model is nevertheless predictive.

Moreover, we will see in section~\ref{sec:UV-completion} from unitarity constraints
that for healthy K-mouflage models that pass the tight 
Solar System tests, there cannot be any standard UV completion.
Therefore, the usual top-down approach cannot be applied and one must turn to
the bottom-up approach presented in this paper.

\subsection{Quantum contributions in highly nonlinear backgrounds}
\label{sec:quantum-nonlinear}

The fact that the classicality condition remains given by the lowest-order result
(\ref{one-loop-small-1}), as seen in (\ref{classical-regime-1}), means that the results
obtained in sections~\ref{sec:one-loop-cosmos} and \ref{sec:one-loop-static} remain
valid when we consider higher-order contributions:
the theory remains classical up to redshifts much higher than $z_{\rm BBN}$ and
far within astrophysical systems and compact objects.
We now derive more precisely the location of the quantum transition in the highly
nonlinear regime in this section.

\subsubsection{Cosmological background}
\label{sec:Higher-orders-cosmological}

We have seen that quantum contributions for the cosmological background are negligible
up to very high redshifts, before BBN, but it is interesting to estimate the redshift when
the K-mouflage theory enters the quantum regime.
We shall find that for the Cubic model the transition from the early-time quantum regime
to the late-time classical regime occurs after the inflation era but very early in the
radiation era, whereas for the Arctan and Sqrt models there is no quantum era
as quantum corrections to the classical action become increasingly negligible at higher $z$.

For the cosmological background the Klein-Gordon equation (\ref{KG-matter})
of the scalar field can be integrated as \cite{Brax:2014a,Brax:2014b}
\be
\sqrt{2\bar{\chi}} \bar{K}' = \frac{\beta\bar\rho_0 t}{M_{\rm Pl} {\cal M}^2 a^3} ,
\label{KG-cosmo}
\ee
where $t$ is the cosmic time, while $\bar p \sim H$.
In the radiation era, where $t \sim a^2$ and $H \sim a^{-2}$ we obtain the scalings
\be
\mbox{radiation era:} \;\;\; \bar{p} \sim a^{-2} , \;\;\; \bar{\chi}^{1/2} \bar{K}' \sim a^{-1} ,
\label{radiation-p-a}
\ee
while in the inflationary era, where $a \sim e^{H_I t}$ with a constant Hubble rate $H_I$
as in an approximate de Sitter phase, we obtain
\be
\mbox{inflation era:} \;\;\; \bar{p} \sim H_I , \;\;\; \bar{\chi}^{1/2} \bar{K}' \sim a^{-3} .
\label{inflation-p-a}
\ee
We now need to consider in turns the Cubic, Arctan, and Sqrt models, as they have
different large-$\chi$ asymptotics.
\newline

\paragraph{Cubic model}
\label{sec:Cubic-aQ}

For the Cubic model, we have at large $\chi$:
\be
\mbox{Cubic model with } \chi \rightarrow \infty : \;\;\; K' \sim \chi^2, \;\; K'' \sim \chi ,
\label{Cubic-large-chi}
\ee
and we obtain from (\ref{radiation-p-a}) in the radiation era, for the behavior of
$\bar\chi$ and the upper bound (\ref{one-loop-small-2}),
\be
\mbox{radiation era:} \;\;\; \bar\chi \sim a^{-2/5} , \mbox{ we need } \bar{p} \ll a^{-3/10} .
\label{radiation-Cubic}
\ee
The comparison with (\ref{radiation-p-a}) shows that $\bar{p} \sim a^{-2}$ grows faster
with redshift than the upper bound (\ref{radiation-Cubic}). Therefore, there exists
a redshift $z_Q$, and scale factor $a_Q$, where the quantum corrections become
of the same order as the classical action and the theory enters the quantum regime.
Numerically, we find for the time where this transition occurs, defined as
$H = {\cal M} \, ( \bar{K}' \bar{\chi} )^{1/4} ( \bar{K}' / \bar{K}''\bar{\chi} )^{1/2}$,
the values
\beqa
&& a_Q \simeq 10^{-19}, \;\; \bar{\chi}_Q \sim 3 \times 10^7 , \;\;
H_Q \sim 10^{-6} \, {\rm GeV} , \nonumber \\
&& T \sim 2 \times 10^6 \, {\rm GeV} .
\label{aQ-Cubic}
\eeqa
We can see that this quantum-classical transition occurs after the end of the inflationary
stage, in the early radiation era.
\newline

\paragraph{Arctan model}
\label{sec:Arctan-aQ}

For the Arctan model, we have at large $\chi$:
\be
\mbox{Arctan model with } \chi \rightarrow \infty : \;\;\; K' \sim K_*, \;\; K'' \sim \chi^{-3} ,
\label{Arctan-large-chi}
\ee
and we obtain from (\ref{radiation-p-a}) in the radiation era, for the behavior of
$\bar\chi$ and the upper bound (\ref{one-loop-small-2}),
\be
\mbox{radiation era:} \;\;\; \bar\chi \sim a^{-2} , \mbox{ we need } \bar{p} \ll a^{-5/2} .
\label{radiation-Arctan}
\ee
In contrast with the Cubic model, the upper bound $\sim a^{-5/2}$ increases faster at
higher redshift than $\bar{p} \sim a^{-2}$. Therefore, we do not come closer to the quantum
regime in the early radiation era. In the inflationary era, we obtain
\be
\mbox{inflation era:} \;\;\; \bar\chi \sim a^{-6} , \mbox{ we need } \bar{p} \ll a^{-15/2} ,
\label{inflation-Arctan}
\ee
and the upper bound increases even faster as compared with the constant
$\bar{p} \sim H_I$.
Thus, for the Arctan model there is no transition to the quantum regime in the early
Universe. In fact, at higher redshift the theory becomes increasingly more classical,
i.e., quantum corrections to the effective action become increasingly negligible.
This is due to the fact that $K'' \chi/K'$ decreases strongly at large $\chi$ for this model,
whereas it remains of order unity in the Cubic model.
\newline

\paragraph{Sqrt model}
\label{sec:Sqrt-aQ}

For the Sqrt model, we have at large $\chi$:
\be
\mbox{Sqrt model with } \chi \rightarrow \infty : \;\;\; K' \sim K_*, \;\; K'' \sim \chi^{-4} ,
\label{Sqrt-large-chi}
\ee
and we obtain from (\ref{radiation-p-a}) in the radiation era, for the behavior of
$\bar\chi$ and the upper bound (\ref{one-loop-small-2}),
\be
\mbox{radiation era:} \;\;\; \bar\chi \sim a^{-2} , \mbox{ we need } \bar{p} \ll a^{-7/2} .
\label{radiation-Sqrt}
\ee
As for the Arctan model, the upper bound $\sim a^{-7/2}$ increases faster at
higher redshift than $\bar{p} \sim a^{-2}$. In the inflationary era, we obtain
\be
\mbox{inflation era:} \;\;\; \bar\chi \sim a^{-6} , \mbox{ we need } \bar{p} \ll a^{-21/2} ,
\label{inflation-Sqrt}
\ee
and the upper bound again increases even faster as compared with $\bar{p} \sim H_I$.
Thus, for the Sqrt model there is no transition to the quantum regime in the early
Universe, and the theory becomes increasingly more classical, as for the Arctan model.

\subsubsection{Astrophysical background}
\label{sec:Higher-orders-astrophysical}

We now investigate the case of astrophysical small-scale situations,
such as in the Solar System. We shall find that there is no transition to the quantum
regime in this context, whether on Solar System scales, on the Earth or in the laboratory.
This is because a quantum regime would only arise in a finite range of radii,
far inside isolated spherically symmetric objects, at a radius $r_Q \sim 10^{-4} \, {\rm m}$
that is also at the transition between the linear and nonlinear regimes.
But such a regime does not exist because of deviations from spherical symmetry and
screening by other objects (e.g., both the Sun and the Earth are within the K-mouflage
radius of each other).

Around a spherical compact object of mass $M$ and K-mouflage radius given by
Eq.(\ref{RK-def}), the Klein-Gordon equation of the scalar field (\ref{KG-matter})
can be integrated as \cite{Brax:2014c}
\be
\sqrt{-2\bar\chi} \bar{K}' = \left( \frac{R_K}{r} \right)^2 ,
\label{KG-static}
\ee
where $r$ is the radial distance from the central object.
This allows us to investigate possible transitions to the quantum regime around
spherically symmetric objects, but as for the cosmological case, we must consider in
turns the Cubic, Arctan and Sqrt models.
\newline

\paragraph{Cubic model}
\label{sec:Cubic-rQ}

For the Cubic model, we obtain $\bar\chi \sim r^{-4/5}$, and the condition
(\ref{one-loop-small-2}) scales as $r^{-1} \ll r^{-3/5}$, with $\bar{p} \sim r^{-1}$.
Thus, as for the cosmological background in the early universe,
at small radii $\bar{p}$ grows faster than the upper bound of Eq.(\ref{one-loop-small-2}),
which means that at a finite radius $r_Q$ we reach the quantum regime.
We obtain
\beq
r_Q \sim {\cal M}^{-1} ( {\cal M} R_K )^{-3/2}
\eeq
Since ${\cal M}^{-1} \sim 10^{-4} \,{\rm m}$ and $R_K \sim 1000 \,{\rm A.U.}$ for the Sun,
we can see that this radius $r_Q$ is tiny and irrelevant for astrophysical purposes.
However, on such small scales, we are inside the astrophysical object, and the
K-mouflage radius (\ref{RK-def}) depends on $r$ through $M(<r)$.
For a constant matter density $\rho$, we obtain
$r_Q \sim {\cal M}^{-1} (\rho/M_{\rm Pl} {\cal M}^3)^{-3/13}$, which gives
$r_Q \sim {\cal M}^{-1} \sim 10^{-4} \,{\rm m}$ for $\rho=1 \, {\rm g.cm}^{-3}$.
However, this is also at the boundary of the linear regime, and at smaller radii
we find that we are moving away from the quantum transition.
Thus, for an isolated and spherically symmetric object, the quantum regime would only
appear at most in a small range of radii $r_Q \sim {\cal M}^{-1} \sim 10^{-4} \, {\rm m}$.
However, this does not apply to practical configurations. In particular, the Earth
being within the K-mouflage radius of the Sun, the local scalar-field background is also
set by the Sun, and we do not enter this regime.
Therefore, in astrophysical or laboratory configurations, we remain in the classical regime,
and quantum corrections to the effective action are negligible.
\newline

\paragraph{Arctan model}
\label{sec:Arctan-rQ}

For the Arctan model, we obtain $\bar\chi \sim r^{-4}$ and the condition
(\ref{one-loop-small-2}) scales as $r^{-1} \ll r^{-5}$, outside of the object.
Thus, as for the cosmological background, at small radii $\bar{p}$ grows more slowly
than the upper bound of Eq.(\ref{one-loop-small-2}). This means that quantum corrections
become increasingly negligible as we move toward the compact object.
However, inside the compact object the K-mouflage radius scales as $R_K \sim r^{3/2}$
and the condition (\ref{one-loop-small-2}) scales as $r^{-1} \ll r^{5/2}$.
Thus, we now obtain a quantum transition at a finite radius
$r_Q \sim {\cal M}^{-1} (\rho/M_{\rm Pl} {\cal M}^3)^{-5/7}$, which gives
$r_Q \sim {\cal M}^{-1} \sim 10^{-4} \,{\rm m}$ for $\rho=1 \, {\rm g.cm}^{-3}$.
As for the Cubic model, this is also at the boundary of the linear regime and at smaller radii
we move away from the quantum transition, but these regimes are not
reached in practice because of deviations from spherical symmetry and screening by
other objects.
\newline

\paragraph{Sqrt model}
\label{sec:Sqrt-rQ}

For the Sqrt model, we obtain $\bar\chi \sim r^{-4}$, and the condition
(\ref{one-loop-small-2}) scales as $r^{-1} \ll r^{-7}$, outside of the object.
As for the Arctan model, quantum corrections become more negligible
closer to the compact object.
Inside the compact object, the condition (\ref{one-loop-small-2}) scales as
$r^{-1} \ll r^{7/2}$, and we obtain a quantum transition at a finite radius,
$r_Q \sim {\cal M}^{-1} (\rho/M_{\rm Pl} {\cal M}^3)^{-7/9}$, which again gives
$r_Q \sim {\cal M}^{-1} \sim 10^{-4} \,{\rm m}$ for $\rho=1 \, {\rm g.cm}^{-3}$.
As for the Cubic and Arctan models,  this is also at the boundary of the linear regime,
and at smaller radii we move away from the quantum transition, but these regimes are not
reached in practice.

\section{Quantum stability}
\label{sec:Stability}

The perturbative analysis presented in section~\ref{sec:Quantum-corrections}
relies on the assumption that $m^2_\phi >0$ in the quadratic Lagrangian
(\ref{L2-def-2}), as in standard Quantum Field Theories.
However, the mass $m_{\phi}^2$ being a nontrivial function of the background,
even in the limit $\bar\chi \bar{K}'' \ll \bar{K}'$ as in Eq.(\ref{m-def}), there is no
guarantee {\it a priori} that it always remains positive.
In fact, for the simple models (\ref{Cubic-def}) and (\ref{Arctan-def}) we shall check
that it is negative in some regimes.
This can be a problem quantum mechanically, as the perturbation theory may no longer
apply (e.g., we are expanding around a false vacuum) or the K-mouflage model itself
may be ill defined.
We first consider the static regime, associated with small-scale astrophysical backgrounds,
and next the time-dependent regime, associated with the cosmological background.

\subsection{Static background}
\label{sec:Radial}

\subsubsection{General analysis}
\label{radial-analysis}

Let us first analyze the problems that may occur when the mass term in the quadratic
Lagrangian can be negative, around a static background such as a compact astrophysical
object.
At the quadratic level, the fluctuations of the quantum field $\phi$ around the background
$\bar\varphi$ are governed by the Gaussian path integral
\beqa
Z & = & \int {\cal D}\phi \; e^{i \int d^4 x \left[ - \frac{1}{2} \partial^{\mu}\phi {\partial}_{\mu}\phi
- \frac{1}{2} m^2_{\phi} \phi^2 \right]} \nonumber \\
& = & \int {\cal D}\phi \; e^{i \int d^4 x \left[
\frac{1}{2} \left( \frac{\partial\phi}{\partial t} \right)^2  - \frac{1}{2} (\nabla \phi)^2
- \frac{1}{2} m^2_{\phi}(\vec{x}) \phi^2 \right]} .
 \label{Zr-def}
\eeqa
In the second expression, we used the fact that we consider a static background,
in the Minkowski metric, so that the mass $m_{\phi}$ only depends on the
3D spatial coordinate $\vec{x}$ and $\nabla$ is the spatial gradient.
As usual, we can interpret this oscillatory path integral as the analytic continuation
of the Euclidian path integral obtained through the Wick rotation $t=-i\tau$.
This gives
\beqa
Z_E & = & \int {\cal D}\phi \; e^{- \int d\tau d^3x \left[
\frac{1}{2} \left( \frac{\partial\phi}{\partial \tau} \right)^2 + \frac{1}{2} (\nabla \phi)^2
+ \frac{1}{2} m^2_{\phi}(\vec{x}) \phi^2 \right]} \nonumber \\
& = & \int {\cal D}\phi \; e^{-S_E} ,
\label{ZE-def}
\eeqa
which yields the Euclidian action $S_E$.
This path integral is well defined if the Euclidian action is bounded from below and
goes to infinity for large fields. This is obviously the case when the squared-mass
$m_{\phi}^2$ is positive, as in standard theories, since $S_E$ is then the sum of three
positive quadratic terms.

If the squared-mass $m_{\phi}^2$ is constant and negative, the action $S_E$ goes to
$-\infty$ for large constant fields $\phi$, and the integral (\ref{ZE-def}) is divergent
and meaningless.
This implies at least that the perturbation theory around $\phi=0$ breaks down.
This could mean that we expand around the wrong background (e.g., a local maximum
or a saddle point) and the path integral might be made convergent by higher-order terms.
However, in our case, there is a unique solution to the classical equations of motion,
by construction of the kinetic function $K$ (we ruled out models with multiple
solutions to avoid jumps of the scalar field, associated with the transitions between
different branches, and small-scale instabilities) \cite{Brax:2014a,Brax:2014c}. Then, the model is meaningless
if this is not a global minimum of the Euclidian action.

In our case, the squared-mass $m_{\phi}^2$ is not constant, and we shall see below that
at large radii, around a spherical compact object, it is positive and goes to zero at infinity,
while it is negative at small radii.
If the range of scales where $m_{\phi}^2$ is negative is sufficiently small,
the path integral (\ref{ZE-def}) will remain well defined.
Indeed, making the field $\phi$ peaked in the region $m_{\phi}^2<0$ leads to
a ``cost'' in the kinetic factor $(\nabla\phi)^2$ that may be dominant.
More precisely, the Euclidian path integral is well defined if the quadratic form
$S_E = (1/2) \int d\tau d^3x \, \phi [ - \partial_{\tau}^2 - \nabla^2 + m_{\phi}^2 ] \phi$ is
positive definite, that is, the eigenvalues $\lambda$ of the operator
$- \partial_{\tau}^2 - \nabla^2 + m_{\phi}^2$ are strictly positive,
\be
\left (-\frac{\partial^2}{\partial \tau^2} - \nabla^2 + m^2_\phi \right) \phi = \lambda \, \phi
\;\;\; \mbox{implies} \;\;\; \lambda > 0 ,
\label{lambda-def}
\ee
for normalized eigenfunctions.
In the static and spherically symmetric case, we can write the eigenvectors as
$\phi = e^{-\ii \omega \tau} \psi(r) Y_{\ell}^m(\vec{\Omega})$ and the eigenvalue equation
becomes
\be
\left[ \omega^2 - \frac{1}{r^2} \frac{d}{dr} \left( r^2 \frac{d}{dr} \right)
+ \frac{\ell (\ell+1)}{r^2} + m_{\phi}^2(r) \right] \psi = \lambda \, \psi .
\ee
The lowest eigenvalue (ground state) is obtained for $\omega=0$ and $\ell=m=0$, and
therefore we only need to check that $\lambda>0$ for the one-dimensional radial
eigenvalue problem with $\omega=0$ and $\ell=m=0$.
Writing $\psi(r)=y(r)/r$, we obtain the one-dimensional eigenvalue problem
\be
- \frac{d^2y}{dr^2} + V(r) \, y = E \, y , \;\;\; \mbox{with} \;\;\; V(r)= m^2_{\phi}(r) , \;\;\;
E = \lambda .
\label{Schrodinger}
\ee
Thus, we look for the ground-state energy $E_0=\lambda_0$ of the one-dimensional
quantum-mechanics wave function $y(r)$, in the potential $V(r)=m^2_{\phi}(r)$,
and the Euclidian path integral (\ref{ZE-def}) is well defined if $E_0>0$.
Since in our case $V(r)$ goes to zero at infinity, this means that there should be no bound
states, with a negative energy.
Then, it is known that the number of bound states $n_{\rm bound}$ is of order
\be
n_{\rm bound}  \sim \frac{1}{\pi} \int_{V(r)<0} dr \sqrt{-V(r)} \; ,
\label{Nbound}
\ee
where we integrate over the domain where $V<0$ \cite{Messiah:1962}. 
This can be obtained from the WentzelÐKramersÐBrillouin (WKB)
approximation and also corresponds to the 
Bohr-Sommerfeld quantization rule.
This means that the estimate of $n_{\rm bound}$ given in Eq.(\ref{Nbound}) must be
somewhat below unity.

It is interesting to consider the simple case where the potential $V(r)$ shows a step profile,
with two constant values, $V=V_1<0$ for $r<L$ and $V=V_2>0$ for $r>L$.
Then, for bound states with $V_1 < E <V_2$, the solution in the two domains reads as
\beqa
r<L : && \;\;\; y = A \, \sin( \sqrt{E-V_1} \, r) , \\
r>L : && \;\;\; y = B \, e^{-\sqrt{V_2-E} \, r} ,
\eeqa
and the matching at $r=L$ of $y$ and $dy/dr$ gives the eigenvalue equation of the
discrete spectrum,
\be
\tan( \sqrt{E-V_1} \, L) = - \sqrt{ \frac{E-V_1}{V_2-E} } .
\ee
Since $V_1 < E < V_2$, in order to obtain at least one ground state it is necessary that
$\sqrt{V_2-V_1}L > \pi/2$. Therefore, a sufficient condition for the absence of bound states
is
\be
\sqrt{V_2-V_1} \, L < \frac{\pi}{2} .
\label{no-bound-state}
\ee
This is consistent with the general estimate (\ref{Nbound}),
$n_{\rm bound} \sim \sqrt{V_2-V_1} L /\pi$, as it corresponds to an estimated number of
bound states of $0$.

\subsubsection{K-mouflage quantum stability in a static background}
\label{radial-analysis-K}

For the K-mouflage theories, the exact quadratic Lagrangian of quantum fluctuations
around a background $\bar\varphi$ is given by Eq.(\ref{L2-exact}).
For a static spherically symmetric background, this gives
\beqa
{\cal L}_2 & = & \frac{1}{2} \left( \frac{\partial\phi}{\partial t} \right)^2
- \frac{\bar{K}'+2\bar\chi \bar{K}''}{2\bar{K}'} \left( \frac{\partial\phi}{\partial r} \right)^2
- \frac{1}{2r^2} \left( \frac{\partial\phi}{\partial\theta}\right)^2 \nonumber \\
&& - \frac{1}{2r^2\sin^2\theta} \left( \frac{\partial\phi}{\partial\varphi}\right)^2
- \frac{\bar{m}_{\phi}^2}{2} \phi^2 ,
\label{L2-static}
\eeqa
with
\beqa
\bar{m}_{\phi}^2 & = & \left( \frac{\bar{K}''}{2\bar{K'}}
+ \frac{\bar{K}''^2\bar\chi}{\bar{K}'^2} \right)
\left( \frac{d^2\bar\chi}{dr^2} + \frac{2}{r} \frac{d\bar\chi}{dr} \right) \nonumber \\
&& \hspace{-1cm} + \left( \frac{\bar{K}'''}{2\bar{K}'} + \frac{3\bar{K}''^2}{4\bar{K}'^2}
+ \frac{2\bar{K}'''\bar{K}''\bar\chi}{\bar{K}'^2}
- \frac{3\bar{K}''^3\bar\chi}{2\bar{K}'^3} \right)
\left( \frac{d\bar\chi}{dr} \right)^2 . \hspace{0.7cm}
\eeqa
The spatial gradient term is modified along the radial direction by the background
factor $X^{\mu}X^{\nu}$ in the kinetic prefactor of Eq.(\ref{L2-exact}), but
the properties (\ref{ghosts}) ensure that this radial term keeps the standard sign.
As in section~\ref{radial-analysis}, for the model to be well defined, we require the Euclidian
action to  have only positive eigenvalues.
Again, we can look for eigenvectors of the form
$\phi = e^{-\ii \omega \tau} \psi(r) Y_{\ell}^m(\vec{\Omega})$, and the lowest eigenvalues
are obtained for $\omega=0$ and $\ell=m=0$. Thus, we are led to the one-dimensional
radial eigenvalue problem
\be
- \frac{1}{r^2} \frac{d}{dr} \left[ r^2 \frac{\bar{K}'+2\bar\chi \bar{K}''}{\bar{K}'}
\frac{d\psi}{dr} \right] + \bar{m}_{\phi}^2(r) \psi = \lambda \, \psi .
\label{eigenvalue-static}
\ee
Because of the modified kinetic factor along the radial direction, this takes the form of
a more general Sturm-Liouville problem, which we can put in the Schrodinger form
(\ref{Schrodinger}) by a change of variable \cite{Morse1953},
\be
- \frac{d^2 y}{dx^2} + V(x) \, y = E \, y , \;\;\; \mbox{with} \;\;\; E = \lambda R_K^2 ,
\label{Schrodinger-K}
\ee
and
\beqa
x & = & \int_0^r \frac{dr}{R_K}
\left( \frac{\bar{K}'+2\bar\chi \bar{K}''}{\bar{K}'} \right)^{-1/2} ,
\label{x-r-Rk-def} \\
\psi & = & \left( \frac{\bar{K}'+2\bar\chi \bar{K}''}{\bar{K}'} \right)^{-1/4}
\frac{y}{r} ,
\label{psi-r-Rk-def} \\
V & = & R_K^2 \bar{m}^2_{\phi} + \frac{1}{r}
\left( \frac{\bar{K}'+2\bar\chi \bar{K}''}{\bar{K}'} \right)^{-1/4} \nonumber \\
&& \times \frac{d^2}{dx^2} \left[ r \left( \frac{\bar{K}'+2\bar\chi \bar{K}''}{\bar{K}'}
\right)^{1/4} \right] .
\label{V-m-Rk-def}
\eeqa
Here, we introduced the K-mouflage radius $R_K$ to obtain dimensionless quantities.

Outside of the compact object, the Klein-Gordon equation obeyed by the K-mouflage
scalar field can be integrated as Eq.(\ref{KG-static})
(we have $\bar\chi<0$ in the static regime).
This means that $\bar\chi$, the radial coordinate $x$, and the potential $V(x)$ defined
in Eqs.(\ref{x-r-Rk-def}) and (\ref{V-m-Rk-def}), are fixed functions of $r/R_K$,
or equivalently of $\bar\chi$ or $x$, independently of the mass $M$ of the object.
Thus, the initial mass term reads as
\be
R_K^2 \bar{m}^2_{\phi}(r) = \frac{- (-2\bar\chi)^{3/2}}{\bar{K}'+2\bar\chi\bar{K}''}
\left[ - 4 \bar\chi \bar{K}''^2 \! + \! \bar{K}' \left( 3 \bar{K}''
\! + \! 4 \bar\chi \bar{K}^{(3)} \right) \right]
\label{m2-x-chi}
\ee
while the full potential can be expressed in terms of $\bar\chi$ as
\beqa
V(x) & = & \frac{- (-2\bar\chi)^{3/2}}{(\bar{K}'+2\bar\chi\bar{K}'')^3}
\biggl \lbrace 2 \bar\chi^3 \bar{K}''^4 + \bar\chi \bar{K}'^2
\left[ -10 (\bar\chi\bar{K}^{(3)})^2 \right . \nonumber \\
&& \left. - 15 \bar{K}''^2 + 4 \bar\chi \bar{K}'' \left( 2 \bar\chi \bar{K}^{(4)}
- \bar{K}^{(3)} \right) \right] \nonumber \\
&& + \bar{K}'^3 \left[ 4 \bar{K}'' + \bar\chi \left( 4 \bar\chi \bar{K}^{(4)}
+ 13 \bar{K}^{(3)} \right) \right] \biggl \rbrace .
\label{V-x-chi}
\eeqa
This means that compact objects with different masses lead to the same eigenvalue problem
(\ref{Schrodinger-K}), expressed in the radial coordinate $x$ and the potential $V(x)$.
This is because the dependence on the mass $M$ of the object is absorbed by the
scaling $r/R_K$ in the Klein-Gordon equation (\ref{KG-static}).
This is no longer valid inside an extended object, where the scalar-field profile depends
on the matter density profile, but in this study, we only check the validity of the model
outside spherically symmetric objects.

\begin{table*}[htb!]
\caption{Behavior at large and small distances of the squared-mass $\bar{m}_{\phi}^2(r)$
and of the potential $V(x)$ for the Cubic, Arctan and Sqrt models.}
\begin{center}
\begin{tabular}{|c||c|c|c|}
\hline
model name & Cubic & Arctan & Sqrt \\
\hline\hline

\rule{0cm}{0.5cm} $r \rightarrow \infty : \;\; R_K^2 \bar{m}_{\phi}^2(r)$ &
$21 K_0 \left( \frac{r}{R_K} \right)^{-10}$  &
$ \frac{7 K_*}{\chi_*^2} \left( \frac{r}{R_K} \right)^{-10}$ &
$ \frac{21 K_* \sigma_*^2 \chi_*}{2(\sigma_*^2+\chi_*^2)^{5/2}}
\left( \frac{r}{R_K} \right)^{-10}$ \\ [0.2cm]

\hline

\rule{0cm}{0.3cm} $r \rightarrow \infty : \;\; x(r)$ & $x \simeq \frac{r}{R_K}$
& $x \simeq \frac{r}{R_K}$ &
$x \simeq \frac{r}{R_K}$ \\ [0.1cm]

\hline

\rule{0cm}{0.5cm} $ x \rightarrow \infty : \;\; V(x)$ &
$51 K_0 \, x^{-10}$ &
$ \frac{17 K_*}{\chi_*^2} \, x^{-10}$ &
$ \frac{51 K_* \sigma_*^2 \chi_*}{2(\sigma_*^2+\chi_*^2)^{5/2}} \, x^{-10}$ \\ [0.2cm]
\hline\hline

\rule{0cm}{0.5cm} $r \rightarrow 0 : \;\; R_K^2 \bar{m}_{\phi}^2(r)$ &
$- \frac{4}{5} \left( \frac{r}{R_K}\right)^{-2}$ &
$ - 144 K_*^4 \chi_*^2 \left( \frac{r}{R_K}\right)^{6}$ &
$ - \frac{624 K_*^6 \sigma_*^2 \chi_*^6}{(\sigma_*^2+\chi_*^2)^{5/2}}
\left( \frac{r}{R_K} \right)^{10}$ \\  [0.2cm]

\hline

\rule{0cm}{0.4cm} $r \rightarrow 0 : \;\; x(r)$ & $x \simeq \frac{r}{\sqrt{5} R_K}$ & $x \simeq \frac{r}{R_K}$ &
$x \simeq \frac{r}{R_K}$ \\  [0.2cm]

\hline

\rule{0cm}{0.5cm} $x \rightarrow 0 : \;\; V(x)$ &
$- \frac{4}{25} \, x^{-2}$ &
$ 208 K_*^4 \chi_*^2 \, x^{6}$ &
$ \frac{1536 K_*^6 \sigma_*^2 \chi_*^6}{(\sigma_*^2+\chi_*^2)^{5/2}} \, x^{10}$ \\  [0.2cm]

\hline

\end{tabular}
\end{center}
\label{table-static}
\end{table*}

At large radii, we obtain $\bar\chi \rightarrow 0^{-}$ and $\bar{K}' \rightarrow 1$,
following the weak-field limit (\ref{low-chi}).
In this regime, expanding Eq.(\ref{V-x-chi}) and using the integrated Klein-Gordon
equation (\ref{KG-static}) we obtain
\beqa
r \gg R_K : && R_K^2 \bar{m}^2_{\phi}(r) = - 3 K''(0) \left(\frac{r}{R_K}\right)^{-6}
\nonumber \\
&& \hspace{-1.5cm} + \frac{-19 K''(0)^2+7 K^{(3)}(0)}{2} \left( \frac{r}{R_K} \right)^{-10}
+ O(r^{-14}) , \hspace{1cm}
\eeqa
\beqa
V(x) & = & - 4 K''(0) \left(\frac{r}{R_K}\right)^{-6}
+ \frac{17}{2} [ -3 K''(0)^2+K^{(3)}(0) ] \nonumber \\
&& \times \left( \frac{r}{R_K} \right)^{-10} + O(r^{-14}) ,
\eeqa
while Eq.(\ref{x-r-Rk-def}) gives
\be
r \gg R_K : \;\;\; x \simeq \frac{r}{R_K} .
\ee
For the three models defined in Eqs.(\ref{Cubic-def})-(\ref{Sqrt-def})
that we study in this paper, we have $K''(0)=0$ [their derivative $K'(\chi)$ is even] and
$K^{(3)}(0)>0$. This gives the behaviors shown in the first three rows in
Table~\ref{table-static}.
Thus, in these models, the potential is positive at large radii but goes to zero very fast
as $x^{-10}$. The radial kinetic prefactor in the Lagrangian (\ref{L2-static}) leads
to a full potential $V(x)$ that follows the same scaling as the initial mass
$R_K^2 \bar{m}^2$ but with a higher amplitude.

At small radii, we have $\bar\chi\rightarrow-\infty$ and $\bar{K}'\rightarrow+\infty$.
This is the screening regime where the behavior of the system depends on the nonlinearities
of the K-mouflage kinetic function $K(\chi)$.
For the three models that we consider in this paper, we obtain the behaviors shown
in the last three rows in Table~\ref{table-static}.
Thus, for the Cubic model, the radial kinetic prefactor in the Lagrangian (\ref{L2-static})
only leads to a change of scale from $r/R_K$ to $x$, and the full potential keeps the same
form as the initial mass $R_K^2 \bar{m}^2$. Both are negative and diverge as $1/x^2$.
For the Arctan and Sqrt models, the radial kinetic prefactor does not lead to a change of scale
but to a change of sign, as the initial squared mass is negative while the full potential
is positive. Both vanish at small radii as $x^6$ for the Arctan model and as $x^{10}$ for the
Sqrt model.
\newline

\paragraph{Cubic model}
\label{Cubic-static}

\begin{figure}
\begin{center}
\epsfxsize=8.5 cm \epsfysize=5.8 cm {\epsfbox{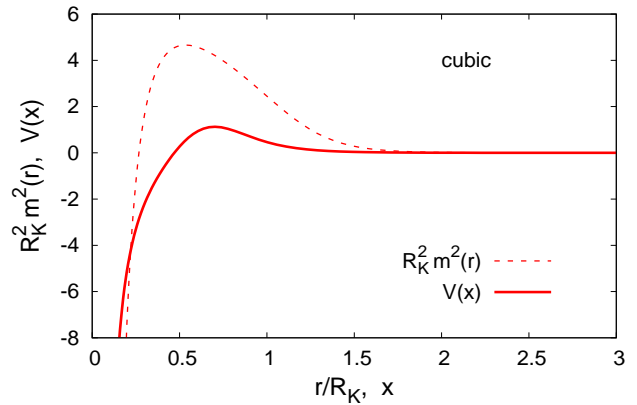}}
\end{center}
\caption{Squared-mass $R_K^2 \bar{m}^2(r)$, as a function of $r/R_K$,
and potential $V(x)$, as a function of $x$, for the Cubic model.}
\label{fig_m2_r_Cubic}
\end{figure}

We show the squared-mass $R_K^2 \bar{m}^2_{\rm Cubic}(r)$ and the potential
$V_{\rm Cubic}(x)$ of the Cubic model in Fig.~\ref{fig_m2_r_Cubic}.
The potential is negative and diverges as $-1/x^2$ at small scales;
see Table~\ref{table-static}.
This implies that the estimated number of bound states from Eq.(\ref{Nbound}) diverges,
which means that there are an infinite number of bound states with a negative energy.
This can be seen as follows.
At small radii, the eigenvalue equation (\ref{Schrodinger-K}) reads as
\be
x \ll 1 : \;\;  - \frac{d^2 y}{dx^2} - \frac{4}{25 x^2} \, y = E \, y ,
\label{Schrodinger-Cubic}
\ee
and for negative energies, the solution with a finite norm is
\be
E<0 : \;\;\; y(x) = \sqrt{x} \, K_{3/10}(\sqrt{-E} \, x) ,
\label{y-E-Cubic}
\ee
where $K_{\nu}(z)$ is the modified Bessel function of the second kind.
Thus, we have a continuous spectrum of bound states of any energy $E<0$, as the
change of energy only corresponds to a rescaling of radii, as could already be seen
in Eq.(\ref{Schrodinger-Cubic}).
Eigenvectors with $E\rightarrow -\infty$ have a typical radius of order
$x_E \sim 1/\sqrt{-E} \rightarrow 0$ and have a negligible weight at distances where
the potential $V(x)$ deviates from $-4/25x^2$. Therefore, the deviation of the true potential
from the $-1/x^2$ shape at $x \gtrsim 1$ only slightly perturbs their energy, and we still
have an infinite number of bound states.
Thus, the Cubic model has an infinite number of negative solutions $\lambda$
to the eigenvalue problem (\ref{eigenvalue-static}) and the quadratic path integral
associated with the Lagrangian (\ref{L2-static}) is ill defined.
Therefore, this model is ruled out as it is not self-consistent at the level 
of quantum fluctuations around a static background.
At the classical level, this model is consistent, but it was already ruled out by Solar System
observations, as it predicts a perihelion of the Moon which is greater than observational
constraints by many orders of magnitude.
\newline

\paragraph{Arctan and Sqrt models}
\label{Arctan-static}

\begin{figure}
\begin{center}
\epsfxsize=8.5 cm \epsfysize=5.8 cm {\epsfbox{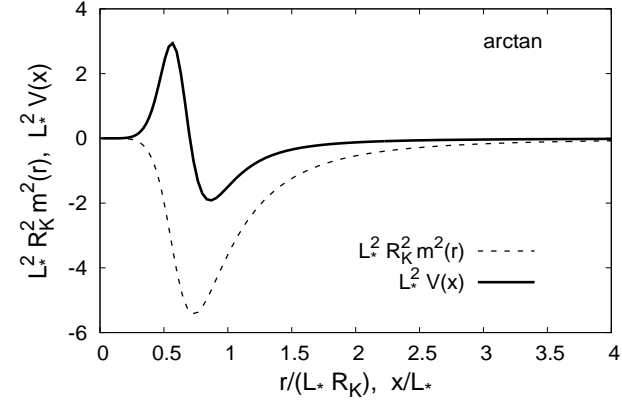}}\\
\epsfxsize=8.5 cm \epsfysize=5.8 cm {\epsfbox{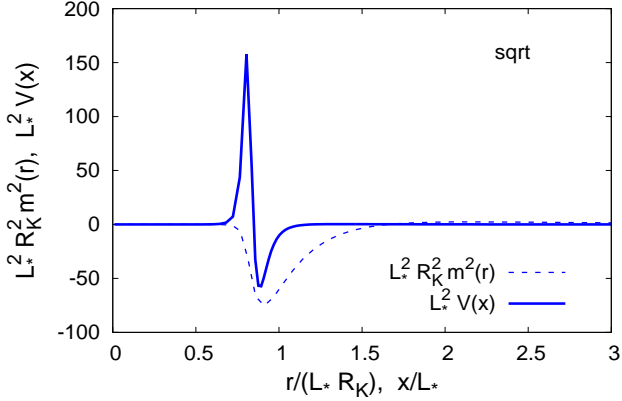}}
\end{center}
\caption{Squared-mass $R_K^2 \bar{m}^2(r)$, as a function of $r/R_K$,
and potential $V(x)$, as a function of $x$, for the Arctan and Sqrt models.}
\label{fig_m2_r_Arctan_Sqrt}
\end{figure}

We show the squared-mass $R_K^2 \bar{m}^2(r)$ and the potential
$V(x)$ of the Arctan and Sqrt models in Fig.~\ref{fig_m2_r_Arctan_Sqrt}.
Because the transition parameters $K_* = 1000$ and $\chi_*=100$ of these
models are significantly greater than unity, the transition to the nonlinear regime
occurs somewhat below $R_K$, at a scale of order $L_* R_K$ with
$L_*=\chi_*^{-1/4} K_*^{-1/2}$, from Eq.(\ref{KG-static}). Thus, in
Fig.~\ref{fig_m2_r_Arctan_Sqrt}, we rescale distances by a factor $L_*$ and potentials
by a factor $L_*^{-2}$.
For the Arctan model, the potential becomes of order unity after this rescaling, but for the
Sqrt model, the amplitude of the potential is significantly greater than unity because
the transition between the linear and nonlinear regime is very sharp; see
Fig.~\ref{fig_Kp_chi}.
In both cases, the effects of the prefactor of the kinetic radial term in the Lagrangian
(\ref{L2-static}) are to reduce the range of radii where the potential deviates from zero,
as compared with $R_K^2 \bar{m}^2(r)$, to decrease its negative part and to make the
potential turn positive at small radii.
This increases the eigenvalues $\lambda$ and makes the Euclidian action more likely
to be positive definite. Indeed, the factor $1+2\bar\chi\bar{K}''/\bar{K}'$ is greater than
unity, which increases the ``cost'' associated with the radial kinetic term in the
Euclidian action.

\begin{figure}
\begin{center}
\epsfxsize=8.5 cm \epsfysize=5.8 cm {\epsfbox{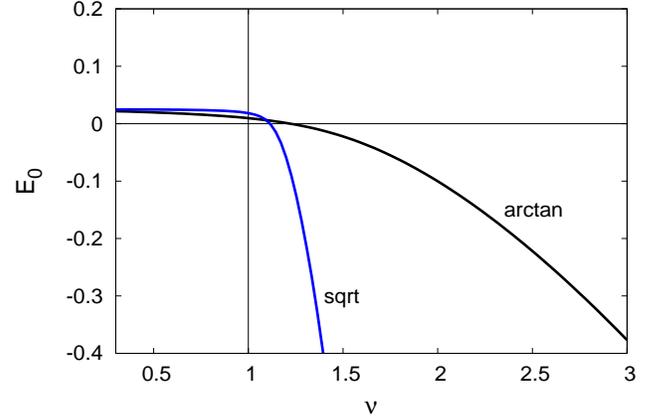}}
\end{center}
\caption{Ground-state energy $E_0$ as a function of the parameter $\nu$ that multiplies
the potential $V(x)$, for the Arctan and Sqrt models. The physical case corresponds to
$\nu=1$, where we find $E_0>0$.}
\label{fig_E_nu}
\end{figure}

For the eigenvalue problem (\ref{Schrodinger-K}), we find that the estimated number of
bound states (\ref{Nbound}) for these two potentials is
\be
n_{\rm Arctan} \sim 0.75 , \;\;\;\;\; n_{\rm Sqrt} \sim 0.36 .
\label{Nbound-Arctan}
\ee
Therefore, we expect that there is no bound state for these models, but that we are
close to the transition associated with the appearance of a ground state of negative energy.
To determine whether there is zero or one bound state, we need to perform
a numerical analysis.
We solve the eigenvalue problem (\ref{Schrodinger-K}) by a numerical spectral
method, where we expand the wave functions on the radial eigenmodes of the isotropic
3D harmonic oscillator (for $\ell=m=0$),
$y_n(x) \propto x L_n^{1/2}(x^2) e^{-x^2/2}$, where $L_n^{1/2}$ are the generalized
Laguerre polynomials.
We also extend the eigenvalue problem (\ref{Schrodinger-K}) by multiplying
the potential $V(x)$ by a real parameter $\nu>0$; i.e., we make the change
$V(x) \rightarrow \nu V(x)$, and we display in Fig.~\ref{fig_E_nu} the dependence
of the ground-state energy $E_0$ on $\nu$.
For small $\nu$, we expect $E_0>0$, as the amplitude of the potential in the negative
domain and the estimate $n_{\rm bound}$ decrease, and for large $\nu$, we expect
$E_0<0$ with an increasing amplitude.
This is consistent with the results found in Fig.~\ref{fig_E_nu}.
(At low $\nu$, we should switch to the continuous spectrum, with $E>0$, but we
obtain a finite positive $E_0$ because of the finite numerical resolution which gives rise
to a finite-size effect. On the other hand,
$E=0$ is not part of the continuous spectrum, which is restricted to $E>0$.)
For both potentials, we have $E_0>0$ for $\nu=1$, but we are not far from the
transition to a negative-energy ground state, in agreement with the estimates
(\ref{Nbound-Arctan}).
Thus, we conclude that for the Arctan and Sqrt models there are no negative
eigenvalues $\lambda$ to the eigenvalue problem (\ref{eigenvalue-static}) and
the Euclidian action $S_E$ is positive definite.
Therefore, these two models are stable with respect to quantum corrections and
correspond to meaningful quantum field theories, around static spherically symmetric
backgrounds.

\subsection{Cosmological instability}
\label{sec:Cosmological-instability}

\subsubsection{Modes of the quadratic action}
\label{sec:modes-cosmo}

For some models, the mass squared can also be negative as we expand around the
cosmological background. For time-varying problems, the Wick rotation cannot be applied.
One has to resort to canonical quantization and the analysis of the mode equations to see
if an instability occurs. This is very similar to the study of cosmological perturbations and its
quantum generation. In particular, because the age of the Universe is finite, we do not
require that the modes are stable but that they do not grow much more than in the standard
cosmological scenario, or at least do not diverge before the present time.
Using the conformal coordinates,
\be
ds^2= a^2 (-d\eta^2 +d\vec{x}^{\,2}) ,
\ee
the action at the quadratic order reads as
\be
S = \int d\eta d^3{x} \; \sqrt{-g} {\cal L}_2 =  \int d\eta d^3{x} \; a^4 {\cal L}_2 ,
\ee
where the quadratic Lagrangian is given by Eq.(\ref{L2-exact}), which yields
\beqa
{\cal L}_2 & = & \frac{1}{2 a^2} \left[ \frac{\bar{K}'+2\bar\chi \bar{K}''}{\bar{K}'}
\left( \frac{\partial\phi}{\partial \eta} \right)^2  - (\nabla \phi)^2 - a^2 \bar{m}_{\phi}^2 \phi^2
\right] , \nonumber \\
\label{L2-cosmo}
\eeqa
with
\beqa
a^2 \bar{m}_{\phi}^2 & = & - \left( \frac{\bar{K}''}{2\bar{K'}}
+ \frac{\bar{K}''^2\bar\chi}{\bar{K}'^2} \right)
\left( \frac{d^2\bar\chi}{d\eta^2} - 2 {\cal H} \frac{d\bar\chi}{d\eta} \right) \nonumber \\
&& \hspace{-1cm} - \left( \frac{\bar{K}'''}{2\bar{K}'} + \frac{3\bar{K}''^2}{4\bar{K}'^2}
+ \frac{2\bar{K}'''\bar{K}''\bar\chi}{\bar{K}'^2}
- \frac{3\bar{K}''^3\bar\chi}{2\bar{K}'^3} \right)
\left( \frac{d\bar\chi}{d\eta} \right)^2 , \hspace{0.7cm}
\label{a2m2-def}
\eeqa
where ${\cal H} = d\ln a/d\eta$ is the conformal expansion rate.

As in the static case, to absorb the kinetic prefactor $(\bar{K}'+2\bar\chi \bar{K}'')/\bar{K}'$,
we make the change of variable
\beqa
\tau & = & \int_0^\eta d\eta \left( \frac{\bar{K}'+2\bar\chi \bar{K}''}{\bar{K}'} \right)^{-1/2} ,
\label{tau-eta-def} \\
\phi & = & \left( \frac{\bar{K}'+2\bar\chi \bar{K}''}{\bar{K}'} \right)^{-1/4}
\frac{v}{a} ,
\label{phi-v-def} \\
m_v^2 & = & a^2 \bar{m}^2_{\phi} - \frac{1}{a}
\left( \frac{\bar{K}'+2\bar\chi \bar{K}''}{\bar{K}'} \right)^{-1/4} \nonumber \\
&& \times \frac{d^2}{d\tau^2} \left[ a \left( \frac{\bar{K}'+2\bar\chi \bar{K}''}{\bar{K}'}
\right)^{1/4} \right] .
\label{V-m-tau-def}
\eeqa
This is the generalization of the Mukhanov-Sasaki variable to our case
\cite{Mukhanov2005}.
With an integration by parts, the action now reads as
\beq
S = \frac{1}{2} \int d\tau d^3x \;  \left[ \left( \frac{\partial v}{\partial \tau} \right)^2
- (\nabla v)^2 - m_v^2(\tau) v^2 \right] .
\label{S-tau-V}
\eeq
For the cosmological background, the Klein-Gordon equation (\ref{KG-matter})
obeyed by the scalar field
can be integrated as Eq.(\ref{KG-cosmo}).
Contrary to the static case, the history of the cosmological expansion [i.e., the functions
$a(\eta)$ or $a(t)$] cannot be absorbed in a model-independent way through the
Klein-Gordon equation (\ref{KG-cosmo}). Whereas in Eq.(\ref{Schrodinger-K}) we obtained
an eigenvalue problem that only depends on the kinetic function $K(\chi)$ (outside of the
compact object), independently of the mass, radius, and profile of the spherical object,
in the action (\ref{S-tau-V}) the mass $m_v^2(\tau)$ shows a small dependence on the
choice of cosmological parameters. In the numerical computations, we consider
a set of cosmological parameters that is consistent with the Planck data.

Using Eq.(\ref{a2m2-def}), the squared mass can be written as
\beqa
m_v^2 & = & - \frac{a''}{a} \frac{\bar{K}'+2\bar{\chi} \bar{K}''}{\bar{K}'}
- \frac{d^2\bar{\chi}}{d\eta^2}
\frac{\bar{\chi} \bar{K}''^2 + \bar{K}' (2 \bar{K}'' \!+\! \bar{\chi} \bar{K}''')}{2 \bar{K}'^2}
\nonumber \\
&& \hspace{-1cm} - \frac{a'}{a} \frac{d\bar{\chi}}{d\eta}
\frac{-4 \bar{\chi} \bar{K}''^2 + \bar{K}' (\bar{K}'' + 2\bar{\chi} \bar{K}''')}{\bar{K}'^2}
+ \left( \frac{d\bar{\chi}}{d\eta} \right)^2 \nonumber \\
&& \hspace{-1cm} \times \left[ 4 \bar{K}'^3 ( \bar{K}' + 2 \bar{\chi} \bar{K}'' ) \right]^{-1}
\left[ 5 \bar{\chi}^2 \bar{K}''^4 + 2 \bar{\chi} \bar{K}' \bar{K}''^2 \right. \nonumber \\
&& \hspace{-1cm} \times (\bar{K}''-3\bar{\chi} \bar{K}''') - 2 \bar{K}'^3
( 3 \bar{K}'''+\bar{\chi} \bar{K}^{(4)} ) + \bar{K}'^2 (2 \bar{K}''^2 \nonumber \\
&& \hspace{-1cm} \left. + \bar{\chi}^2 \bar{K}'''^2 - 4 \bar{\chi} \bar{K}'' (3 \bar{K}''' + \bar{\chi} \bar{K}^{(4)} ) ) \right] ,
\label{mv2-chi}
\eeqa
where we note $a'=da/d\eta$ and $a''=d^2a/d\eta^2$.
In the standard cosmological case, where $K'=1$, we recover the usual result
$m_v^2=-a''/a$.

As the cosmological background is homogeneous and isotropic, we can decompose the
modes of the quadratic action (\ref{S-tau-V}) in Fourier space \cite{Peter2013}, as
$v_{\vec k}(\tau,\vec{x})= e^{\ii \vec{k} \cdot \vec{x}} v_k(\tau)$ with
\beq
\frac{d^2 v_k}{d\tau^2} + ( k^2 + m_v^2 ) v_k = 0 ,
\label{modes-cosmo-KG}
\eeq
and $k= | \vec{k} |$.
In terms of creation and annihilation operators, the quantum field $\hat{v}$ can be written as
\begin{equation}
\hat{v}(\tau,\vec{x}) = \int \frac{d^3k}{(2\pi)^{3/2}}
\left[ v_{k}(\tau) e^{\ii {\vec k} \cdot {\vec x}} \hat{c}_{\vec{k}}
+ v_{k}^*(\tau) e^{-\ii {\vec k} \cdot {\vec x}} \hat{c}_{\vec{k}}^{\dagger} \right] .
\label{v-vk-def}
\end{equation}
The conjugate momentum of $v$ is $\Pi=\dot{v}$, where we note $\dot{v}=dv/d\tau$,
and the operators $\hat{v}$ and $\hat{\Pi}$ must satisfy canonical commutation relations
on constant-time slices,
\be
\left[\hat{v}(\tau,{\vec x}),\hat{\Pi}(\tau,{\vec y})\right] = \ii \, \delta_D({\vec x}-{\vec y}) .
\ee
Using the commutation rules of the creation and annihilation operators,
$[ \hat{c}_{\vec{k}} , \hat{c}_{\vec{p}}^{\dagger}] = \delta_D({\vec k}-{\vec p})$,
one finds that the Wronskian $W_k$ must be normalized as
\begin{equation}
W_k \equiv v_{k} \dot{v}^*_{k} - v_{k}^* \dot{v}_{k} = \ii .
\label{Wronskian}
\end{equation}

\subsubsection{Squared-mass $m_v^2$}
\label{sec:mv2}

\begin{table}[htb!]
\caption{Behavior in the early radiation era of the squared-mass $m_v^2(a)$
for the Cubic, Arctan and Sqrt models.}
\begin{center}
\begin{tabular}{|c||c|c|c|}
\hline
Model name & Cubic & Arctan & Sqrt \\
\hline\hline

\rule{0cm}{0.5cm} $\!\! a \ll a_{\rm eq}, \;\; m_v^2$ &
$ - \frac{34}{5} \frac{a_{\rm eq} H_0^2 \Omega_{\rm m0}}{a^2} $  &
$ - \frac{H_0^2 \Omega_{\rm m0}}{2 a} \!+\! {\cal O}(a^2) $ &
$ - \frac{H_0^2 \Omega_{\rm m0}}{2 a} \!+\! {\cal O}(a^4) $ \\ [0.2cm]

\hline

\end{tabular}
\end{center}
\label{table-cosmo}
\end{table}

\begin{figure}
\begin{center}
\epsfxsize=8.5 cm \epsfysize=5.8 cm {\epsfbox{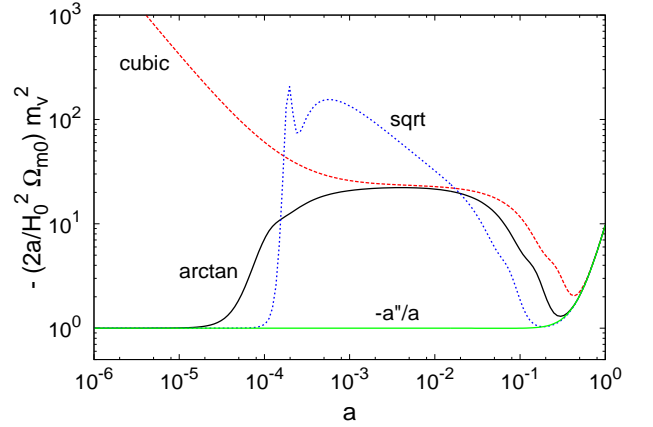}}
\end{center}
\caption{Squared mass $m_v^2(a)$ from Eq.(\ref{mv2-chi}) as a function of the
scale factor, for the Cubic, Arctan and Sqrt models.
We show the combination $-(2a/H_0^2\Omega_{\rm m0}) m_v^2$, which is positive
(i.e. $m_v^2<0$). We also plot the case $m_v^2=-a''/a$ (lower line).}
\label{fig_mv2}
\end{figure}

To understand the dynamics of the modes $v_k$ we need the evolution with time
of the squared-mass $m_v^2$.
It is useful to obtain some analytic expressions in the early radiation and matter eras,
when the dark energy is negligible. Then, the Friedmann equation reads as
\beq
z \gg 1 : \;\;\;  \frac{H^2}{H_0^2} = \Omega_{\rm m0} a^{-4} (a+a_{\rm eq}) ,
\label{Hubble-rad}
\eeq
where $a_{\rm eq}$ is the scale factor at equality between the matter and radiation
energy densities.
This yields
\beqa
z \gg 1 : \;\;\; \eta & = & \frac{2}{H_0 \sqrt{\Omega_{\rm m0}}}
\left( \sqrt{a+a_{\rm eq}} - \sqrt{a_{\rm eq}} \right)
\label{eta-a-rad} \\
& \simeq & \frac{a}{H_0 \sqrt{\Omega_{\rm m0} a_{\rm eq}}} ,
\eeqa
\beq
t = \frac{2}{3 H_0 \sqrt{\Omega_{\rm m0}}}
\left[ (a-2a_{\rm eq}) \sqrt{a+a_{\rm eq}} + 2 a_{\rm eq}^{3/2} \right] ,
\eeq
and
\beq
\frac{a'}{a} = \frac{H_0 \sqrt{\Omega_{\rm m0}} \sqrt{a+a_{\rm eq}}}{a} , \;\;\;
\frac{a''}{a} = \frac{H_0^2 \Omega_{\rm m0}}{2 a} .
\eeq
The functions $\bar\chi(a)$, $d\bar\chi/d\eta(a)$, and $d^2\bar\chi/d\eta^2(a)$
depend on the kinetic function $K(\chi)$ and are obtained from
Eq.(\ref{KG-cosmo}). This allows us to obtain the analytic expression of $m_v^2(a)$
at early times from Eq.(\ref{mv2-chi}), and we show our results in the early radiation era
in Table~\ref{table-cosmo}, for the three models studied in this paper.
As expected, we find that for the Arctan and Sqrt models $m_v^2 \simeq - a''/a$ at
early times, because they satisfy $\chi K'' \ll K'$ at large $\chi$ so that the squared
mass $m_v^2$ is dominated by the first term in Eq.(\ref{mv2-chi}), which is almost equal
to $-a''/a$.
The subleading terms are very small and vanish as $a^2$ or $a^4$ for the Arctan
and Sqrt models, depending on the rate of convergence to zero of $\chi K''/K'$
at large $\chi$.
For the Cubic model, we have $\chi K'' / K' \simeq 2$ in the highly nonlinear regime
so that the first term in Eq.(\ref{mv2-chi}) is subdominant while the other terms scale
as $1/\eta^2 \sim 1/a^2$.

We show the squared mass $m_v^2$ and $-a''/a$ in Fig.~\ref{fig_mv2}.
At all redshifts, we have $m_v^2<0$ and $-a''/a<0$.
At low redshifts, when $\bar\chi \rightarrow 0$, for all three models, $m_v^2 \simeq -a''/a$
and it is dominated by the first term in Eq.(\ref{mv2-chi}).
As explained above, at high redshifts, we also recover $m_v^2 \simeq -a''/a$
for the Arctan and Sqrt models, in the highly nonlinear regime where $K'$ has reached
its constant high-$\chi$ value, whereas for the Cubic model, $m_v^2$ keeps growing
as $-1/a^2$.
As for the static problem studied in Sec.~\ref{sec:Radial}, the Sqrt model, which
has a sharper transition between the linear and highly-nonlinear regimes in terms
of $K'(\chi)$ than the Arctan model, see Fig.~\ref{fig_Kp_chi},
shows a narrower deviation (i.e. over a smaller range of $a$) from the standard behavior
($-a''/a$) but with a greater amplitude.

\subsubsection{Large-scale growing and decaying modes $v_0^{\pm}(a)$}
\label{sec:v0p}

\begin{figure}
\begin{center}
\epsfxsize=8.5 cm \epsfysize=5.8 cm {\epsfbox{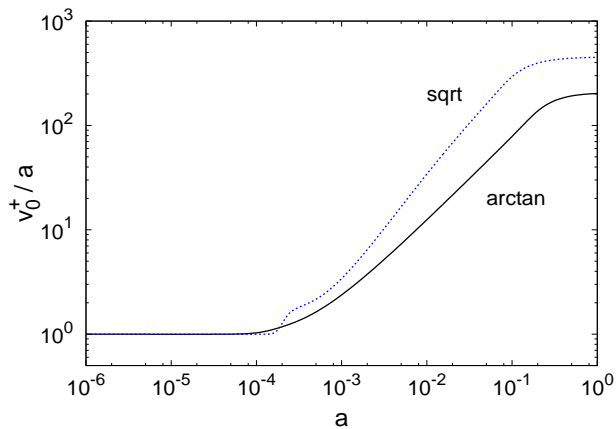}}
\end{center}
\caption{Growing mode $v_k^+(\tau)$, as a function of the scale factor $a$,
for the Arctan and Sqrt models at $k=0$.
We show the ratio $v_0^+/a$, which is constant at both high and low redshifts for these
models.}
\label{fig_v0p}
\end{figure}

\paragraph{Arctan and Sqrt models}
\label{Arctan-cosmo}

Let us now consider the  evolution of the modes $v_k(\tau)$ of Eq.(\ref{modes-cosmo-KG}),
for low wave numbers $k^2 \ll |m_v^2|$.
The standard case corresponds to $\tau=\eta$ and $m_v^2=-a''/a$. Then,
Eq.(\ref{modes-cosmo-KG}) becomes $v_0'' = (a''/a) v_0$, for $k=0$, with the growing
solution $v_0^+(\eta) = a(\eta)$ and a decaying solution $v_0^-(\eta)$.
In the K-mouflage models, $-m_v^2$ is greater than $a''/a$ over some range of
redshift, which implies that $v_0^+$ grows faster than $a$ in this range, with
an amplification that is somewhat reduced by the change of variable from $\eta$ to
$\tau$ in Eq.(\ref{tau-eta-def}).
We show our results for the Arctan and Sqrt models in Fig.~\ref{fig_v0p}.
At early times, we can again normalize the growing mode as $v_0^+(\tau) = a$
because we recover the standard case with $m_v^2 \simeq -a''/a$.
In the intermediate range, $10^{-4} \lesssim a \lesssim 10^{-1}$, where
$-m_v^2$ is significantly greater than $a''/a$ as seen in Fig.~\ref{fig_mv2},
$v_0^+$ grows faster than $a$ and is amplified by a factor of order 200 at $a \sim 0.1$.
At low redshifts, where $m_v^2$ again converges to $-a''/a$, we resume a linear
growth $v_0^+ \propto a$.
Thus, in both models, the growing mode $v_0^+(\tau)$ mainly behaves as the scale factor
$a(\tau)$, with a normalization factor that grows from unity to about 200 over the
range $10^{-4} \lesssim a \lesssim 10^{-1}$, which leads to
\beq
\mbox{Arctan and Sqrt:} \;\;\; v_0^+(a=1) \sim 10^2 .
\eeq
The decaying mode is given by
\beq
v_0^-(a) \propto v_0^+ \int_{\tau}^{\infty} \frac{d\tau}{(v_0^+)^2} .
\eeq

In the radiation era and before $m_v^2$ departs from $-a''/a$, we have
$v_0^+=a$ and $\tau=\eta$. Using (\ref{eta-a-rad}), we obtain
\beq
a < 10^{-4} : \;\;\; v_0^+(a)= a , \;\;\;
\dot{v}_0^+(a) = H_0 \sqrt{\Omega_{\rm m0} a_{\rm eq}} ,
\label{v0p-rad-Arctan}
\eeq
and
\beq
v_0^-(a)= 1 + \frac{a}{2a_{\rm eq}} \ln  \frac{a}{a_{\rm eq}}  ,
\;\;\; \dot{v}_0^-(a) = \frac{H_0 \sqrt{\Omega_{\rm m0}}}{2\sqrt{ a_{\rm eq}}}
\ln \frac{a}{a_{\rm eq}} .
\label{v0m-rad-Arctan}
\eeq
\newline

\paragraph{Cubic model}
\label{Cubic-cosmo}

For the Cubic model, in the nonlinear regime, we have $(K'+2\chi K'')/K' \simeq 5$
so that $\tau \simeq \eta/\sqrt{5}$ in the radiation era, while
$a \simeq H_0 \sqrt{\Omega_{\rm m0} a_{\rm eq}} \eta$ and $m_v^2$ is given by the
expression in Table~\ref{table-cosmo}.
Therefore, the equation of motion (\ref{modes-cosmo-KG}) reads for $k=0$ as
\beq
\frac{d^2 v_0}{da^2} - \frac{34}{25 a^2} v_0 = 0 .
\eeq
This gives the growing and decaying solutions
\beqa
a<10^{-4} : && \;\;\; v_0^+(a) = a_Q (a/a_Q)^{1/2+\sqrt{161}/10} , \nonumber \\
&& v_0^-(a) = (a/a_Q)^{1/2-\sqrt{161}/10} ,
\label{v0-Cubic}
\eeqa
which we normalized as $v_0^+(a_Q)=a_Q$ and $v_0^-(a_Q)=1$, where $a_Q$ is
the scale factor at the transition to the quantum regime given in (\ref{aQ-Cubic}).
We obtain today
\beq
\mbox{Cubic:} \;\;\; v_0^+(a=1) \sim 10^{14} ,
\eeq
taking into account an additional amplification by a factor $10^6$ between
$10^{-4} \leq a \leq 1$, as $m_v^2$ becomes of the same order as for the Arctan model
in this range.
Thus, as compared with the standard case $v_0^+=a$, the Cubic model gives
an amplification by a factor $10^{14}$ today, as compared with the factor $10^2$
obtained for the Arctan and Sqrt models, because of the stronger divergence of
$m_v^2$ at high redshift.

\subsubsection{Initial conditions}
\label{sec:initial-vk}

In the Minkowski background with a constant positive squared mass,
the modes $v_{\vec k}$ are the standard plane waves
$e^{\ii (\vec{k}\cdot\vec{x}-\omega_k\tau)}$ with $\omega_k= \sqrt{k^2+m_v^2}$,
see Eqs.(\ref{u-uk-def})-(\ref{uk-t}) in appendix~\ref{app:one-loop}.
There, a constant negative squared mass must be ruled out because it yields an exponential
instability, as large-scale modes have a negative $\omega_k^2$, unless we restrict
to finite times.

In the cosmological context \cite{Peter2013}, the standard case corresponds 
to $\tau=\eta$ and $m_v^2 = - a''/a$, which is negative (this case is recovered for 
$K' \equiv 1$).
Then, small-scale modes below the horizon, $k\tau\gg 1$, are governed by the $k^2$ term in
Eq.(\ref{modes-cosmo-KG}) and oscillate as in the Minkowski case over time scales
that are short as compared with the Hubble time.
In this regime, one can use the WKB approximation.
Large-scale modes beyond the horizon, $k\tau \ll 1$, obey $v_0''/v_0=a''/a$, with the growing
solution $v_0^+=a$. Thus, the modes outside the horizon do not oscillate but show
a secular growth as $a(\tau)$, which in fact corresponds to a constant amplitude for the
scalar field $\phi=v/a$, see Eq.(\ref{phi-v-def}) with $K''=0$.
\newline

\paragraph{Arctan and Sqrt models}
\label{initial-inflation-Arctan-Sqrt}

We first consider the Arctan and Sqrt models, which can be analyzed following the
same approach as for the production of quantum
fluctuations of the inflaton field (or test fields) during the inflationary stage and
their evolution during later stages through the radiation, matter, and dark-energy eras.
Indeed, we have seen in section~\ref{sec:Higher-orders-cosmological} that there is no
quantum transition in the early Universe for these scenarios as quantum corrections to
the effective action $\Gamma$ are increasingly negligible at higher redshift.
This means that the classical background obtained from the classical integrated
Klein-Gordon equation (\ref{KG-cosmo}) and the quadratic action (\ref{S-tau-V}) for the
quantum fluctuations can be used at all redshifts and far in the early inflationary era.
In particular, as in the standard study of cosmological perturbations
\cite{Peter2013},
relevant physical scales $a/k$ that can be observed today were well
within the horizon in the early inflation stage, next were stretched beyond the horizon
because of the exponential inflationary expansion until the end of the inflation, and
then grew more slowly than the Hubble scale $c t \sim c/H$ in the radiation and matter
eras and became smaller than the horizon at redshifts $z \lesssim 10^6$.
The inflationary stage ensures that these scales were initially deep inside the horizon,
at which time they actually experienced an effective Minkowski background and one
can use standard Quantum Field Theory procedures. In particular, this provides
a well-defined Bunch-Davies vacuum. This corresponds to choosing the mode function
$v_k(\tau)$ such that at early time, in the subhorizon limit, it converges to the usual
Minkowski limit $v_k(\tau) \rightarrow e^{-\ii k \tau}/\sqrt{2 k}$.

Let us recall that this choice of vacuum, or initial conditions, can be derived in this
framework as the minimization of the expected value of the Hamiltonian
\cite{Mukhanov2007}.
In terms of the mode expansion (\ref{v-vk-def}), the Hamiltonian reads as
\beq
\hat{H} = \frac{1}{2} \int d^3k \left[ F_k \hat{c}_{\vec k} \hat{c}_{-\vec k}
+ F_k^* \hat{c}^{\dagger}_{\vec k} \hat{c}^{\dagger}_{-\vec k}
+ E_k ( \delta_D(0) + 2 \hat{c}^{\dagger}_{\vec k} \hat{c}_{\vec k} ) \right] ,
\label{H-cc-def}
\eeq
with
\beq
F_k = \dot{v}_k^2 + \omega_k^2 v_k^2 , \;\;\; E_k = | \dot{v}_k |^2 + \omega_k^2 | v_k |^2 ,
\label{Fk-Ek-def}
\eeq
where we defined
\beq
\omega_k^2 = k^2 + m_v^2 .
\label{omega-k-def}
\eeq
Since the vacuum is defined by $c_{\vec k}|0 \rangle =0$, this gives
\beq
\langle 0 | \hat{H} | 0 \rangle = \frac{\delta_D(0)}{2} \int d^3k \, E_k ,
\label{0H0-Ek}
\eeq
and the vacuum energy density reads as
\beq
\epsilon = \frac{1}{2} \int \frac{d^3k}{(2\pi)^3} \, E_k .
\label{epsilon-Ek}
\eeq

To minimize the Hamiltonian we simply need to minimize $E_k$ for each $k$-mode,
with the constraint (\ref{Wronskian}).
In the standard field-theory case, with a constant squared mass $m_v^2>0$
in the Minkowski background, there is a well-defined minimum that gives
the usual result $v_k(\tau) = e^{-\ii \omega_k \tau}/\sqrt{2\omega_k}$, as in
Eq.(\ref{uk-t}), and the energy density (\ref{epsilon-Ek}) takes the form of
Eq.(\ref{rho-V}).

In a time-dependent spacetime, or more generally when the mass $m_v(\tau)$
depends on time, the frequencies $\omega_k(\tau)$ depend on time.
This implies that the choice of vacuum depends on the time where we choose to minimize
the Hamiltonian, $|0\rangle_{\tau_1} \neq |0\rangle_{\tau_2}$ for $\tau_1\neq\tau_2$
\cite{Mukhanov2007}.
In the usual study of the quasi-de Sitter inflationary era, this ambiguity is solved
by noticing that at sufficiently early times all modes were deep inside the horizon
and behaved as in the standard Minkowski spacetime with time-independent frequencies
$\omega_k^2 \rightarrow k^2$.
Therefore, we can take the Minkowski initial conditions,
$v_k(\tau) \rightarrow e^{-\ii k\tau}/\sqrt{2k}$ for $\tau\rightarrow-\infty$, to define the
modes $v_k(\tau)$. This defines a set of mode functions $v_k(\tau)$ and a unique
vacuum, the Bunch-Davies vacuum.
More explicitly, this corresponds to minimizing the vacuum energy density $\epsilon$
of Eq.(\ref{epsilon-Ek}), for each mode $k$, in the inflationary era when the mode
$k$ was deep inside the horizon. This removes the ambiguity of the choice of vacuum
as the result does not depend on the initial time as long as it is in this early subhorizon
stage. This means that the mode functions $v_k(\tau)$ are defined by the initial condition
\be
\tau \rightarrow -\infty : \;\; v_k(\tau) = \frac{e^{-\ii k\tau}}{\sqrt{2k}} ,
\label{vk-inflation}
\ee
as in the quasi-de Sitter era the past corresponds to $\tau \rightarrow -\infty$ and
the scale factor is given by
\be
a = - \frac{1}{H_I\eta} \;\;\; \mbox{and} \;\;\; \tau = \eta .
\ee
As we recalled above, this procedure also applies to the Arctan and Sqrt models,
where we can go back in time far into the inflationary stage and $m_v^2=-a''/a$
at early times.

In the inflationary era, using
\be
m_v^2 = - \frac{a''}{a} = - \frac{2}{\eta^2} = - \frac{2}{\tau^2} = - 2 a^2 H_I^2 ,
\ee
the equation of motion (\ref{modes-cosmo-KG}) takes the standard form
\be
\frac{d^2 v_k}{d\tau^2} + \left( k^2 - \frac{2}{\tau^2} \right) v_k = 0 ,
\ee
with the exact solution
\be
v_k(\tau) = \alpha \frac{e^{-\ii k\tau}}{\sqrt{2k}} \left( 1 - \frac{i}{k\tau} \right)
+ \beta \frac{e^{\ii k\tau}}{\sqrt{2k}} \left( 1 + \frac{i}{k\tau} \right) .
\ee
The initial condition (\ref{vk-inflation}) sets $\alpha=1$ and $\beta=0$ and determines
the mode function as
\be
v_k(\tau) = \frac{e^{-\ii k\tau}}{\sqrt{2k}} \left( 1 - \frac{i}{k\tau} \right) .
\label{mode-vk-inflation}
\ee
We can see that modes below this horizon oscillate as in Minkowski spacetime,
\be
| k \tau | = \frac{k}{a H_I} \gg 1 : \;\;\; v_k(\tau) \simeq \frac{e^{-\ii k\tau}}{\sqrt{2k}} ,
\label{vk-inflation-subhorizon}
\ee
while modes beyond the horizon grow as $v_0^+(\tau) = a$,
\be
| k \tau | = \frac{k}{a H_I} \ll 1 : \;\;\; v_k(\tau) \simeq \frac{-\ii}{\sqrt{2}k^{3/2}\tau}
= \frac{\ii H_I}{\sqrt{2} k^{3/2}} \, a .
\label{vk-inflation-superhorizon}
\ee

These results allow us to compute the vacuum energy density $\epsilon_f$ as the end
of the inflationary era, at scale factor $a_f$.
From Eqs.(\ref{vk-inflation-subhorizon}) and (\ref{vk-inflation-superhorizon}) we obtain
\be
\frac{k}{a H_I} \gg 1 : \;\; E_k \simeq k , \;\;\;
\frac{k}{a H_I} \ll 1 : \;\; E_k \simeq - \frac{H_I^4 a_f^4}{2 k^3} ,
\ee
and
\be
\epsilon(k>|m_v|) = \int_{|m_v|}^{\infty} \frac{dk}{(2\pi)^2} \, k^3
\sim H_I^4 a_f^4,
\ee
where the ultraviolet divergence is renormalized as described in
Appendix~\ref{app:one-loop}, see Eqs.(\ref{rho-V}) and (\ref{rhov-renorm}),
and we disregard the logarithmic factors, while
\be
\epsilon(k<|m_v|) = - \frac{a_f^4 H_I^4}{8\pi^2} \int_{|m_v(a_i)|}^{|m_v(a_f)|} \frac{dk}{k}
\sim H_I^4 a_f^4 \, N ,
\ee
where we regularized the infrared divergence at the modes that left the horizon at the
beginning of the inflationary era, $a_i$, and $N=\ln(a_f/a_i)$ is the number of $e$-folds
of the inflation stage. This gives
\be
\mbox{at } a_f : \;\;\; \epsilon_f \sim H_I^4 a_f^4 .
\label{epsilon-af}
\ee
\newline

\paragraph{Cubic model}
\label{initial-Cubic}

For the Cubic model, the standard approach described above
is not possible because the quadratic Lagrangian (\ref{L2-cosmo})
only applies up to the redshift $z_Q$, or scale factor $a_Q$, obtained in
(\ref{aQ-Cubic}), as the effective action is dominated by high-order
loop contributions at earlier times and the classical background
 (\ref{KG-cosmo}) and the quadratic action (\ref{S-tau-V}) for the quantum
 fluctuations are no longer valid.
Because this transition occurs after the inflationary stage, this means that
we cannot go back in the past up to the inflationary era and we must set the initial
conditions for the modes $v_k$ at redshift $z_Q$, when the scales $a_Q/k$
are outside of the horizon. This means that we can no longer use the early Minkowski
limit to define our vacuum.
If the squared mass $m_v^2$ were positive, we could still choose a well-defined
vacuum by minimizing the vacuum energy density (\ref{epsilon-Ek}) at the initial
time $a_Q$. However, this is not possible in our case because $m_v^2<0$ and
there is no minimum for large-scale modes $k<|m_v^2|$.
Nevertheless, we can consider a ``natural'' choice of vacuum and orders of magnitudes
of the initial modes at $a_Q$ as follows.

First, at the initial time or scale factor $a_Q$, obtained in (\ref{aQ-Cubic}),
we expect the field $\hat{v}$ to show a strong quantum character because this time
marks the transition between the quantum and classical regimes of the K-mouflage
theory, as described in section~\ref{sec:Quantum-corrections}.
This means that in the Wronskian identity (\ref{Wronskian}) both terms in the left-hand
side should be of order unity.
At later times, for superhorizon modes, the mode function $v_k(\tau)$ grows
as the homogeneous growing solution $v_0^+(\tau)$ so that the two terms in
the left-hand side grow as $v_0^+ \dot{v}_0^+ \sim a^{\sqrt{161}/5}$.
This implies that as time increases the right-hand side in the Wronskian identity
(\ref{Wronskian}) becomes negligible, which means that the noncommutativity
can be neglected and we recover classical fields.
The same quantum-to-classical transition also appears in the usual study of inflationary
fluctuations and for the Arctan and Sqrt models, the transition occurring when the modes
exit the Hubble radius; see Eqs.(\ref{mode-vk-inflation})-(\ref{vk-inflation-superhorizon}).
In the case of the Cubic model, we can expect that this transition appears at redshift
$z_Q$ for all modes.
Second, even though the Hamiltonian expectation value (\ref{0H0-Ek}) has no well-defined
minimum for large-scale modes, in the absence of any further information on the
behavior of the theory at $a<a_Q$, we can expect that both terms in $E_k$ in
Eq.(\ref{Fk-Ek-def}) have the same order of magnitude.
Having one of these terms much greater than the other is likely to correspond to some
fine-tuning.
Therefore, we assume the order-of-magnitude relations
\beq
\mbox{at } a_Q : \;\;
v_k \dot{v}_k \sim 1, \;\;\; \dot{v}_k^2 \sim \omega_k^2 v_k^2 .
\eeq
This gives
\beq
\mbox{at } a_Q : \;\;
v_k \sim \omega_k^{-1/2} , \;\;\; \dot{v}_k \sim \omega_k^{1/2} , \;\;\;
E_k \sim \omega_k .
\label{vk-Ek-omegak}
\eeq
Notice that the usual choice of vacuum in Minkowski spacetime and positive squared mass
gives the same scalings with $\omega_k=\sqrt{k^2+m^2}$.
This gives on small and large scales
\be
k \gg | m_v | : \;\;\; v_k \sim k^{-1/2} , \;\;\; \dot{v}_k \sim k^{1/2} , \;\;\;
E_k \sim k ,
\label{Ek-subhorizon-cubic-aQ}
\ee
\be
k \ll | m_v | : \;\;\; v_k \sim |m_v|^{-1/2} , \;\;\; \dot{v}_k \sim |m_v|^{1/2} , \;\;\;
E_k \sim |m_v| .
\label{Ek-superhorizon-cubic-aQ}
\ee
On small scales we recover the usual Minkowski normalization
(\ref{Ek-subhorizon-cubic-aQ}), which also applied to Eq.(\ref{vk-inflation-subhorizon}).
On large scales we obtain a different scaling (\ref{Ek-superhorizon-cubic-aQ})
than in Eq.(\ref{vk-inflation-superhorizon}), because we cannot match these scales
to a Minkowski-like era and we assume that they all remain quantum until the
redshift $z_Q$ when the theory starts being well described by its classical action.
Assuming again that the ultraviolet divergence of the vacuum energy density
(\ref{epsilon-Ek}) is regularized in the Minkowski regime as described in
Appendix~\ref{app:one-loop}, see Eqs.(\ref{rho-V}) and (\ref{rhov-renorm}), we obtain
\be
\mbox{at } a_Q : \;\;\; \epsilon_Q \sim m_v^4(a_Q) \sim \frac{H_0^4 a_{\rm eq}^2}{a_Q^4} ,
\label{epsilon-aQ-cubic}
\ee
where we used the result given in Table~\ref{table-cosmo}.

\subsubsection{Evolution with time of $\epsilon$}
\label{sec:evolution-epsilon}

From the initial conditions obtained in section~\ref{sec:initial-vk}, we can now compute
the evolution with the time of the vacuum energy density $\epsilon$ associated with the
quantum fluctuations of the K-mouflage scalar field.
\newline

\paragraph{Arctan and Sqrt models}
\label{Arctan-modes-cosmo}

As we shall only need crude order-of-magnitude estimates, we neglect the reheating
phase between the end of the inflation era and the beginning of the radiation era.
Using the continuity of the total energy density and of the Hubble expansion rate
at the transition and Eq.(\ref{Hubble-rad}), we write
$H_I^2 = H_0^2 \Omega_{\rm m0} a_{\rm eq}/a_f^4$ and we obtain
\be
a_f = H_I^{-1/2} H_0^{1/2} a_{\rm eq}^{1/4} .
\label{af-def}
\ee
Here and in the following, we neglect factors such as $\Omega_{\rm m0}$
that are of order unity.
Then, the squared mass $m_v^2$ drops at the transition,
\be
m_v^2(a_f^-) = - 2 H_I^2 a_f^2 = - 2 H_I H_0 a_{\rm eq}^{1/2} ,
\label{mv2-af-}
\ee
\be
m_v^2(a_f^+) = - \frac{H_0^2 \Omega_{\rm m0}}{2 a_f} \sim
- H_I^{1/2} H_0^{3/2} a_{\rm eq}^{-1/4} .
\label{mv2-af+}
\ee
Next, $|m_v|$ decreases with time as $a^{-1/2}$, from Table~\ref{table-cosmo},
and modes gradually enter the Minkowski-like regime where $k \gg |m_v|$ and
$\omega_k \simeq k$.

At any time $a_f < a < a_{\rm eq}$ in the radiation era, we have several regimes
for the modes $\hat{v}_k$.
First, high wave numbers $k>|m_v(a_f^-)|$ were already above $|m_v|$ before the end
of the inflationary era, and they obey
\be
k > | m_v(a_f^-) | \sim H_I a_f : \;\; v_k \simeq \frac{e^{-\ii k \tau}}{\sqrt{2k}} , \;\;\;
E_k \simeq k .
\label{k>mv(af)}
\ee
Next, we find that both modes that entered the Minkowski-like regime at the transition
$a_f$ ($H_0 a_f^{-1/2} < k < H_I a_f$), because of the sudden drop of $|m_v|$,
and at a later time $a_f < a' < a$ ($H_0 a^{-1/2} < k < H_0 a_f^{-1/2}$)
obey
\beqa
H_0 a^{-1/2} < k < H_I a_f : && \;\; v_k \sim \ii \frac{H_I H_0 a_{\rm eq}^{1/2}}{k^{5/2}}
\sin(k\tau) ,  \nonumber \\
&& E_k \sim \frac{H_I^2 H_0^2 a_{\rm eq}}{k^3} ,
\eeqa
where we omitted an irrelevant phase.
Here, we used the fact that in the regime $k \ll |m_v|$ the modes keep evolving
as the growing mode,
\be
k \ll |m_v| : \;\; v_k(\tau) = \frac{\ii H_I}{\sqrt{2} k^{3/2}} a(\tau) ,
\label{k<mv-radiation}
\ee
because of the initial condition (\ref{vk-inflation-superhorizon}), which is already the
general large-scale growing solution of Eq.(\ref{modes-cosmo-KG}),
$v_0(\tau) = a(\tau)$, irrespective of the detailed form of $a(\tau)$
(and $m_v^2 \simeq -a''/a$ at high redshift), whereas modes that enter the small-scale
regime evolve as
\be
k \gg |m_v| : \;\; v_k(\tau) = \alpha \, e^{-\ii k\tau} + \beta \, e^{\ii k\tau} ,
\label{k>mv-radiation}
\ee
and we compute the coefficients $\alpha$ and $\beta$ by matching
$v_k$ and $\dot{v}_k$ at the transition $k=|m_v|$.
Finally, the modes that are still in the large-scale regime (\ref{k<mv-radiation}) are still
given by Eq.(\ref{k<mv-radiation}) with
\be
k < H_0 a^{-1/2} : \;\; E_k \sim \frac{H_I^2 H_0^2 a_{\rm eq}}{k^3} .
\ee
Collecting these results, we obtain
\be
a_f < a < a_{\rm eq} : \;\; \epsilon(a) \sim H_I^2 H_0^2 a_{\rm eq} = H_I^4 a_f^4
\sim \epsilon(a_f) ,
\label{epsilon-a-Arctan}
\ee
and we recover the scale set by the initial vacuum energy density
(\ref{epsilon-af}) at the end of the inflationary era.
Note, however, that $\epsilon$ is not exactly constant because the factor
$\omega_k^2$ in Eq.(\ref{Fk-Ek-def}) depends on time, for the large-scale modes
that still obey $k<|m_v|$. The growth of these large-scale modes,
as $v_0^+(\tau)$, is counterbalanced by the decrease of their factor $\omega_k^2$
and by the fact that an increasingly large number of modes gradually leaves this regime
to enter the oscillatory regime $k>|m_v|$.
Moreover, all regimes, $H_I a_f<k$, $H_0 a_f^{-1/2} < k <  H_I a_f$,
$H_0 a^{-1/2} < k < H_0 a_f^{-1/2}$ and $k<H_0 a^{-1/2} $ contribute to $\epsilon$
with the same magnitude of order $H_I^2 H_0^2 a_{\rm eq} = H_I^4 a_f^4$.
\newline

\paragraph{Cubic model}
\label{Cubic-modes-cosmo}

For the Cubic model, the initial conditions
(\ref{Ek-subhorizon-cubic-aQ})-(\ref{Ek-superhorizon-cubic-aQ}) are set in the
early radiation era. Again, $|m_v|$ decreases with time, although at a faster
rate $1/a$ than for the Arctan and Sqrt models, and modes gradually enter the
Minkowski-like regime $k>|m_v|$.
From the initial conditions (\ref{Ek-superhorizon-cubic-aQ}) and the large-scale
growing and decaying solutions (\ref{v0-Cubic}), we find that the large-scale modes
evolve as
\be
k \ll | m_v | : \;\; v_k(\tau) \sim H_0^{-1/2} a_{\rm eq}^{-1/4} a_Q^{1/2}
(a/a_Q)^{1/2+\sqrt{161}/10} .
\label{k-large-cubic-rad}
\ee
Small-scale modes, $k \gg |m_v|$, evolve as in the Minkowski case and
Eq.(\ref{k>mv-radiation}), with coefficients $\alpha$ and $\beta$ that are obtained
from the matching at the transition $k=|m_v|$.
Again, we have several regimes for the modes $\hat{v}_k$.
High wave numbers from Eq.(\ref{Ek-subhorizon-cubic-aQ})  that were already above
$|m_v|$ at the initial time $a_Q$ obey
\be
k > |m_v(a_Q)| \sim H_0 a_{\rm eq}^{1/2} a_Q^{-1} : \;\;
v_k \sim \frac{\sin(k\tau)}{\sqrt{k}} , \;\; E_k \sim k .
\ee
Modes that entered the small-scale regime at a later time, $a_Q < a' < a$, behave as
\beqa
H_0 a_{\rm eq}^{1/2} a^{-1} < k < H_0 a_{\rm eq}^{1/2} a_Q^{-1} : && \nonumber \\
&& \hspace{-4cm} v_k \sim k^{-1/2}
\left( \frac{H_0 a_{\rm eq}^{1/2}}{k a_Q} \right)^{\sqrt{161}/10}
\sin(k\tau) ,  \\
&& \hspace{-4cm} E_k \sim k
\left( \frac{H_0 a_{\rm eq}^{1/2}}{k a_Q} \right)^{\sqrt{161}/5} .
\eeqa
Finally, the modes that are still in the large-scale regime (\ref{k-large-cubic-rad})
give
\be
k < H_0 a_{\rm eq}^{1/2} a^{-1} : \;\; E_k \sim H_0 a_{\rm eq}^{1/2} a^{-1}
(a/a_Q)^{\sqrt{161}/5} .
\label{Ek-large-scale-a-Cubic}
\ee
Collecting these results we obtain
\be
a_Q < a < a_{\rm eq} : \;\; \epsilon(a) \sim H_0^4 a_{\rm eq}^2 a_Q^{-4} \
\sim \epsilon(a_Q) .
\label{epsilon-a-Cubic}
\ee
We again recover the scale set by the initial vacuum energy density, which for the Cubic
model is given by Eq.(\ref{epsilon-aQ-cubic}).
In contrast with the Arctan and Sqrt models, the energy density is not equally distributed
over all regimes as it is dominated by the small-scale modes
$k \gtrsim H_0 a_{\rm eq}^{1/2} a_Q^{-1}$ and the contribution from the large-scale
modes (\ref{Ek-large-scale-a-Cubic}) is negligible.

\subsubsection{Self-consistent background}
\label{sec:density-vk}

In this study of the quantum stability of the theory in the cosmological
context, we wish to check whether the negative squared mass $m_v^2<0$
rules out the K-mouflage scenarios.
We have recalled that a negative squared mass is not specific to the K-mouflage context,
as it already appears in the standard study of cosmological perturbations
\cite{Mukhanov2005,Peter2013},
but the greater value of $|m_v^2|$ associated with the K-mouflage scalar field leads
to a faster growth of the mode functions $v_k$ with time.
Then, we must check that this enhanced growth does not lead to serious problems.
The minimal constraint that we consider here is that the energy density associated with
these fluctuations does not dominate the energy density of the Universe,
or the full action (\ref{S-def}) where we include matter and radiation components.
The energy density $\epsilon$ defined in Eq.(\ref{epsilon-Ek}) refers to the quadratic
action (\ref{S-tau-V}), with the time and space coordinates $(\tau,\vec x)$.
To compare with physical-coordinates energy densities, we must change
to the physical time and space coordinates $(t,\vec r)$. Using $d\tau \sim d\eta = dt/a$
and $\vec x = \vec r/a$, we write the physical energy density associated with the
quantum fluctuations $\hat{v}_k$ as
\beq
\rho_v (a) \sim \frac{\epsilon(a)}{a^4} .
\label{rhov-epsilon}
\eeq

\paragraph{Arctan and Sqrt models}
\label{Arctan-modes-vacuum-density}

In the inflationary era, the vacuum energy density is given by Eq.(\ref{epsilon-af}).
In this expression, $a_f$ was taken as the scale-factor at the end of  inflation,
but this result also holds at any time during the inflationary era by replacing $a_f$
by the current scale factor $a$. Therefore, using Eq.(\ref{rhov-epsilon}), we can see that
the vacuum energy density $\rho_v$ is constant during this era with
\be
\mbox{inflation era:} \;\;\; \rho_v \sim H_I^4 .
\ee
The Friedmann equation gives $3 M_{\rm Pl}^2 H_I^2= \bar\rho$, where $\bar\rho$ is
the background energy density (in a flat universe), which at this stage is dominated by the
inflaton. Thus, we obtain
\be
\mbox{inflation era:} \;\;\;
\frac{\rho_v}{\bar\rho} \sim \left( \frac{H_I}{M_{\rm Pl}} \right)^2 \sim 10^{-10} ,
\ee
where we used $H_I \lesssim 10^{-5} M_{\rm Pl}$.
Next, in the radiation era we obtain from Eq.(\ref{epsilon-a-Arctan})
\be
\mbox{radiation era:} \;\;\;
\rho_v \sim \frac{H_I a_f^4}{a^4} , \;\;\; \frac{\rho_v}{\bar\rho} \sim 10^{-10} .
\label{rhov-rho-radiation-era-Arctan}
\ee
The ratio $\rho_v/\bar\rho$ remains constant in time as both $\rho_v$ and
$\bar\rho$ (which is now dominated by the radiation) decrease as $a^{-4}$.
Therefore, until $a_{\rm eq}$ the ratio $\rho_v/\bar\rho$ is set to its
initial value during the inflationary stage.
During the matter and dark-energy eras, the evolution of the vacuum energy density
is no longer given by Eq.(\ref{epsilon-a-Arctan}), because the relation $a(\eta)$ is modified
and the effective mass $m_v^2(a)$ and the large-scale growing mode $v_0^+(a)$ are also
sensitive to the K-mouflage effects.
From Figs.~\ref{fig_mv2} and \ref{fig_v0p}, we can see that, as compared with the
radiation-era behavior ($m_v^2 \propto a^{-1}$),
the squared mass is typically multiplied by a factor 10 and the growing
mode by a factor 100, at $a=1$.
Therefore, the vacuum energy density is roughly multiplied by a factor $10^5$.
On the other hand, the background density is multiplied by a factor $10^3$ as it is mostly
dominated by the matter component, which decreases as $a^{-3}$ instead of $a^{-4}$.
Thus, the ratio $\rho_v/\bar\rho$ is multiplied by a factor $10^2$ from the constant
value (\ref{rhov-rho-radiation-era-Arctan}), which gives today the estimate
\be
a=1: \;\;\;  \frac{\rho_v}{\bar\rho} \sim 10^{-8} .
\label{rhov-rho-today-Arctan}
\ee
Thus, we find that from the inflationary era until today the vacuum energy density associated
with the quantum fluctuations of the K-mouflage scalar field is negligible as compared with
the mean background density. Therefore, the faster growth of the large-scale modes
shown in Fig.~\ref{fig_v0p}, due to nonlinear K-mouflage effects, does not have important
consequences and does not rule out the model.

\paragraph{Cubic model}
\label{Cubic-modes-vacuum-density}

For the Cubic model, we start the analysis in the radiation era, at time $a_Q$.
From Eq.(\ref{epsilon-a-Cubic}), we obtain
\beqa
a_Q < a < a_{\rm eq} & : & \;\;\; \rho_v \sim \frac{H_0^4 a_{\rm eq}^2}{a_Q^4 a^4} ,
\nonumber \\
&& \;\;\; \frac{\rho_v}{\bar\rho} \sim \frac{H_0^2 a_{\rm eq}}{M_{\rm Pl}^2 a_Q^4}
\sim 10^{-48} ,
\eeqa
where we used the result (\ref{aQ-Cubic}).
After the radiation era, the Cubic model behaves roughly as the Arctan model, as can be
seen from the squared mass $m_v^2(a)$ shown in Fig.~\ref{fig_mv2}.
Therefore, we can again consider that the ratio $\rho_v/\bar\rho$ is multiplied by a factor
$10^2$ from $a=a_{\rm eq}$ to $a=1$, which yields
\be
a=1: \;\;\;  \frac{\rho_v}{\bar\rho} \sim 10^{-46} .
\label{rhov-rho-today-Cubic}
\ee
Therefore, we find that until today the vacuum energy density $\rho_v$ is again negligible
as compared with the mean background density.
It is even much smaller than for the Arctan and Sqrt models, because the quantum fluctuations
are set at the time $a_Q$, inside the radiation era, instead of the end of the inflationary era.
The very small amplitude (\ref{rhov-rho-today-Cubic}) suggests that $\rho_v \ll \bar\rho$
independently of the details of the initial conditions, set when the theory
goes through the quantum-to-classical transition, and that this K-mouflage model is not ruled
out by the behavior of its small quantum fluctuations.

\section{Unitarity and classicalization: two-body scattering
$\varphi\varphi \rightarrow \varphi\varphi$}
\label{sec:unitarity}

\subsection{Perturbative regime and unitarity for low-energy scalar collisions}
\label{sec:2-body-perturbative}

\subsubsection{Earth surface background}
\label{sec:Earth-surface}

Although the K-mouflage theory is well defined on astrophysical and cosmological backgrounds,
one may wonder what happens in the context of particle physics at high energy.
A natural way of probing the theory on particle-physics scales and at high energy is to consider
the two-body scattering problem between asymptotic states representing scalar excitations of the
Minkowski vacuum. This may occur in particle colliders on Earth well inside its K-mouflage
radius, where the geometry of spacetime is well approximated by Minkowski space and
the scalar field admits a background value $\bar\varphi$, with a nonzero gradient
$\nabla\bar\varphi$. Thus, in this section, we consider relatively low-energy events
where the scalar field is set at lowest order by the astrophysical background due to the Earth.

Let us first describe this background.
For a K-mouflage coupling $\beta=0.1$, the K-mouflage radius
(\ref{RK-def}) of the Sun is $R_K({\rm Sun})\simeq 1097 {\rm A.U.}$ while the
K-mouflage radius of the Earth is $R_K({\rm Earth}) \simeq 1.9 {\rm A.U.}$.
Therefore, at the surface of the Earth, the screening parameters $R_K/r$ due to the
Sun and to the Earth are
\be
\frac{R_K({\rm Sun})}{d_{\rm Sun-Earth}} \simeq 1097 , \;\;
\frac{R_K({\rm Earth})}{R_{\rm Earth}} \simeq 44622 .
\label{RK-Sun-Earth}
\ee
Therefore, we are far within the K-mouflage screening regime, and the background scalar
field is dominated by the mass due to the Earth, as the Sun only makes a small contribution.
In this spherically symmetric approximation, the background is set by
Eq.(\ref{KG-static}). Moreover, for both the Arctan and Sqrt models, which satisfy
Solar System constraints, we have $K' \sim K_* =10^3$ for large negative $\chi$.
This gives at the surface of the Earth,
\be
\mbox{Earth surface:} \;\; \bar K' \simeq K_* =10^3 , \;\;\; \bar\chi_{\oplus} \simeq -2 \times 10^{12} ,
\label{Earth-background}
\ee
\be
\bar{K}_{\oplus \rm Arctan}'' \simeq -2.5 \times 10^{-30} , \;\;\;
\bar{K}_{\oplus \rm Sqrt}'' \simeq -1.9 \times 10^{-42} .
\label{Earth-Kss}
\ee

\subsubsection{Scattering amplitude for $\varphi\varphi \rightarrow \varphi\varphi$}
\label{sec:scatt-amplitude}

Expanding in
$\varphi=\bar\varphi +\tilde \phi$ with $\tilde\phi=\phi/\sqrt{\bar{K}'}$, as in
Eqs.(\ref{background-split}), (\ref{phi-def}), and (\ref{deltaL-phi-n}),
the leading Lagrangian part which contributes to the two-body scattering is
\beq
{\cal L}_{\rm 2 body} = - \frac{1}{2} \partial^{\mu}\phi \partial_{\mu}\phi +
{\cal L}_3 + {\cal L}_4
\label{L-2body}
\eeq
with
\beqa
{\cal L}_3 & = & \frac{1}{{\cal M}^2 \bar{K}'^{3/2}}
\left[ \frac{\bar{K}'' \sqrt{-\bar\chi}}{\sqrt{2}} \, (\vec{n}\cdot\nabla\phi) \;
\partial^{\mu}\phi \partial_{\mu}\phi \right. \nonumber \\
&& \left. - \frac{\bar{K}'''(-\bar{\chi})^{3/2}\sqrt{2}}{3}
(\vec{n}\cdot\nabla\phi)^3 \right]
\label{L-3-2body}
\eeqa
and
\beqa
{\cal L}_4 & = & \frac{1}{{\cal M}^4 \bar{K}'^2}
\left[ \frac{\bar{K}''}{8} ( \partial^{\mu}\phi \partial_{\mu}\phi)^2
+ \frac{\bar{K}'''\bar\chi}{2} (\vec{n}\cdot\nabla\phi)^2 \;
\partial^{\mu}\phi \partial_{\mu}\phi \right. \nonumber \\
&& \left. + \frac{\bar{K}^{(4)}\bar{\chi}^2}{6} (\vec{n}\cdot\nabla\phi)^4 \right] .
\label{L-4-2body}
\eeqa
Here $\vec{n}$ is the unit vector along the gradient of the background scalar field,
\be
\nabla\bar\varphi = \sqrt{-2{\cal M}^4 \bar\chi} \; \vec{n} ,
\ee
and we used the fact that for such small-scale astrophysical backgrounds
(such as on the Earth or in the Solar System) time derivatives are negligible as we relax
to the quasistatic solution, $|\partial\bar\varphi/\partial t| \ll | \nabla\bar\varphi |$.
The mass of the scalar field is negligible for such high-energy
experiments, since from Eq.(\ref{mphi-pbar}) we have $m_{\phi} \sim
(\bar{K}''\bar{\chi}/\bar{K}')^{1/2} \bar{p} \ll \bar{p} \ll p$.
The hierarchy $\bar{K}''\bar{\chi} \ll \bar{K}'$ also implies that the cubic vertex ${\cal L}_3$
gives a negligible contribution to the $2\rightarrow 2$ scattering (through the diagram
$>\!\!\! - \! - \!\!\! <$), as compared with the quartic vertex ${\cal L}_4$
(through the diagram $> \! <$).

The anisotropic background breaks Lorentz invariance in the Lagrangian (\ref{L-2body}),
as the various terms in ${\cal L}_4$ are typically of the same order.
However, to simplify the analysis, we consider two-body scattering in the plane orthogonal
to the background gradient $\vec{n}$.
As the background is set by the Earth and the Sun only makes a small contribution
from Eq.(\ref{RK-Sun-Earth}), such scattering events with $\vec{p}_i\cdot\vec{n}=0$
correspond to momenta that are within the tangent plane to the surface of the Earth
(as $\vec{n}$ is the unit radial vector).

In the quantum regime associated with small occupation numbers, the field $\phi$
satisfies commutation rules as in Eq.(\ref{Wronskian}) with $\phi_k \dot{\phi}_k \sim 1$.
Then, we consider massless wave packets of size $1/\omega$ and wave number $k=\omega$,
with
\be
\mbox{low occupation number:} \;\;\; \phi_k \sim \omega^{-1/2} , \;\;
\phi \sim \omega ,
\label{phi-quantum-N=1}
\ee
which correspond to the elementary asymptotic states of the collision
$\varphi\varphi \rightarrow \varphi\varphi$.
These correspond to small fluctuations as compared with the Earth background
(\ref{Earth-background}) if $| \nabla\tilde\phi | \ll | \nabla\bar\varphi |$, which gives
\be
\mbox{small scalar fluctuations:} \;\;\; \omega \ll \left | \bar{K}' \bar{\chi} \right |^{1/4} {\cal M} ,
\label{small-wrt-background}
\ee
whence from Eq.(\ref{Earth-background}),
\be
\mbox{small scalar fluctuations:} \;\;\; \omega \ll 15 \, {\rm eV} .
\label{small-wrt-background-eV}
\ee
This corresponds to rather low-energy excitations, as compared with usual particle-physics
scales.

In this regime, the tree-level scattering amplitude defined by the Lagrangian
(\ref{L-2body}), associated with the diagram $> \! <$ set by ${\cal L}_4$,
is given in terms of the Mandelstam variables, $s= -(p_1+p_2)^2$ and $t= -(p_1-p_3)^2$,
by
\be
{\cal A}^{(4\perp)}_{\rm tree}(s,t) = \frac{\bar K''}{2 \bar K'^2}
\frac{s^2 + t^2 + s t}{\cM^4} ,
\label{A-perp-st}
\ee
where we have used the near masslessness of the scalar field and the superscript
``$(4\perp)$'' recalls that this expression holds for $\vec{p}_i\cdot\vec{n}=0$ and comes
from the quartic part ${\cal L}_4$.
The tree-level amplitude (\ref{A-perp-st}) is real so that the imaginary part
$\Im{\cal A}$ is given at leading order by the one-loop diagram ${\rm > \!\! O \!\! <}$,
which yields the scaling
\be
{\cal A}^{(4)}_{\rm 1loop} \sim \ii \left( \frac{\bar{K}'' s^2}{\bar{K}'^2 {\cal M}^4} \right)^2
\sim \ii \left( {\cal A}^{(4\perp)}_{\rm tree}(s) \right)^2 .
\label{A-s-1loop}
\ee

The perturbative regime, where the scattering cross section can be computed from the
low-order Feynman diagrams, is given by the condition
$| {\cal A}_{\rm 1loop} | \ll | {\cal A}_{\rm tree} |$, which yields
$|  {\cal A}_{\rm tree} | \ll 1$ and
\be
\mbox{perturbative regime:} \;\;\; \sqrt{s} \ll \frac{\bar K'^{1/2}}{\bar K''^{1/4}} \; {\cal M} .
\label{s-perturbative}
\ee
Higher-order $L$-loop diagrams also scale as
\beq
{\cal A}^{(4)}_{L} \sim \left( \frac{\bar{K}'' s^2}{{\bar K}'^2{\cal M}^4} \right)^{L+1}
\sim \left( {\cal A}^{(4)}_{\rm tree} \right)^{L+1} ,
\label{A-L-A-tree}
\eeq
and become large at the same scale as the one-loop diagram.
From Eqs.(\ref{Earth-background})-(\ref{Earth-Kss}), the perturbative regime
(\ref{s-perturbative}) corresponds to
\be
\mbox{Arctan:} \;\; \sqrt{s} \lesssim 2 \; {\rm MeV} , \;\;\;\;
\mbox{Sqrt:} \;\; \sqrt{s} \lesssim 2 \; {\rm GeV} .
\label{s-perturbative-Arctan-Sqrt}
\ee
This happens to be much above the upper bound (\ref{small-wrt-background-eV}), because
$\bar{K}''\bar{\chi} \ll \bar{K}'$ in highly nonlinear astrophysical backgrounds.
Therefore, all events where the scalar excitations are small as compared with the
astrophysical background are far within the perturbative regime for two-body collisions.

Finally, we check that the cubic vertex ${\cal L}_3$ in the Lagrangian (\ref{L-2body}) gives
a negligible contribution to the two-body scattering.
This vertex vanishes for $\vec{n}\cdot\vec{p}_i=0$, but if we consider generic orientations,
we obtain the estimate
\beq
{\cal A}^{(3)}_{\rm tree} \sim \frac{\bar{K}''^2 \bar{\chi} s^2}{{\bar K}'^3{\cal M}^4}
\sim \frac{\bar{K}'' \bar\chi}{\bar{K}'} {\cal A}^{(4)}_{\rm tree}
\label{A-3-A-tree}
\eeq
for the amplitude of the tree diagram $>\!\!\! - \! - \!\!\! <$.
This is much smaller than the amplitude (\ref{A-perp-st}) associated with the diagram
$> \! <$, by a factor $10^{-21}$ or $10^{-33}$ for the Arctan or Sqrt models,
see Eqs.(\ref{Earth-background})-(\ref{Earth-Kss}).

\subsubsection{Perturbative unitarity}
\label{sec:pert-unitarity}

Scattering amplitudes must obey nonperturbative unitary bounds,
which can be derived as follows \cite{Schwartz2013}.
The optical theorem relates the imaginary part of the forward scattering amplitude,
where $t=0$, to the total cross section,
\beqa
\Im {\cal A}(s,0) & = & 2 E_{\rm CM} | \vec{p} | \sum_{X}
\sigma_{\rm tot}(\{p_1,p_2\} \rightarrow X) \nonumber \\
& \geq & 2 E_{\rm CM} | \vec{p} | \sigma_{\rm tot}(2 \rightarrow 2) . \;\;\;
\label{optical-theorem}
\eeqa
Using $\sigma_{\rm tot}(2 \rightarrow 2) = \int d\vec\Omega |{\cal A}|^2/(64\pi^2 E_{\rm CM}^2)$ and $|\Im{\cal A}| \leq | {\cal A}|$  this gives
\be
|{\cal A}(s,0)| \geq | \Im{\cal A}(s,0) | \geq
\int \frac{d\vec{\Omega}}{64\pi^2} |{\cal A}(\vec\Omega)|^2 .
\label{inequality-unitary}
\ee
Because of the anisotropic terms in Eq.(\ref{L-2body}), the amplitude ${\cal A}$
depends on both the spherical coordinates angles $(\theta,\varphi)={\vec\Omega}$.
Nevertheless, because the anisotropic terms in the Lagrangian (\ref{L-2body})
are of the same order as the Lorentz-invariant term, the angular average of
${\cal A}$ is of the same order as ${\cal A}^{\perp}$, which yields
\be
\mbox{nonperturbative unitarity:} \;\; |{\cal A}^{\perp}(s,0)| \lesssim 16\pi .
\label{non-pert-unitarity}
\ee
From the perturbative expression (\ref{A-perp-st}) or from the scaling
(\ref{A-L-A-tree}) of loop contributions, we can see that the tree-level amplitude
reaches the upper bound (\ref{non-pert-unitarity}) at the limit of validity (\ref{s-perturbative})
of the perturbative regime. This means that nonperturbative effects must completely
stop the growth of $|{\cal A}|$.
Using the one-loop expression (\ref{A-s-1loop}) for the imaginary
part of the amplitude and the tree-level expression (\ref{A-perp-st}) for the real part,
we find that both terms in the second inequality (\ref{inequality-unitary}) are of the same order
in the perturbative regime, whatever the energy scale $\sqrt{s}$ as long as it remains
below the upper bound (\ref{s-perturbative-Arctan-Sqrt}).
This is consistent with the fact that the inequality is satisfied as the theory is unitary
(the Hamiltonian is Hermitian).
However, as we noticed in Eq.(\ref{small-wrt-background-eV}), it happens that, before the
energy of the scalar excitations reaches this nonperturbative threshold, they become of the
same order as the astrophysical background, and the scattering analysis presented above
no longer applies.

\subsection{Unitarity and UV completion}
\label{sec:UV-completion}

Models of particle physics are often understood as effective field theories, which only apply
below some energy cutoff $\Lambda_c$. In our context, this is also the picture assumed
in \cite{deRham2014}, which uses a renormalization group approach to investigate the
magnitude of the quantum corrections generated by modes $k<\Lambda_c$ below this cutoff;
see the discussion in section~\ref{sec:de-Rham}.
As pointed out in \cite{Adams:2006aa}, the form of the low-energy scattering amplitudes
can be related to the possibility to build such UV completions of the theory,
where we embed the low-energy
effective field theory in a more complete theory where the scalar field would appear
as a low-energy manifestation of a more complex situation.
In this case, the forward amplitude would be analytic in the complex
s-plane up to possible branch cuts signaling the creation of massive states or poles
where intermediate states would be created. Then, as long as the Froissard
bound is satisfied for the UV completion, one can write a dispersion relation,
\be
\left. \frac{d^2}{ds^2} {\cal A}^{\perp}(s,0) \right|_0 = \oint \frac{ds}{\ii\pi}
\frac{{\cal A}^{\perp}(s,0)}{s^3} =
\frac{4}{\pi} \int_{s_\star}^\infty ds \frac{\Im {\cal A}^{\perp}(s,0)}{s^3}
\label{amplitude-branch-cut}
\ee
where $s_\star$ is the branch point on the positive axis and we used the crossing
symmetry ${\cal A}(-s,0)={\cal A}(s,0)$. As recalled in Eq.(\ref{optical-theorem}),
the optical theorem relates the imaginary part of the amplitude to the total cross section,
$\Im {\cal A}^{\perp}(s,0) = s\sigma^{\perp}(s) > 0$,
implying that the K-mouflage theory can be embedded in a UV completion if
$\left. \frac{d^2{\cal A}}{ds^2}(s,0) \right|_0 > 0$.
Using the low-energy expression (\ref{A-perp-st}) gives
\be
\mbox{UV completion:} \;\;\; \bar K'' >0 .
\label{K2-UV-completion}
\ee
For the models considered in this paper, this condition is violated around astrophysical
backgrounds (it is satisfied around the cosmological background where $K''>0$).
However, the constraint on $\bar{K}''$ in the highly nonlinear regime is due to the
anomalous perihelion of the Moon orbit and reads as
\be
\frac{\beta^2 |\bar\chi \bar K''|}{\bar K'^2} < 8 \times 10^{-13} ,
\label{Moon-perihelion}
\ee
for the background $\bar\chi_{\rm m.e.}$ associated with the Earth-Moon system.
This simply requires that $K''$ becomes very small in the nonlinear regime.
For the Arctan and Sqrt models,  where $K'$ is a monotonic decreasing function from
$K'=K_*=10^3$ at $\chi=-\infty$ to $K'=1$ at $\chi=0$, $K'' <0$ on the negative real
axis $\chi<0$. This is because we considered simple models that can satisfy
Solar System and cosmological constraints.
However, we could also build kinetic functions $K(\chi)$ with $K''>0$ over the range
$\chi<\chi_*$, where $K' \simeq K_*$ and $K'' \simeq 0$. Then, $K''$ would have to
change sign in the range $[\chi_*,0]$ so that $K'$ again decreases down to $K'=1$ at
$\chi=0$.
Therefore, it is possible to build models such that $K'' >0$ in the highly nonlinear
small-scale regime, associated with astrophysical and Solar System backgrounds,
and the UV completion condition (\ref{K2-UV-completion}) is satisfied on the Earth.
However, this condition cannot be satisfied in all backgrounds as there must exist
a range of backgrounds, $\chi \sim -1$, where $K'' <0$.
Thus, in all these K-mouflage models, there is no standard quantum field-theory UV completion.

However, we do not need to interpret the K-mouflage model (\ref{S-def}) as an effective
field theory, which should be embedded in a more complete theory beyond a cutoff scale
$\Lambda_c$. The condition (\ref{K2-UV-completion}) shows that this is not possible,
but the discussion in section~\ref{sec:de-Rham} also suggests that this is not necessary.
Indeed, we found that there exists a classical regime (\ref{classical-regime-1})
where higher-order operators associated with quantum contributions and counterterms
are negligible. As checked in sections~\ref{sec:classical-one-loop} and
\ref{sec:quantum-nonlinear}, this regime applies to all astrophysical and cosmological
configurations.
This also applies to scalar collisions at low energies (\ref{small-wrt-background-eV}) with
respect to the Earth background, which are also in the standard perturbative regime
(\ref{s-perturbative}), and as we shall find below, at high energies
(\ref{isolated-events-N}), in the nonstandard nonlinear but classical regime
(\ref{classical-N})-(\ref{classical-Sqrt-N}).
In all these regimes, which cover all energy scales,
the theory is predictive as we can trust tree-level or classical computations.

Then, the violation of the condition $\bar{K}''>0$ means that we do not interpret the 
K-mouflage Lagrangian (\ref{L0-def}) as a low-energy effective field theory but 
as a gravitational-like theory in the sense of \cite{Keltner:2015xda}, where the 
assumptions leading to Eq.(\ref{amplitude-branch-cut}) and the condition 
(\ref{K2-UV-completion}) are violated.

We can note that in curved spacetime the optical theorem and the analytic
structure of the Green functions and amplitudes are modified.
For instance, the refractive index $n(\omega)$ may no longer be analytic in the 
upper-half plane, and dispersion relations are modified 
\cite{Hollowood:2008a,Hollowood:2008b,Hollowood:2012a}, so that the relation
(\ref{amplitude-branch-cut}) and the constraint (\ref{K2-UV-completion}) may no longer
apply. 
However, in this section, we consider the Minkowski limit that is relevant for most
practical applications of scalar-scalar scattering and that is sufficient to show that
K-mouflage scenarios are not standard effective field theories with UV completion.

\subsection{Quantum and classical regimes for high-energy collisions}
\label{sec:quantum-collision}

In the previous sections, we considered relatively low-energy scalar collisions, where the
zeroth-order scalar field remains set by the astrophysical background due to the Earth.
These events were described by the Lagrangian (\ref{L-2body}), based on the expansion
around this astrophysical background. We now consider high-energy events where
this astrophysical background can be neglected as compared with the values of the
scalar field generated by the collisions.
From Eq.(\ref{small-wrt-background-eV}) this corresponds to energies
$\omega \gg 15 \, {\rm eV}$.
Then, we no longer use the perturbative
Lagrangian (\ref{L-2body}) as the collision can be treated as an isolated event described
by the full K-mouflage Lagrangian (\ref{S-def}).
We analyze the behavior of K-mouflage models at these high energies in view of the
nonrenormalization theorem described in section~\ref{sec:Non-renormalization}.
We want to establish that at high energy the classical
theory can be trusted and no quantum corrections to the K-mouflage action should be
considered, as in Eq.(\ref{classical-regime-1}). In other words, at high energies, we recover
the classical nonlinear regime discussed in section~\ref{sec:Quantum-corrections}
that was found to apply to the cosmological and astrophysical cases.

At tree level, for the K-mouflage Lagrangian (\ref{L0-def}),
the equation of motion (\ref{KG-matter}) of the scalar field in the vacuum reads as
$\partial_{\mu} \left( K' \partial^{\mu} \varphi \right) = 0$, which also reads as
\be
\frac{\partial}{\partial t} \left( K' \frac{\partial\varphi}{\partial t} \right)
- \nabla \cdot \left( K' \nabla\varphi \right) = 0 ,
\label{KG-vacuum}
\ee
with
\be
\chi = \frac{1}{2{\cal M}^4} \left[ \left( \frac{\partial\varphi}{\partial t} \right)^2
- ( \nabla\varphi)^2 \right] .
\ee
Let us consider wavefronts that propagate along the direction ${\vec e}$
(with ${\vec e}^{\;2}=1$) with constant velocity $v$,
\be
\varphi({\vec x},t) = \varphi({\vec e}\cdot{\vec x} - v t) .
\label{phi-nonlinear-wave}
\ee
For $v^2=1$, they are exact solutions of the nonlinear Klein-Gordon equation
(\ref{KG-vacuum}) with
\be
v^2=1 : \;\;\; \chi = \frac{\varphi'^2}{2{\cal M}^4} (v^2-1) = 0 ,
\ee
as $K'=1$ is now constant and Eq.(\ref{KG-vacuum}) takes the same form as the usual
linear Klein-Gordon equation.
Thus, these nonlinear solutions, of arbitrary amplitude and shape $\varphi$,
propagate at the speed of light.

They are also linearly stable with respect to high frequencies and wave numbers.
Indeed, if we consider small perturbations $\phi$ around such solutions propagating
along the $x$-axis,
\beq
\varphi = \bar\varphi(x - t) + \phi({\vec x},t) ,
\eeq
they obey at linear order the wave equation
\beq
\frac{\pl^2\phi}{\pl t^2} - \nabla^2 \phi + \frac{\bar{K}'' \bar\varphi'^2}{{\cal M}^4} \left(
\frac{\pl^2\phi}{\pl t^2} + 2 \frac{\pl^2\phi}{\pl t \pl x} + \frac{\pl^2\phi}{\pl x^2} \right) = 0 .
\eeq
Here we considered high-frequency perturbations and neglected derivatives of
$\bar\varphi'^2$.
Looking for plane-wave solutions $e^{\ii\omega t - \ii {\vec k}\cdot{\vec x}}$, we obtain the
dispersion relation
\beq
\left( 1 + \frac{\bar{K}'' \bar\varphi'^2}{{\cal M}^4} \right) \omega^2
- 2 \frac{\bar{K}'' \bar\varphi'^2}{{\cal M}^4} k_x \omega - k^2
+ \frac{\bar{K}'' \bar\varphi'^2}{{\cal M}^4} k_x^2 = 0 ,
\eeq
and the solutions $\omega$ are real, provided the discriminant is positive,
\beq
\omega^2 \geq 0 : \;\;\; k^2 + \frac{\bar{K}'' \bar\varphi'^2}{{\cal M}^4} (k_y^2+k_z^2) \geq 0 .
\eeq
Therefore, the nonlinear solutions (\ref{phi-nonlinear-wave}) are linearly stable if
$K''(0) \geq 0$. For the models that we consider in this paper, this is indeed the case,
as $K''(0)=0$ because we chose simple examples where $K'(\chi)$ is even.

The conjugate momentum is $\pi = K' (\partial\varphi/\partial t)$, and the Hamiltonian
density is
${\cal H}=K' (\partial\varphi/\partial t)^2 - {\cal M}^4 K =
(\partial\varphi/\partial t)^2 + {\cal M}^4$.
We consider scalar excitations that are greater than the Earth background,
that is, $(\pl\varphi/\pl t)^2/{\cal M}^4 \gg | \bar\chi_{\oplus} |$.
From Eq.(\ref{Earth-background}), this yields
\beq
\mbox{isolated wave:} \;\;\; \left| \frac{\pl\varphi}{\pl t} \right|^{1/2} \gg 10^3 {\cal M}
\sim 1 \; {\rm eV} .
\label{isolated-wave}
\eeq
This also implies ${\cal H} \simeq (\partial\varphi/\partial t)^2$.
Writing the width of $\varphi({\vec x},t)$ along the direction ${\vec e}$ as $\Delta x \sim 1/k$
and $k=\omega$, the characteristic energy density $\epsilon$ at the position
of the wave is
\be
\epsilon \simeq \left( \frac{\partial\varphi}{\partial t} \right)^2 = ( \nabla\varphi)^2
\simeq N \omega^4 , \;\;\; \varphi \sim \sqrt{N} \omega ,
\label{N-quanta-def}
\ee
where $N$ is the number of quanta of the scalar excitation. For $N=1$, we recover
$(\partial\varphi/\partial t) \sim \omega^2$ and $\varphi \sim \omega$,
as for the elementary quantum states (\ref{phi-quantum-N=1}).
Then, the criterion (\ref{isolated-wave}) reads as
\beq
\mbox{isolated event:} \;\;\; \omega \gg N^{-1/4} \; \rm{eV} .
\label{isolated-events-N}
\eeq

We now consider the collision of two such solutions $\varphi_1$ and $\varphi_2$, which
propagate with the velocities ${\vec v}_1$ and ${\vec v}_2$.
We assume these wave packets actually have a finite extent and propagate freely
before the collision, in a fashion similar to Eq.(\ref{phi-nonlinear-wave}).
At the collision, $\chi$ is no longer zero as the two wave packets overlap, and we have
\be
\chi = - \frac{1}{{\cal M}^4} \partial^{\mu}\varphi_1 \partial_{\mu}\varphi_2
\sim \frac{\varphi'^2}{{\cal M}^4} ( 1 - {\vec v}_1 \cdot {\vec v}_2 )
\sim \frac{N\omega^4}{{\cal M}^4} ,
\ee
where we take the two packets to have the same amplitude and width,
as in Eq.(\ref{N-quanta-def}).
The collision is well described by the classical Klein-Gordon equation (\ref{KG-vacuum})
if the classicality criterion (\ref{classical-regime-1})-(\ref{classical-regime-2}) is satisfied,
$\omega \ll {\cal M} (K'\chi)^{1/4} (K'/K''\chi)^{1/2}$.
In the highly nonlinear regime, we write
\beq
| \chi | \rightarrow \infty: \;\;\; K'' \simeq K_2 \left( \frac{\chi}{\chi_*} \right)^{-\nu} , \;\;\; \nu>1 .
\label{Ks-NL}
\eeq
For the Arctan and Sqrt models, we have
\beq
\mbox{Arctan:} \;\;\; K_2=\frac{2K_*}{\chi_*} = 20 , \;\;\; \nu=3 ,
\eeq
\beq
\mbox{Sqrt:} \;\;\; K_2=\frac{3K_*\sigma_*^2}{\chi_*^3} = 0.3 , \;\;\; \nu=4 .
\eeq
Then, the classicality criterion reads as
\beq
\left( \frac{\omega}{{\cal M}} \right)^{\nu-1} \gg K_*^{-3/8} K_2^{1/4} \chi_*^{\nu/4} N^{(1-2\nu)/8} ,
\label{classical-N}
\eeq
which gives
\beq
\mbox{classical for Arctan:} \;\;\; \omega \gg N^{-5/16} 10^{-3} \; \rm{eV} ,
\label{classical-Arctan-N}
\eeq
\beq
\mbox{classical for Sqrt:} \;\;\; \omega \gg N^{-7/24} 10^{-3} \; \rm{eV} .
\label{classical-Sqrt-N}
\eeq
Thus, we find that for high-energy events $\omega \gg 1 \; \rm{eV}$,
the Earth background is negligible from Eq.(\ref{isolated-events-N}), and we are in the 
classical regime from Eqs.(\ref{classical-Arctan-N})-(\ref{classical-Sqrt-N}).

We can conclude that for scalar collisions the K-mouflage theory defined by the classical
Lagrangian (\ref{L0-def}) is predictive both at low energies $\omega \ll 1 \; {\rm eV}$
and at high energies $\omega \gg 1 \; {\rm eV}$.
In the low-energy case, the scalar excitations are small fluctuations on the Earth background,
Eq.(\ref{small-wrt-background-eV}), and we are in a standard perturbative regime,
Eq.(\ref{s-perturbative-Arctan-Sqrt}), where higher-order quantum corrections 
(Feynman diagrams) are small and we are dominated by the lowest-order terms of the 
tree-level Lagrangian (\ref{L-2body}).
In the high-energy case, the Earth background is negligible, and the scalar excitations can be
considered as isolated events in vacuum, Eq.(\ref{isolated-events-N}), and we are
in a nonstandard classical regime (as compared with usual particle-physics quantum field
theories), which also applies to other gravitational-like theories displaying
the Vainshtein mechanism \cite{Keltner:2015xda}, where quantum corrections
become increasingly negligible at higher energies but  are sensitive to the nonlinearities
of the classical Lagrangian (\ref{L0-def}).

\section{Interactions between fermions and scalars at high energy}
\label{sec:fermions-scalars}

\subsection{Classical scalar cloud around fermions}
\label{sec:scalar-cloud}

We have already seen that at high energy the quantum collision of scalars is characterized
by the convergence of the quantum action for K-mouflage to its classical regime.
Here, we will study what happens in the case of fermion antifermion annihilation such
as it happened at LEP or is currently under scrutiny at the LHC.
To be specific, we shall be interested in the case of the two-fermion annihilation 
process $f \bar f\to \varphi$. 
This process takes place as matter couples to the scalar field through the conformal 
mapping in the action (\ref{S-def}), as matter particles follow the geodesics of the 
Jordan metric $\tilde{g}_{\mu\nu} = A^2(\varphi) g_{\mu\nu}$.
Using the linear approximation (\ref{A-def}) for the coupling function, as
$\beta\varphi/M_{\rm Pl} \ll 1$ in all practical cases, this corresponds to the interaction
Lagrangian
\beq
{\cal L}_{\rm int} = \frac{\beta}{M_{\rm Pl}} \; \varphi \;T ,
\label{Lint-def}
\eeq
which must be added to the scalar-field Lagrangian (\ref{L0-def}), where
$T = T^{\mu}_{\mu}$ is the trace of the matter energy-momentum tensor.
This gives back the nonlinear Klein-Gordon equation (\ref{KG-matter}) (here, we work
in the Minkowski metric).
As for astrophysical objects such as the Sun and the Earth, elementary particles with
a nonvanishing trace $T$ act as a source for the scalar field, through the right-hand side
in the nonlinear Klein-Gordon equation (\ref{KG-matter}) at the classical level, or through
the interaction Lagrangian (\ref{Lint-def}).
Before studying the annihilation process $f \bar f\to \varphi$, we first investigate the scalar
excitation generated by a free propagating fermion $f$.
In this section we follow a classical approach, where both the scalar field and the fermionic
source are treated at the classical level.

The trace of the energy-momentum tensor of massive fermions on shell is
\be
\hat{T}_\psi = - m_\psi : \bar\psi \psi : \;
\label{T-psi-def}
\ee
which is a quantum operator. When $\vert \chi \vert \gg 1$ and for the models where $K'$
goes to a constant in the highly nonlinear regime, the Klein-Gordon equation can be written
as
\be
\Box \hat{\phi} = - \frac{\beta}{M_{\rm Pl} \sqrt{K'}} \hat{T}_\psi , \;\;\;
\hat\varphi = \frac{\hat\phi}{\sqrt{K'}} ,
\label{KG-Tpsi}
\ee
and this should be understood as an equality between quantum operators.
Here and in the following, we take $K'$ to be constant, and we introduced the canonically
normalized scalar field $\phi$, as we also did in Eqs.(\ref{phi-def}) or (\ref{L-2body})
for instance.
We expand as usual the four-component Dirac spinor $\psi$ over the creation and annihilation
operators as
\be
\hat{\psi} = \int \frac{d^3 k}{(2\pi)^{3/2} \sqrt{2\omega_k}} \sum_{\alpha=1}^2 \left[ u^\alpha(\vec{k})
e^{\ii k \cdot x} \hat{b}^{\alpha}_{\vec k} + v^\alpha(\vec{k}) e^{-\ii k \cdot x}
\hat{d}^{\alpha\dagger}_{\vec k} \right]
\label{psi-fermion-def}
\ee
where $k \cdot x=\eta_{\mu\nu} k^\mu x^\nu=-\omega_k t+\vec{k}\cdot\vec{x}$
and the creation and annihilation operators satisfy
$\{ \hat{b}^{\alpha}_{\vec k} , \hat{b}^{\beta\dagger}_{\vec p} \} = \delta_{\alpha,\beta}
\delta_D( {\vec k} - \vec{p} )$ and
$\{ \hat{d}^{\alpha}_{\vec k} , \hat{d}^{\beta\dagger}_{\vec p} \} = \delta_{\alpha,\beta}
\delta_D( {\vec k} - \vec{p} )$.
The Greek indices $\alpha,\beta=1,2$ represent the two spins of the fermions.
We have $\omega_k = \sqrt{ \vec{k}^2 + m_\psi^2}$ and the Dirac spinors are defined by
\begin{eqnarray}
&& u^\alpha (\vec{k}) = \sqrt{\omega_k +m_\psi} \left( \begin{array}{c} \chi^\alpha \\
\frac{\vec{\sigma} \cdot \vec{k}}{\omega_k +m_\psi} \chi^\alpha \\
\end{array} \right ) \nonumber \\
&& v^\alpha (\vec{k}) = \sqrt{\omega_k +m_\psi} \left( \begin{array}{c}
\frac{\vec{\sigma} \cdot \vec{k}}{\omega_k +m_\psi} \chi^\alpha \\
\chi^\alpha \\ \end{array} \right )
\end{eqnarray}
where $\chi^\alpha$ is a two-component spinor with
$\chi^1= \left ( \begin{array}{c}  1\\0\\ \end{array}\right)$ and
$\chi^2= \left ( \begin{array}{c}  0\\1\\ \end{array}\right)$.
We also have $(\gamma^0)^\dagger=\gamma^0$, $(\gamma^0)^2=1$, and
$\{\gamma^{\mu},\gamma^{\nu}\} = - 2 \eta^{\mu\nu}$.
The propagating free fermion $f$ is represented by the state vector
\be
\vert f \rangle = \int \frac{d^3 k}{\sqrt{\omega_k}} f(\vec{k}) \hat{b}^{1\dagger}_{\vec{k}}
\vert 0 \rangle ,
\label{fermion-f-def}
\ee
where we have chosen a given spin.
The number operator $\hat{Q}$ reads as
\beq
\hat{Q} = \int d^3k \; \sum_{\alpha} \left( \hat{b}^{\alpha\dagger}_{\vec k} \hat{b}^{\alpha}_{\vec k}
- \hat{d}^{\alpha\dagger}_{\vec k} \hat{d}^{\alpha}_{\vec k} \right) ,
\label{Q-charge}
\eeq
and the wave function $f({\vec k})$ is normalized according to
\be
\int \frac{d^3 k}{\omega_k} \left | f({\vec k}) \right |^2 = 1 ,
\ee
so that $\langle f \vert f \rangle = 1$ and $\langle f \vert \hat{Q} \vert f \rangle = 1$.

In a classical approach, we approximate the Klein-Gordon equation (\ref{KG-Tpsi}) by
the classical equation
\beq
\Box \phi = - \frac{\beta T_{\psi}}{M_{\rm Pl} \sqrt{K'}} \;\;\; \mbox{with} \;\;\;
T_{\psi} = \langle f \vert \hat{T}_{\psi} \vert f \rangle ,
\label{KG-Tpsi-class}
\eeq
that is, we treat both the scalar field $\phi$ and the source
$T_{\psi}$ as classical fields.
From Eqs.(\ref{T-psi-def}), (\ref{psi-fermion-def}), and (\ref{fermion-f-def}), we obtain
\beqa
T_{\psi}(x) & = & - m_{\psi} \int \frac{d^3k_1 d^3k_2}{(2\pi)^3 2\omega_1\omega_2}
f({\vec k}_1)^* f({\vec k}_2) \bar{u}^1({\vec k}_1) u^1({\vec k}_2) \nonumber \\
&& \times \; e^{\ii (\omega_1-\omega_2) t - \ii ({\vec k}_1-{\vec k}_2) \cdot {\vec x} } .
\label{Tpsi-f}
\eeqa
Integrating over space gives
\beq
\int d^3x \; T_{\psi}(t,{\vec x}) = - m_{\psi}^2 \int \frac{d^3k}{\omega_k^2} \left| f({\vec k}) \right|^2
\simeq - \frac{m_{\psi}^2}{\omega} ,
\label{Tpsi-space}
\eeq
where we assumed the wave function is peaked around some value of ${\vec k}$ and $\omega$
and we used the relation
$\bar{u}^{\alpha}({\vec k}) u^{\beta}({\vec k}) = 2 m_{\psi} \delta_{\alpha,\beta}$.
It is convenient to work in the rest frame of the particle ($\omega=m_{\psi}$).
From Eqs.(\ref{Tpsi-f}) and (\ref{Tpsi-space}), we approximate the trace $T_{\psi}$ as a Dirac
peak,
\beq
T_{\psi}(t,{\vec x}) \simeq - m_{\psi} \, \delta_D({\vec x}) ,
\eeq
and the solution of the classical equation (\ref{KG-Tpsi-class}) is
\beq
\phi(t,{\vec x}) = - \frac{\beta m_{\psi}}{4\pi\sqrt{K'} M_{\rm Pl} |{\vec x}|} .
\label{phi-cloud}
\eeq
Then, the Lorentz-invariant quantity $\chi$ of Eq.(\ref{chi-def}) reads as
\beq
\chi = - \frac{1}{2{\cal M}^4 K'} \pl^{\mu}\phi\pl_{\mu}\phi =
-  \frac{1}{2K'^2} \left( \frac{R_K}{r} \right)^4 ,
\label{chi-fermion}
\eeq
where $r=|{\vec x}|$ is the the distance from the particle in the rest frame and the K-mouflage
radius is given by
\be
R_K = \left( \frac{\beta m_\psi}{4\pi M_{\rm Pl} \cM^2} \right)^{1/2} .
\label{RK-fermion}
\ee
Eq.(\ref{chi-fermion}) shows that we are in the negative branch regime, $\chi<0$, associated
with static events (because we consider a stationary state that is static in the rest frame).
We find that Eq.(\ref{chi-fermion}) agrees with Eq.(\ref{KG-static}), associated with spherically
symmetric compact objects, and that Eq.(\ref{RK-fermion}) agrees with Eq.(\ref{RK-def}),
where the mass $M$ associated with a nonrelativistic macroscopic object (such as the Sun)
must be replaced by $m_{\psi}$.
Note that, even for relativistic particles, it is $m_{\psi}$ rather than the energy $\omega$
that enters the K-mouflage radius (\ref{RK-fermion}), or the source term of the Klein-Gordon
equation.
This is related to the fact that the trace $T^{\mu}_{\mu}$ of the energy-momentum tensor
vanishes for radiation and massless particles.

This gives
\be
R_K = \sqrt{\frac{\beta}{0.1} \frac{m_\psi}{m_e}} \; 0.1 \; {\rm fm}
\label{RK-cloud}
\ee
where $m_e$ is the electron mass.
Taking $K'\sim 10^3$ and $\beta \sim 0.1$, the terrestrial background can be neglected in
experiments on Earth provided $\vert \chi \vert \gg \vert \chi_\oplus \vert \sim 10^{12}$;
see Eq.(\ref{Earth-background}).
From Eq.(\ref{chi-fermion}), this is reached for distances
\beq
\mbox{negligible Earth background:} \;\;\; r \ll \sqrt{\frac{m_\psi}{m_e}} \; 10^{-6} \; {\rm fm} .
\label{isolated-cloud}
\eeq
This is much smaller than the Compton radius of order $1/m_\psi$ for particles like the 
electron.
This implies that these particles evolve in the terrestrial background and give rise to
a scalar cloud that is a very small perturbation to this background value.
This also implies that in a particle collision at high energy, it is only at very high energy 
that the collision probes distances $r \sim 1/E$ where the effect of the particles on the 
background field is not negligible.

\subsection{Scalar Bremsstrahlung}
\label{sec:Bremsstrahlung}

In the previous subsection, we considered the scalar cloud that surrounds a free moving fermion,
in the stationary case, and we used a classical approach for both the scalar field and the
fermionic source term.
We now study the scalar emission by a moving fermion, which may accelerate.
For this scalar Bremsstrahlung process, we adopt a semiclassical approach,
where we again use a classical approximation for the fermionic source term but now
keep the scalar field as a quantum field.
Thus, as in Eqs.(\ref{v-vk-def}) or (\ref{u-uk-def}), the free scalar field before the
source has been turned on is expanded as
\begin{equation}
\hat{\phi}_{\rm in}(t,\vec{x}) = \int \frac{d^3k}{(2\pi)^{3/2}\sqrt{2\omega}}
\left[ e^{\ii k \cdot x} \hat{c}_{\vec{k}} + e^{-\ii k \cdot x} \hat{c}_{\vec{k}}^{\dagger} \right] ,
\label{phi-ck-def}
\end{equation}
where $\hat\phi$ is the canonically normalized field as in Eq.(\ref{KG-Tpsi}),
$K'$ is set by the Earth background and
$[\hat{c}_{\vec k},\hat{c}_{\vec p}^{\dagger}]=\delta_D({\vec k}-{\vec p})$.
Instead of the classical approximation (\ref{KG-Tpsi-class}), we write the Klein-Gordon
equation (\ref{KG-Tpsi}) as the semiclassical equation
\beq
- \Box \hat\phi = j , \;\;\; j = \frac{\beta}{M_{\rm Pl}\sqrt{K'}} \langle f \vert \hat{T}_{\psi}
\vert f \rangle ,
\label{KG-Brem-1}
\eeq
where we keep the scalar field quantum.
Thus, we study the emission of ``scalarons'' by the classical current $j$, in a fashion similar
to the emission of photons by a classical current $j^{\mu}$ (bremsstrahlung).

Introducing the retarded and advanced Green functions,
\be
G_{\rm ret,adv} = \frac{1}{2\pi} \theta (\pm x^0) \delta_D(x^2) ,
\label{G-ret-dav}
\ee
which satisfy $- \Box_x G_{\rm ret,adv}(x-y) = \delta_D^4(x-y)$,
we can write the solution of Eq.(\ref{KG-Brem-1}) as
\beqa
\hat\phi (x) & = & \hat\phi_{\rm in}(x) + \int d^4 x' \; G_{\rm ret} (x-x') j(x') \\
& = & \hat\phi_{\rm out}(x) + \int d^4x' \; G_{\rm adv} (x-x') j(x') .
\eeqa
This allows us to obtain the relation between the in- and out-fields
\cite{ItzyksonZuber1980},
\beq
\hat\phi_{\rm out}(x) = \hat\phi_{\rm in}(x) + \phi_{\rm clas}(x) ,
\label{phi-clas-1}
\eeq
with
\beq
\phi_{\rm clas}(x) \! = \! \int d^4 x' \; G^{(-)}(x-x') j(x')
\label{phi-class-def}
\eeq
and
\beq
G^{(-)}(x) = G_{\rm ret}(x) - G_{\rm adv}(x) = \frac{1}{2\pi} \epsilon(x^0) \delta_D(x^2) ,
\label{G(-)-def}
\eeq
where $\epsilon(x^0)=\pm 1$ depending on the sign of $x^0$.
The field $\phi_{\rm clas}(x)$ is also a solution of the homogeneous Klein-Gordon equation
and it is the classical field radiated by the moving particle associated with the current $j$.
Using the fact that $G^{(-)}(x) = - \Delta(x)$, where $\Delta(x)$ is the commutator of
scalar massless free fields,
\beq
[ \hat\phi_{\rm in}(x) , \hat\phi_{\rm in}(x') ] = \ii \Delta(x-x') ,
\label{Delta-commutator}
\eeq
we can rewrite Eq.(\ref{phi-clas-1}) as
\beq
\hat\phi_{\rm out}(x) = \hat\phi_{\rm in}(x) + \ii \int d^4x' [ \hat\phi_{\rm in}(x) , \hat\phi_{\rm in}(x') j(x') ] .
\eeq
This can also be written as \cite{ItzyksonZuber1980}
\beq
\hat\phi_{\rm out} = S^{-1} \, \hat\phi_{\rm in} \, S ,
\eeq
where the $S$ matrix reads
\be
S = \exp \left( \ii \int d^4x \; \hat\phi_{\rm in}(x) j(x) \right) .
\label{S-matrix-def}
\ee
It is convenient to rewrite $S$ in normal order. Splitting the quantum field $\hat\phi_{\rm in}$
of Eq.(\ref{phi-ck-def}) as a sum of annihilation $\hat\phi^{(-)}_{\rm in}$ and creation
$\hat\phi^{(+)}_{\rm in}$ operators,
$\hat\phi_{\rm in} = \hat\phi^{(-)}_{\rm in} +\hat\phi^{(+)}_{\rm in}$,
we can write the $S$ matrix (\ref{S-matrix-def}) as
\beqa
S & = & e^{\ii \int d^4x \; \hat\phi_{\rm in}^{(+)}(x) j(x)} \;
e^{\ii \int d^4x \; \hat\phi_{\rm in}^{(-)}(x) j(x)} \nonumber \\
&& \times \; e^{- \frac{1}{2} \int \frac{d^3k}{(2\pi)^3 2\omega_k} \; \vert j(k) \vert^2 } .
\label{S-matrix-normal-ordering}
\eeqa
The Hermitian conjuguate $S^{\dagger}$ is given by the same expression where the factors
$\ii$ are replaced by $-\ii$ and we defined the Fourier transform of the current by
\be
j(k) = \int d^4 x \; e^{\ii k \cdot x} j(x) .
\ee

The S-matrix transforms incoming states into outgoing ones. As a result, the incoming vacuum
state $\vert {\rm in} \rangle \equiv \vert 0 \rangle_{\rm in}$ of the scalar field becomes
\beqa
\vert {\rm out} \rangle & = & S^{\dagger} \vert {\rm in} \rangle \nonumber \\
& = & e^{- \frac{1}{2} \int \frac{d^3k}{(2\pi)^3 2\omega_k} \; \vert j(k) \vert^2 }
e^{-\ii \int d^4x \; \hat\phi_{\rm in}^{(+)}(x) j(x)} \vert 0 \rangle_{\rm in} \hspace{1cm}
\label{out-in-def}
\eeqa
and we can check that it is an eigenvector of the annihilation operator,
\beq
\hat{c}_{\vec k} \vert {\rm out} \rangle =  \frac{- \ii \; j(k)^*}{(2\pi)^{3/2}\sqrt{2\omega_k}}
\vert {\rm out} \rangle .
\eeq
This means that it is a coherent state created by the classical source $j(x)$.
This is the analog result to the creation of a photon coherent state by a classical source
\cite{ItzyksonZuber1980}, and this describes bremsstrahlung in a quantum way.
The mean energy $\bar{E}_{\rm brem}$ emitted by the particle in this process can be
obtained from the Hamiltonian of the scalar field,
\be
\hat{H}(\phi_{\rm in}) = \int d^3 k \; \omega_k \; \hat{c}_{\vec k}^\dagger \hat{c}_{\vec k} .
\ee
We obtain
\beqa
{\bar E}_{\rm brem} & = & \langle 0 \vert \hat{H}(\phi_{\rm out}) \vert 0 \rangle_{\rm in}
= \langle 0 \vert S^{-1} \hat{H}(\phi_{\rm in}) S \vert 0 \rangle_{\rm in} \nonumber \\
& = & \frac{1}{2} \int \frac{d^3 k}{(2\pi)^3} \vert j(k) \vert^2 .
\label{E-brem-def}
\eeqa
This coincides with the classical energy of the classical radiated scalar field $\phi_{\rm class}$
\be
E_{\rm clas} = \frac{1}{2} \int d^3 x \left[ \left( \frac{\pl\phi_{\rm clas}}{\pl t} \right)^2
+ (\nabla \phi_{\rm clas})^2 \right] .
\ee
(We have seen in previous sections that the scalar field is nearly massless
in the nonlinear Earth background.)
This emphasizes the classicality of the emitted radiation.
Similarly, the number of emitted scalars is on average
\be
\bar{N} = \langle 0 \vert S^{-1} \hat{N} (\phi_{\rm in}) S \vert 0 \rangle_{\rm in} =
 \int \frac{d^3 k}{(2\pi)^3 2\omega_k} \vert j(k) \vert^2
\ee
and the probability of emitting $n$ scalars is given by a Poisson law of average $\bar N$.
Here, contrary to electrodynamics, the fact that the scalars are massive (even though the mass
is small) implies that there is no infrared catastrophe; i.e., there is not an infinite number
of emitted scalars of nearly vanishing energy from a source of finite energy.

We have seen in the previous section from Eqs.(\ref{Tpsi-f}) and (\ref{Tpsi-space})
that we have
\beq
\langle f \vert \hat{T}_{\psi} \vert f \rangle \sim - \frac{m_{\psi}^2}{\omega}
\delta_D({\vec x} - {\vec x}_f(t)) = - m_{\psi} \frac{d\tau}{dt} \delta_D({\vec x} - {\vec x}_f(t)) ,
\eeq
for a wave packet that has an energy peaked around $\omega$.
In the last equality we made the identification with the trace of the energy-momentum
tensor of a point particle ``$f$'' of mass $m_{\psi}$ and trajectory ${\vec x}_f(t)$,
using $\omega= p_f^0 = m_\psi \frac{dx_f^0}{d\tau}$ for a particle with proper time $\tau$.
This gives for the current $j(x)$ of Eq.(\ref{KG-Brem-1})
\beq
j(x) = - \frac{\beta m_{\psi}}{M_{\rm Pl}\sqrt{K'}} \int d\tau' \; \delta_D(x-x_f(\tau')) ,
\label{j-T-tau}
\eeq
and we recover for the classical field given by Eq.(\ref{phi-class-def})
the classical Lienard-Wiechert potential
\be
\phi_{\rm clas}(x) = \frac{\beta m_{\psi}}{4\pi M_{\rm Pl} \sqrt{K'}} \left[
\frac{1}{u_+ \cdot (x-x_+)} + \frac{1}{u_- \cdot (x-x_-)} \right ] ,
\label{phi-clas-x+}
\ee
where $x_{\pm}=x_f(\tau_{\pm})$ are the $x$-dependent retarded and advanced
points on the particle trajectory $x_{f}(\tau)$ such that
\beq
(x - x_{\pm})^2 = 0 , \;\;\; x^0_+ < x^0 , \;\;\; x^0_- > x^0 ,
\eeq
and $u_{\pm} = \frac{dx_f}{d\tau}(\tau_{\pm})$.
A particle with a constant velocity is such that one can always go to its rest frame globally.
In this case, one has $u^\mu_\pm= (1,0,0,0)$, and we find that $\phi_{\rm clas} = 0$;
i.e., there is no emitted energy by a particle in constant velocity motion.
As for standard bremsstrahlung, radiation only occurs when the particle accelerates or
decelerates.
Quantum mechanically, this implies that scalar particles are only created in this case.
This is consistent with the result that a free moving particle, with constant velocity,
is associated with the stationary scalar cloud (\ref{phi-cloud}) obtained in the previous 
section.
The ratio $m_{\psi}/M_{\rm Pl}$ in Eq.(\ref{phi-clas-x+}) is very small,
of the order of $10^{-22}$ for $m_{\psi}=m_e$, whereas electromagnetic 
bremsstrahlung comes with the dimensionless coupling $e \sim 1$ 
(in Lorentz-Heaviside units).
Thus, the dissipation associated with the emission of the K-mouflage scalar is negligible
as compared with the emission of photons, for accelerating charged fermions.

\subsubsection{Cross section $f\bar{f} \to \varphi \to f\bar{f}$}
\label{sec:cross-section}

We have seen that a single particle at constant speed does not radiate scalars in the terrestrial
environment of particle physics experiments and that
the classical field generated by this particle, which is static in the particle rest frame, is a small
perturbation to the large background created by the Earth.
In this context, the Lagrangian of the K-mouflage theories that we consider, i.e., the ones which
pass the tight tests of gravity in the Solar System, converges to a free scalar theory with a
normalized scalar $\phi$ and a rescaled coupling to matter $\beta_K= \beta/\sqrt {K'}$,
where $K'$ is a constant in this regime.

We still have to check that the classical description of the collision is valid. We have seen
in Eq.(\ref{isolated-cloud}) that at energies  such that $r=1/E$ is
$r \gtrsim  (\frac{m_{\psi}}{m_e})^{1/2} \,10^{-6} {\rm fm}$, which correspond to energies less
than $100$ TeV, i.e., energies of present and future particle experiments, the terrestrial background
dominates over the perturbation created by the particles. In the terrestrial background, we have
also seen that high-order quantum corrections are negligible.
Hence, we can consider that the fermion annihilation process takes place within this
classical terrestrial background. The canonically normalized scalar field $\phi$
of Eq.(\ref{KG-Tpsi}) is described by the free massless Lagrangian and the interaction
term (\ref{Lint-def}), where $\beta$ is renormalized to $\beta_K= \beta/\sqrt {K'}$
as in Eq.(\ref{KG-Tpsi}) and $K'$ is a large positive constant set by the Earth background.

The three-body annihilation $f \bar f\to \phi$ is forbidden by kinematic rules. Then, the
leading process is the tree-level scattering $f\bar{f} \to \phi \to f\bar{f}$,
the cross section of which reads
\be
\sigma = \left( \frac{\beta_K m_\psi}{M_{\rm Pl}} \right)^4  \frac{1}{256 \pi s}
\left( 1 - 10 \frac{m_{\psi}^2}{s} \right)^2 ,
\ee
where $s$ is the Mandelstam variable. As $m_\psi \ll M_{\rm Pl}$, this is a tiny scattering
cross section, and it does not lead to discrepancies with particle-physics data.

In conclusion, we have seen that the high-energy behavior of K-mouflage models that pass the
Solar System tests is essentially given, in the terrestrial background, by a free scalar-field
theory coupled to matter.
This is valid at energies where the scalar field created by the individual particles remains a
perturbation compared to terrestrial background.
At much higher energies, e.g. larger than $100$ TeV for electrons, a particle collision probes
regions of spacetime where the terrestrial background becomes negligible.
In this regime, the collisions probe well into the K-mouflage radius of
Eq.(\ref{RK-cloud}).
However, the scalar sector of the theory is well described by its classical Lagrangian inside
the classical radius $r \ll R_{\rm clas} $ with
\beqa
R_{\rm clas} & = & K_*^{-1/2} \chi_*^{-1/4}
\left( \frac{K_*}{K_2^2 \chi_*^2} \right)^{1/[8(\nu-1)]} \nonumber \\
&& \times \left( \frac{\beta m_\psi}{M_{\rm Pl}} \right)^{1/[4(\nu-1)]} R_K(m_{\psi}) ,
\eeqa
where we used the classicality criterion (\ref{classical-regime-2}), with $\bar{p} \sim r^{-1}$,
and the form (\ref{Ks-NL}) of the kinetic function in the highly nonlinear regime.
For the Arctan model (\ref{Arctan-def}), we obtain
\be
R_{\rm clas} \simeq \left( \frac{{m_\psi}}{m_e} \right)^{5/8} \; 10^{-6} \; {\rm fm} ,
\label{Rclass-Arctan-mpsi}
\ee
and for the Sqrt model (\ref{Sqrt-def}),
\be
R_{\rm clas} \simeq \left( \frac{m_\psi}{m_e} \right)^{7/12}
10^{-5} \; {\rm fm} ,
\label{Rclass-Sqrt-mpsi}
\ee
which are of the order or above the radius (\ref{isolated-cloud}).
This means that for electrons or quark-antiquark pairs in proton collisions, the classical 
regime is achieved before the distances probed by the collision are such that the 
terrestrial background becomes subdominant.
In this case, even at such very high energies, the fermion-antifermion annihilation is well
described by the free scalar-field Lagrangian coupled to matter.

\section{Conclusion}
\label{sec:conclusion}

In this work, we have considered K-mouflage models from a quantum field-theoretic point 
of view. In a traditional sense, these theories are not renormalizable as they involve 
any number of powers of the scalar kinetic terms, i.e., an arbitrary function of the 
kinetic terms. All these higher-order operators are not renormalizable, and one may worry 
that quantum corrections should alter the classical Lagrangian in an uncontrollable way. 
To have a better handle on this issue, we set up the renormalization program
for these theories and provide a recursive  algorithm to construct the renormalized 
effective action. We do this using background quantization around a solution of the 
classical equations of motion, which could be either a static or a cosmological 
time-dependent configuration for instance, and perturbation theory, where 
Feynman diagrams are regularized by dimensional regularization. 
In usual renormalizable theories, starting from a bare Lagrangian containing bare 
(and formally infinite) couplings, the renormalized action can be obtained in one step as 
the bare action contains counterterms that keep the same form as the bare Lagrangian
while exactly canceling the divergences obtained by calculating the quantum corrections 
using Feynman diagrams. 
For K-mouflage, this does not work similarly as the original classical action does not 
contain all the operators which are generated quantum mechanically. 
So, one must proceed stepwise by first producing a first set of counterterms to cancel all 
divergences induced quantum mechanically from the classical Lagrangian, and then 
recalculate the divergences that the new vertices brought by the counterterms entail. This 
generates a new set of counterterms, and {\it a priori} this recursive construction must be carried 
on indefinitely. Of course, there is no guarantee that the associated series of recursively 
constructed Lagrangians converges at all, and therefore one may cast a doubt on the validity 
of K-mouflage models. Despite all, one can confirm that the corrections to the classical 
Lagrangian always involve extra derivatives of the kinetic terms and that therefore the 
classical action is not renormalized \cite{deRham2014}. We then define a calculability 
criterion whereby all the quantum corrections in the renormalized effective action are 
negligible compared to the classical action. In this case and in this ``classical regime'', 
it is sufficient to consider K-mouflage theories at the classical level and quantum corrections 
can be safely neglected. We apply this criterion to healthy K-mouflage models, i.e., with 
no ghosts and no gradient instabilities, that pass the tight tests of gravity in the Solar System, 
and we show that in all astrophysical and cosmological situations of interest the classicality 
criterion is satisfied. Moreover, we find that for such theories the quantum regime is never 
reached even in the early inflationary Universe or in the very short distance regime on Earth.

These results rely on the assumption that perturbation theory is well defined around 
a given background, and this could be violated when the mass squared of the scalar 
becomes negative, because this may render the path integral ill defined. 
We examine two situations: the astrophysical one around a static spherical source 
and the cosmological one up to very early times. 
In the former, stability is guaranteed when the Sturm-Liouville problem associated with 
the linear perturbations around the static background has no negative eigenvalues. 
We confirm that this is indeed the case for the healthy models that pass the Solar System 
tests. 
The latter is more subtle as negative square masses for linear perturbations are 
a standard feature of cosmological perturbations, and we use a different criterion there. 
We impose that the extra instability induced by the negative mass squared due to the 
K-mouflage Lagrangian (and not only the $\frac{a''}{a}$ term of cosmological perturbations) 
does not lead to an explosion of the energy density of the K-mouflage field. Indeed, 
this would disrupt the evolution of the Universe and we confirm that this is far from being 
the case for the healthy models passing the solar tests. 

Despite the existence of the classicality regime, one may wonder what happens at very 
high energy way beyond the cutoff scale of the theory. Traditionally, in a top-down 
approach, one may advocate that some type of UV-completion must exist at high energy 
and that the K-mouflage models should emerge naturally from the renormalization 
group evolution. This approach was followed in particular in \cite{deRham2014}. 
Here, we find out that positivity and unitarity constraints imposed on scattering amplitudes 
\cite{Adams:2006aa} cannot be met by healthy models that pass the Solar System tests, 
and that therefore no such UV completion can be hoped to be constructed. 
This invalidates the approach of \cite{deRham2014} in our case, and one is left with the 
only prospect of having to deal with K-mouflage in a bottom-up approach and using 
the renormalization program that we have set in order to study high-energy properties 
of such models. This negative result might have invalidated the usefulness of 
K-mouflage models beyond the realm of very low-energy astrophysics and cosmology. 
To tackle this issue, we consider two relevant situations. The first one concerns 
the high-energy collision of scalars as may happen in colliders. In this case, we show that 
the collision occurs more and more in the classical regime as the energy increases. 
Hence, the K-mouflage models ``classicalize'' in this high-energy setting,
and we can always use the classical Lagrangian.
We also consider the interaction of the K-mouflage scalar with fermions.
We find that free fermions are dressed by a negligible classical scalar cloud.
Fermions also radiate scalars when they are accelerated, as for standard
bremsstrahlung, but this is again negligible as compared with the standard
electromagnetic bremsstrahlung because of the coupling factor $m_{\psi}/M_{\rm Pl}$.
In a similar fashion, the scattering $f\bar{f} \to \varphi \to f \bar{f}$, which corresponds to
the annihilation of a fermion pair into another one via an intermediate scalar,
comes with a factor $(m_{\psi}/M_{\rm Pl})^4$ that yields a negligible cross
section.
As for scalar collisions, this process can be described from the bare K-mouflage 
Lagrangian, both at low and high energies.

Hence, we have seen that healthy K-mouflage models that pass the stringent tests of gravity 
in the Solar System have remarkable properties quantum mechanically. 
They are not renormalized, and the finite corrections of higher order can be neglected 
in a ``classical regime'' which applies to astrophysical and cosmological systems of interest. 
Moreover, even pushed to high energy such as in collider experiments, the classicality 
criterion still applies, implying that trustworthy calculations for associated cross sections 
can be performed. K-mouflage models of the type considered here are most likely 
to be tested by cosmological and astrophysical means in the near future, and it is 
reassuring that such nonlinear models of dark energy/modified gravity can be simply 
used at the classical level.

\begin{acknowledgments}
This work is supported in part by the French Agence Nationale de la Recherche under 
Grant ANR-12-BS05-0002. 
\end{acknowledgments}

\appendix

\section{One-loop contribution to the vacuum energy density}
\label{app:one-loop}

In this Appendix, we derive the one-loop contribution to the vacuum energy due to a massive
scalar $u$ with the action
\be
S = \frac{1}{2} \int d^4x \left[ \left(\frac{\partial u}{\partial t}\right)^2 - (\nabla u)^2
- m^2 u^2 \right] .
\label{Su-def}
\ee
This is the analog of the cosmological quadratic action (\ref{S-tau-V}), but here we work
in Minkowski space with a constant mass $m$.
As in Eq.(\ref{v-vk-def}), the quantum field can be expanded in modes
\be
\hat{u}(t,{\vec x}) = \int \frac{d^3k}{(2\pi)^{3/2}} \left[ u_k(t) e^{\ii{\vec k}\cdot{\vec x}}
\hat{c}_{\vec k} + u_k^*(t) e^{-\ii{\vec k}\cdot{\vec x}} \hat{c}^\dagger_{\vec k} \right]
\label{u-uk-def}
\ee
with the commutation rules $[\hat{c}_{\vec k}, \hat{c}^\dagger _{\vec p}]=
\delta_D( \vec k -\vec p)$.
In Minkowski space with a constant mass, the mode functions are simply
\be
u_k(t) = \frac{e^{-\ii\omega_k t}}{\sqrt{2\omega_k}} \;\;\; \mbox{with} \;\;\;
\omega_k= \sqrt{{\vec k}^2+m^2} .
\label{uk-t}
\ee
The energy-momentum tensor of the scalar field is
\be
T_{\mu\nu} = \partial_{\mu}u\partial_{\nu}u - \eta_{\mu\nu} \left[
\frac{1}{2} \partial^{\alpha}u\partial_{\alpha}u+\frac{1}{2} m^2 u^2 \right] ,
\ee
which gives for the $T_{00}$ component
\be
T_{00} = \frac{1}{2} \dot{u}^2 + \frac{1}{2} (\nabla u)^2 + \frac{1}{2} m^2 u^2 ,
\ee
where we note $\dot u=\partial u/\partial t$, and for the trace of the energy-momentum
tensor,
\be
T \equiv T^{\mu}_{\mu} = \dot{u}^2 -(\nabla u)^2 - 2 m^2 u^2 .
\ee
Using the mode expansion (\ref{u-uk-def})-(\ref{uk-t}), we obtain for the vacuum
energy density $\rho_v$ \cite{Martin2012,Akhmedov2002,Koksma2011},
\be
\rho_v \equiv \langle 0 \vert \hat{T}_{00} \vert 0 \rangle = \int \frac{d^3k}{(2\pi)^3}
\; \frac{\omega_k}{2} ,
\label{rho-V}
\ee
for the trace $T_v$,
\be
T_v \equiv \langle 0 \vert \hat{T} \vert 0 \rangle = - \int \frac{d^3k}{(2\pi)^3}
\; \frac{m^2}{2\omega_k} ,
\label{T-V}
\ee
and for the pressure $p_v$,
\be
p_v \equiv \frac{\rho_v+T_v}{3} = \int \frac{d^3k}{(2\pi)^3}
\; \frac{\vec k^2}{6\omega_k} .
\label{p-V}
\ee

It is interesting to note that these vacuum energy density and pressure can be related
to the Feynman propagator evaluated at coincident points \cite{Martin2012}.
Indeed, using again the mode expansion (\ref{u-uk-def})-(\ref{uk-t}) and denoting
$T\{...\}$ the time-ordered product, we have \cite{Schwartz2013}
\beqa
G_F(x_1 - x_2) & \equiv & \langle 0 \vert T \{ u(x_1) u(x_2) \} \vert 0 \rangle \nonumber \\
& = & i\Delta_F(x_1-x_2) \\
& = & \lim_{\varepsilon\rightarrow 0^+} \int \frac{d^4k}{(2\pi)^4} \;
\frac{-\ii \; e^{\ii k \cdot (x_1-x_2)}}{k^2+m^2-\ii\varepsilon} , \;\;\;
\label{GF-def}
\eeqa
where $k^2\equiv -\omega^2+{\vec k}^2$ and $\omega=k^0$.
This gives
\be
G_F(0) = \int \frac{d^4k}{(2\pi)^4} \; \frac{-\ii}{k^2+m^2-\ii\varepsilon}
= \int \frac{d^3k}{(2\pi)^3} \; \frac{1}{2\omega_k} ,
\label{GF-0-def}
\ee
where in the second equality we used the residue theorem and we let the limit
on $\varepsilon$ be implicit.
In a similar fashion, we write
\be
\rho_v = \int \frac{d^4k}{(2\pi)^4} \; \frac{-\ii \omega^2}{k^2+m^2-\ii\varepsilon} ,
\ee
\be
p_v = \int \frac{d^4k}{(2\pi)^4} \; \frac{-\ii {\vec k}^2/3}{k^2+m^2-\ii\varepsilon} ,
\ee
\be
T_v = - m^2 G_F(0)
= \int \frac{d^4k}{(2\pi)^4} \; \frac{\ii m^2}{k^2+m^2-\ii\varepsilon} .
\ee
Performing a Wick rotation, we find
\be
p_v = - \rho_v ,
\label{pv-rhov}
\ee
as expected for the equation of state of the vacuum. Using $T_v = - m^2 G_F(0)$, this
yields
\be
\rho_v = \frac{m^2}{4} G_F(0) .
\label{rhov-GF0}
\ee
In terms of Feynman diagrams, Eq.(\ref{rhov-GF0}) also shows that the vacuum energy
density can be written in terms of bubble diagrams (here at one-loop order).

All these quantities diverge at large $k$ and must be regularized. As is well known
\cite{Martin2012,Akhmedov2002},
introducing a high-energy cutoff, $k<\Lambda_c$, leads to incorrect results because
it breaks the symmetries of the system (Lorentz invariance). Therefore, we use
dimensional regularization \cite{Hooft1972,Peskin1995}.
Working with $G_F(0)$ and the first equality (\ref{GF-0-def}), we write
\be
G_F(0) = \mu^{4-d} \int \frac{d^d k_E}{(2\pi)^d} \frac{1}{k_E^2+m^2} ,
\ee
where we performed a Wick rotation to Euclidean space and we introduced the
sliding scale $\mu$ to keep $G_F(0)$ dimensionally correct.
This gives
\be
G_F(0) = \frac{m^2}{16\pi^2} \left( \frac{4\pi\mu^2}{m^2} \right)^{\epsilon/2}
\Gamma(-1+\epsilon/2) ,
\ee
where we introduced $\epsilon=4-d$, and the expansion around $\epsilon=0$
gives
\be
G_F(0) = - \frac{m^2}{16\pi^2} \left[ \frac{2}{\epsilon} + \ln(4\pi) - \gamma + 1
- \ln(m^2/\mu^2) + ... \right] ,
\ee
where $\gamma \simeq 0.577$ is the Euler-Mascheroni constant.
From Eq.(\ref{rhov-GF0}), this also gives
\be
\rho_v = - \frac{m^4}{64\pi^2} \left[ \frac{2}{\epsilon} + \ln(4\pi) - \gamma + 1
- \ln(m^2/\mu^2) + ... \right] .
\ee
Using a $\overline{\rm MS}$ renormalization scheme, where we subtract the
$2/\epsilon$ term together with the accompanying constant terms, we obtain
the renormalized vacuum energy density
\cite{Martin2012,Koksma2011}
\be
\rho_v^{\rm renorm} = \frac{m^4}{64\pi^2} \ln \left( \frac{m^2}{\mu^2} \right) .
\label{rhov-renorm}
\ee
The divergence has been cancelled by introducing a counterterm in the action
(a pure cosmological constant),
\be
\Delta\rho_v = \frac{m^4}{64\pi^2}  \left[ \frac{2}{\epsilon} + \ln(4\pi) - \gamma
+ 1 \right] .
\label{rhov-counterterm}
\ee
The vacuum pressure follows from (\ref{pv-rhov}), which leads to
$p_v^{\rm renorm} = - \rho_v^{\rm renorm}$,
and shows that the vacuum energy behaves like a cosmological constant of equation
of state $-1$.
From Eq.(\ref{rhov-renorm}), we also recover the one-loop effective action
(\ref{Gamma-1-1loop}), since for a time- and scale-independent system we have
$\Gamma=-E=-TV\rho_v$ \cite{Peskin1995}.

\section{Matter loops}
\label{sec:matter-loops}

Matter loops can also contribute to the effective action. We show here that they can 
be neglected. Let us consider the insertion of a matter loop on a scalar propagator 
corresponding to the only cubic vertex between one scalar and two fermions. 
In any of the integrals involved in the evaluation of the effective action, this implies 
the replacement
\be
\frac{1}{p^2 + m_\phi^2} \to \frac{1}{(p^2 +m_\phi^2)^2} F(p^2) ,
\ee
where the fermion loop is simply
\be
F(p^2) = \left( \frac{\beta m_\psi}{M_{\rm Pl} \sqrt{K'}} \right)^{\! 2} 
\! \int \!\!  \frac{d^4 q}{(2\pi)^4}
\frac{{\rm Tr} [ ( \slashed{p}+\slashed{q}+m_\psi ) ( \slashed{q}+m_\psi ) ]}
{[ (p+q)^2-m_\psi^2] [q^2-m_\psi^2]}
\ee
which is proportional to $p^2$ in dimensional regularization.
As a result, the fermion loop does not change the UV behavior of the scalar propagator 
and involves a tiny correction of order $(\frac{\beta m_\psi}{M_{\rm Pl}})^2\ll 1$ which can 
be neglected. In the main text, we have neglected the contributions from the fermion loops
to the effective action.

\bibliography{ref1}   

\end{document}